\documentclass[twoside,12pt]{article}
\usepackage{epsfig}
\usepackage{amsmath,amssymb}

\def\Journal#1#2#3#4{{#1} {#2} (#4) #3 }
\def\NCA{{\em Nuovo Cimento} A}

\def\NPA{{\em Nucl. Phys.} A}

\def\NPB{{\em Nucl. Phys.} B}

\def\PLB{{\em Phys. Lett.} B}

\def\PRL{\em Phys. Rev. Lett.}
\def\PREV{\em Phys. Rev.}
\def\PREP{\em Phys. Rep.}

\def\PRD{{\em Phys. Rev.} D}
\def\PRC{{\em Phys. Rev.} C}

\def\RMP{{\em Rev. Mod. Phys.}}

\def\INT{{\em Int. J. Mod. Phys.} E}

\newcommand{\be}{\begin{equation}}
\newcommand{\ee}{\end{equation}}
\newcommand{\bea}{\begin{eqnarray}}
\newcommand{\eea}{\end{eqnarray}}

\begin{document}

\title{ \vspace{1cm} Double Beta Decay Experiments at Canfranc Underground Laboratory}
\author{S.\ Cebri\'{a}n$^{1,2}$\\
\\
$^1$Centro de Astropart\'{i}culas y F\'{i}sica de Altas Energ\'{i}as, Facultad de Ciencias,\\Universidad de Zaragoza, 50009 Zaragoza, Spain\\
$^2$Laboratorio Subterr\'{a}neo de Canfranc, 22880 Canfranc Estaci\'{o}n, Spain}
\maketitle
\begin{abstract}
The first activities of the Canfranc Underground Laboratory (``Laboratorio Subterr\'{a}neo de Canfanc'', LSC) started in the mid-eighties in a railway tunnel located under the Spanish Pyrenees; since then, it has become an international multidisciplinary facility equipped with different services for underground science. The research activity at LSC is about Astroparticle Physics, dark matter searches and neutrino Physics; but also activities in Nuclear Astrophysics, Geophysics, and Biology are carried out. The investigation of the neutrinoless double beta decay has been one of the main research lines of LSC since the beginning. Many unknowns remain in the characterization of the basic neutrino properties and the study of this rare decay process requiring Physics beyond the Standard Model of Particle Physics can shed light on the lepton number conservation, the nature of the neutrinos as Dirac or Majorana particles and the absolute scale and ordering of the masses of the three generations. Here, the double beta decay searches performed at LSC for different emitters and following very different experimental approaches will be reviewed: from the very first experiments in the laboratory including the successful IGEX (``International Germanium EXperiment'') for $^{76}$Ge, which released very stringent limits to the effective neutrino mass at the time, to the present NEXT experiment (``Neutrino Experiment with a Xenon Time-Projection Chamber'') for $^{136}$Xe and future project CROSS (``Cryogenic Rare-event Observatory with Surface Sensitivity'') for $^{130}$Te and $^{100}$Mo, both implementing innovative detector technologies to discriminate backgrounds. For the neutrinoless double beta decay channel and at 90\% C.L., IGEX derived a limit to the half-life of $^{76}$Ge of $T_{1/2}^{0\nu}>1.57 \times 10^{25}$~y while the corresponding expected limits are $T_{1/2}^{0\nu}>1.0\times 10^{26}$~y for $^{136}$Xe from NEXT-100 (for an exposure of 500~kg$\cdot$y) and $T_{1/2}^{0\nu}>2.8 \times 10^{25}$~y for $^{100}$Mo from CROSS (for 5~y and 4.7~kg of isotope). Activities related to double beta decays searches carried out in other underground laboratories have also been developed at LSC and will be presented too, like the operation of the BiPo-3 detector for radiopurity measurements of thin sheets with very high sensitivity. For each one of these experiments, the concept, the experimental set-ups and relevant results will be discussed.
\end{abstract}
\tableofcontents
\section{Introduction} \label{intro}

Even after the impressive confirmation of the non-zero mass of neutrinos following the observation of flavor oscillations in neutrinos from different sources, neutrino Physics is a hot and exciting research area since many questions on the basic features of neutrinos are still open \cite{nupdg}. Double beta decay (DBD) is a rare nuclear process that some nuclei can undergo where a nucleus changes into an isobar with the emission of two electrons and two antineutrinos; although rare, this process has been observed for several nuclei. A non-standard version of this decay without the emission of antineutrinos has been proposed and great efforts are being devoted to its observation due to the outstanding implications of its occurrence, which can inform on some of the pending questions in the field of neutrino Physics: violation of leptonic number, confirmation of the Majorana nature of neutrinos, that is, if neutrinos and antineutrinos are the same particle, and even an estimate of the neutrino mass scale and CP phases. Processes of DBD with the emission of positrons or by electron capture are also possible and are being investigated.

The Canfranc Underground Laboratory, in Spain, joined since the very beginning of its history the worldwide efforts to detect the neutrinoless DBD. The first experiments carried out in the Somport railway tunnel connecting Spain and France, not in use, were followed by the operation of the IGEX experiment in improved facilities using High Purity Germanium (HPGe) detectors enriched in $^{76}$Ge. After the consolidation of the laboratory as an international multidisciplinary facility, the NEXT experiment started the operation of a high-pressure xenon gas Time Projection Chamber (TPC) with electroluminescent amplification to investigate the decay of $^{136}$Xe. More recently, the CROSS project has become an approved experiment to install in Canfranc arrays of TeO$_{2}$ and Li$_{2}$MoO$_{4}$ bolometers enriched in $^{130}$Te and $^{100}$Mo, respectively. In addition, the BiPo-3 detector was built and is being operated in Canfranc for radiopurity measurements of very thin sheets achieving very high sensitivity; it was conceived and has been mainly used to screen the DBD source foils of the SuperNEMO DBD experiment operating in the Modane Underground Laboratory, in France.

The goal of this paper is to review all these DBD experiments developed at the Canfranc Underground Laboratory. For each one of them, the concept and motivation will be firstly presented, the experimental set-ups will be described and the relevant results obtained will be discussed. The article is organized as follows: the basic concepts related to the DBD process are summarized in Sec.~\ref{secdbd} and the main features of the Canfranc Underground Laboratory described in Sec.~\ref{seclsc}; then, the first DBD experiments performed in the laboratory are commented in Sec.~\ref{firstdbd} while Secs.~\ref{igex}, \ref{nextsec}, \ref{crosssec} and \ref{biposec} present in detail the IGEX, NEXT and CROSS experiments and the BiPo-3 detector, respectively; a brief summary and outlook is given in Sec.~\ref{sumsec} to close the review.

\section{Double Beta Decay} \label{secdbd}

This section is intended to present the motivation and the status of DBD searches in a general context. In particular, DBD processes will be firstly described, emphasizing their connection with the neutrino mass problem. Then, the experimental work to investigate this nuclear process will also be briefly revised, describing their stringent requirements and comparing the different experimental approaches and techniques followed.

The Standard Model of Particle Physics includes three species or ``flavors'' of neutrinos, linked to electrons, muons and tauons; in addition, neutrinos are considered massless and different than antineutrinos. However, many different experiments studying neutrinos from different sources have confirmed in the last twenty years the non-zero mass of neutrinos thanks to the observation of the so-called neutrino oscillations: neutrinos having a certain flavor change into another species during their propagation. The oscillations have been observed and characterized for neutrinos produced in the atmosphere \cite{skresult,antaresresult,icecuberesult}, coming from the Sun \cite{snoresult}, emitted in nuclear reactors in power plants \cite{kamlandresult,doublechoozresult,dayabayresult,renoresult} or generated in accelerators \cite{k2kresult,t2kresult,operaresult}. Oscillations are explored either in terms of neutrino disappearance or in terms of the appearance of a different neutrino flavor. This effect can only be explained if neutrinos are massive particles and mixing among the three mass eigenstates ($m_{j}$, for $j=$1-3) happens \cite{vogelnu,mohapatrarep,concharep,zubernu}. The Pontecorvo-Maki-Nakagawa-Sakata matrix (PMNS) describes the mixing of neutrinos and is parameterized by three angles $\theta_{12}$, $\theta_{23}$ and $\theta_{13}$ and one/three CP violating phases for Dirac/Majorana neutrinos:
\be
{U_{i,j}}=\left(
\begin{array}{ccc}
c_{12}c_{13}	& s_{12}c_{13} 	 & s_{13}e^{-i\delta} \\
-s_{12}c_{23}-c_{12}s_{23}s_{13}e^{i\delta} 	& c_{12}c_{23}-s_{12}s_{23}s_{13}e^{i\delta}	 & s_{23}c_{13} \\
s_{12}s_{23}-c_{12}c_{23}s_{13}e^{i\delta}		& -c_{12}s_{23}-s_{12}c_{23}s_{13}e^{i\delta}	 & c_{23}c_{13}
\end{array}
\right)
\times \rm{diag} (1, e^{i\alpha_{1}/2}, e^{i\alpha_{2}/2}),
\label{eqPMNS}
\ee

\noindent where $c_{ij} \equiv \cos\theta_{ij}$, $s_{ij} \equiv \sin\theta_{ij}$, $\delta$ is the Dirac CP violation phase and $\alpha_1$ and $\alpha_2$ are two Majorana CP violation phases. Oscillation experiments have constrained neutrino square mass differences of the three eigenstates $(m_{i}^{2}-m_{j}^{2})$ and most of the PMNS mixing parameters within rather narrow bands. Table \ref{pdgdata} summarizes the results for these parameters as compiled by the Particle Data Group \cite{nupdg}. The measured square mass differences have proven that one neutrino state is much more split than the other two; the missing of the sign of $\Delta m^{2}_{32}$ allows for two different mass orderings, the so-called direct hierarchy ($m_{1} \lesssim m_{2} \ll m_{3}$) and the inverted hierarchy ($m_{3} \ll  m_{1} \lesssim m_{2}$). The global analysis of all the available data, including oscillation results and cosmological data, gives an evidence of the preference for the normal neutrinos mass ordering versus the inverted scenario, with a significance of 3.5 standard deviations \cite{nuordering}.

\begin{table}
\begin{center}
\begin{minipage}[t]{16.5 cm}
\label{pdgdata}
\caption{Results for measured square mass differences and mixing angles of neutrinos from~\cite{nupdg}. The convention with $m_2>m_1$ and $m_3$ the most split state is followed. For $\Delta m^2_{32}$ and $\theta_{23}$ different values for normal (NH) or inverted hierarchy (IH) are presented.}
\end{minipage}
\vskip 0.5 cm
{\begin{tabular}{@{}l|c@{}}
\hline
$\Delta m^{2}_{21}$ (eV$^2$) & $(7.53\pm0.18)\times$10$^{-5}$ \\
$\Delta m^{2}_{32}$ (eV$^2$) &  $(2.51\pm0.05)\times$10$^{-3}$ for NH\\
&   $-(2.56\pm0.04)\times$10$^{-3}$ for IH\\
$\sin^{2}\theta_{12}$ &   $0.307^{+0.013}_{-0.012}$ \\
$\sin^{2}\theta_{23}$ &  ($0.417^{+0.025}_{-0.028}$) for NH\\
& ($0.421^{+0.033}_{-0.025}$) for IH \\
$\sin^{2}\theta_{13}$ &  $ (2.12\pm0.08)\times10^{-2}$ \\

\hline
\end{tabular}}
\end{center}
\end{table}

Neutrino detection is very hard, due to the extremely low interaction cross sections; then, a great effort had to be devoted in the last decades to study the intrinsic properties of neutrinos. But being very elusive particles, neutrinos are excellent surveys of phenomena like supernovae or gamma ray bursts, processes in the Sun or even cosmological studies \cite{kimnu,mohapatranu,pastorrep,vallenu}. The results of oscillation experiments are outstanding for the determination of some neutrino properties \cite{nuoscana}. But in order to fix the absolute scale of neutrino masses, other kind of experiments are needed and only upper limits have been derived up to date. Neutrino mass could be directly estimated analyzing the shape of the end of the beta spectrum in nuclei with a low transition energy like tritium. The use of low temperature detectors for calorimetric measurements of the beta spectrum is being also considered. The KATRIN experiment operating in Karlsruhe (Germany) has presented an improved limit on the neutrino mass from a direct kinematic method using the first four-week science run made in spring 2019. Beta-decay electrons from a high-purity gaseous molecular tritium source have been energy analyzed by a high-resolution filter. The upper bound for the observable related to the electronic neutrino mass is of 1.1~eV (90\% C.L.) \cite{katrin2019}. Other limits come from the analysis of supernovae emissions or from cosmological observations like those of the Planck satellite, by fitting the observations to complex cosmological and astrophysical models; the physical quantity probed by cosmological surveys is the sum of the masses of the three light neutrinos. In this article the focus will be on DBD, which can provide information not only on the absolute mass scale but also on the neutrino nature (Dirac or Majorana). The general view of the investigation of DBD can be obtained from \cite{libroklapdor} and from detailed reviews on the topic like those of Refs.~\cite{reviewfaessler,suhonen98,reviewelliot,reviewejiri,reviewavignone,vergados,reviewgiuliani,reviewgomez,reviewcremonesi,revieworo,reviewdolinski,appec}. Theoretical and experimental aspects specifically for the two-neutrino DBD are reviewed in \cite{reviewsaakyan}.

\subsection{The Double Beta Decay process}

Double beta decay is a second-order standard weak process, where a nucleus (with mass number $A$ and atomic number $Z$) changes into an isobar with the spontaneous and simultaneous emission of two electrons and two antineutrinos, keeping the conservation of the leptonic number:
\be
(2\beta^{-})_{2\nu}: (A,Z)\rightarrow (A,Z+2) + 2 e^{-} + 2\overline{\nu}_{e} \;. \label{eqdbdnu}
\ee

\begin{figure}[tb]
\centerline{\includegraphics{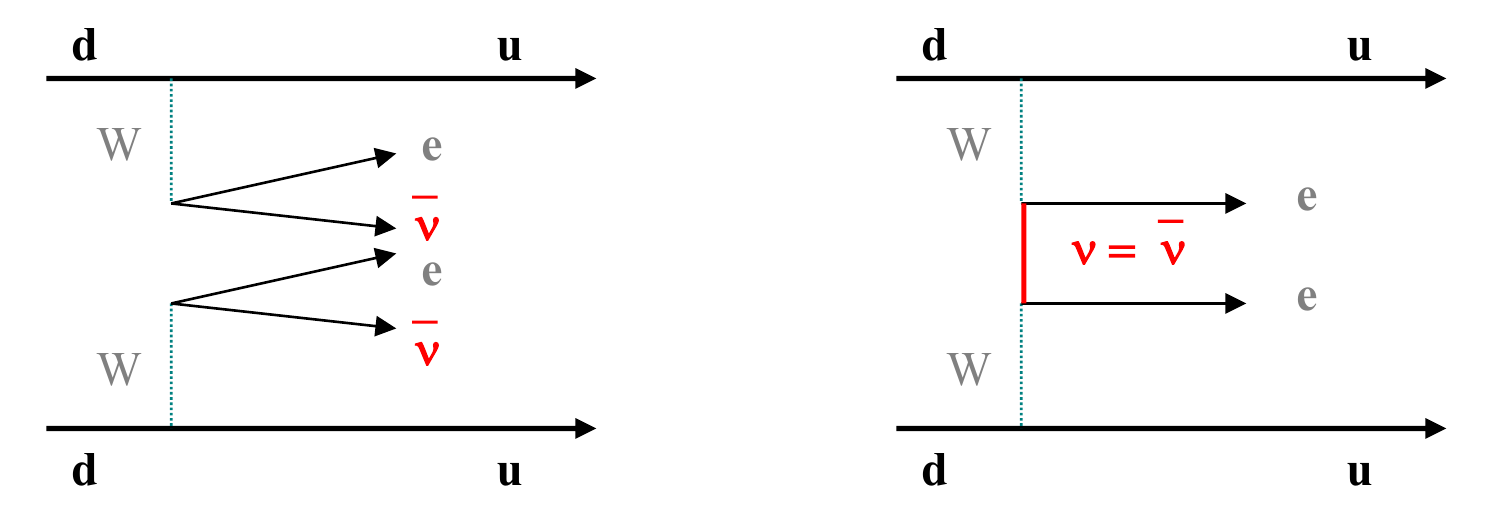}}
\begin{center}
\begin{minipage}[t]{16.5 cm}
\caption{Representation at the quark level of the DBD process with emission of antineutrinos (left) and without this emission (right) due to the exchange of massive Majorana neutrinos.\label{doblebeta}}
\end{minipage}
\end{center}
\end{figure}

This decay is detectable for nuclei (A,Z) with an even number of both protons and neutrons for which the beta transition to the isobar (A,Z+1) is energetically forbidden or at least strongly suppressed by a big change of angular momentum. At the nucleon level, the DBD corresponds to the conversion of two neutrons into protons, and at the quark level, two quarks d change into quarks u. Figure~\ref{doblebeta} (left) shows the DBD at this quark level. But in addition to this standard channel, a process without the emission of antineutrinos has been proposed admitting the violation of the leptonic number, as shown in Fig.~\ref{doblebeta} (right):
\be
(2\beta^{-})_{0\nu}: (A,Z)\rightarrow (A,Z+2) + 2 e^{-} \;. \label{eqdbd}
\ee

The DBD with emission of neutrinos has been observed using different techniques for several nuclei: $^{48}$Ca, $^{76}$Ge, $^{82}$Se, $^{96}$Zr, $^{100}$Mo, $^{116}$Cd, $^{128}$Te, $^{130}$Te, $^{136}$Xe, $^{150}$Nd and $^{238}$U. Table~\ref{dbdinfo} shows the natural isotopic abundance and transition energy of some of these DBD emitters. Measured half-lives vary from approximately $10^{19}$ to $10^{24}$~y (see recommended values at \cite{barabash19}). No evidence has been reported up to now for the neutrinoless DBD, being currently the most stringent half-life limits of the order of 10$^{25}$–10$^{26}$~y.

\begin{table}
\begin{center}
\begin{minipage}[t]{16.5 cm}
\caption{Natural isotopic abundance and transition energy $Q$ for some of the observed emitters of DBD with two neutrinos, as in Ref. \cite{revieworo}.}
\label{dbdinfo}
\end{minipage}
\vskip 0.5 cm
\begin{tabular}{l|cc}
\hline
Isotope & Isotopic abundance (\%) &  $Q$ (MeV) \\ \hline
$^{48}$Ca &0.187& 4.263 \\
$^{76}$Ge &7.8 &2.039\\
$^{82}$Se &9.2 &2.998\\
$^{96}$Zr &2.8 &3.348\\
$^{100}$Mo& 9.6 &3.035\\
$^{116}$Cd& 7.6 &2.813\\
$^{130}$Te& 34.08& 2.527\\
$^{136}$Xe& 8.9 &2.459\\
$^{150}$Nd& 5.6 &3.371\\ \hline
\end{tabular}
\end{center}
\end{table}

\begin{figure}[tb]
\centerline{\includegraphics[width=10cm]{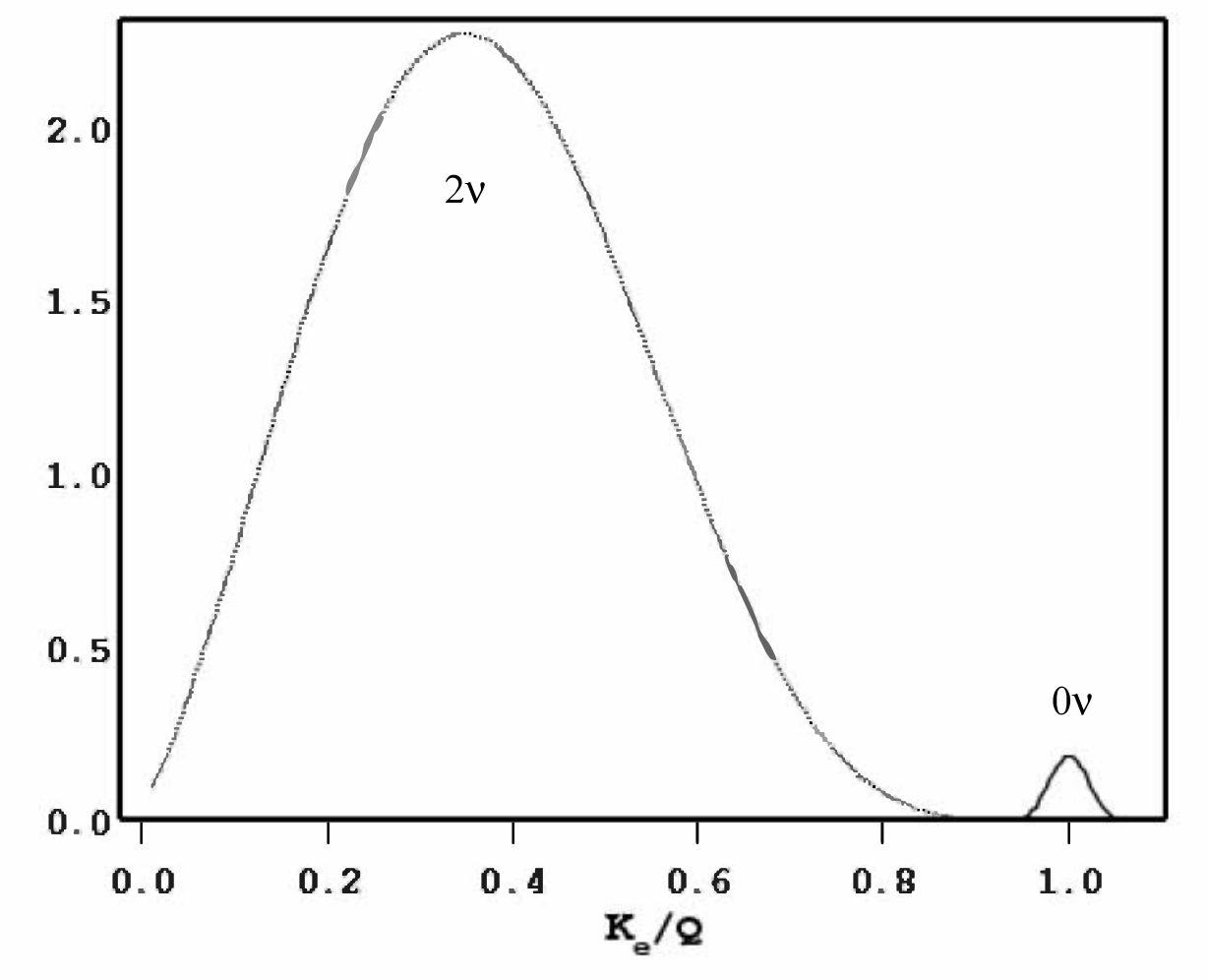}}
\begin{center}
\begin{minipage}[t]{16.5 cm}
\caption{Shape of the spectrum of the kinetic energy $K_{e}$ of the two emitted electrons (divided by the transition energy $Q$) for the two channels of DBD. The one with the emission of neutrinos is continuous up to $Q$ while the neutrinoless process should appear as a peak at $Q$. The energy resolution of the detection system determine the width of the neutrinoless DBD peak.\label{espbeta}}
\end{minipage}
\end{center}
\end{figure}

An important difference between the two DBD channels indicated in Eqs.~\ref{eqdbdnu} and \ref{eqdbd} is the energy spectrum of the emitted electrons. In the neutrinoless process, the two electrons share all the transition energy $Q$, since they can be considered as a single particle (two-body decay) and the recoil of the daughter nucleus is negligible. As shown in Fig.~\ref{espbeta}, the spectrum of the sum energy of the electrons in this case is a peak on the transition energy, while when the antineutrinos are emitted the energy spectrum of electrons is continuous, due to the additional particles in the final state. This difference, as commented later on, is important for the direct detection of both processes.

The observation of a neutrinoless DBD transition would be an evidence of an explicit violation of the number of leptons. This would support that leptons were relevant in the creation of the matter-antimatter asymmetry in the universe \cite{leptogenesis}. No process where the number of leptons (nor the number of baryons) varies has been observed yet, suggesting that the lepton $L$ and the baryon $B$ numbers follow conservation  laws, even if there are not fundamental arguments for that \cite{Bnumber}. The status of the investigations on the total lepton number in the Standard Model and some extensions has been reviewed in \cite{revieworo}. The relevance of observing the DBD without the emission of neutrinos comes also from the fact that it would be an evidence that neutrinos are Majorana particles with a non-zero mass. In extensions of the Standard Model of Particle Physics, several  mechanisms have been proposed for the occurrence of neutrinoless DBD: through heavy right-handed neutrino exchange, SUSY particle processes, light sterile neutrinos of Majorana type,\dots); but the simplest one is based on the exchanging of a light Majorana neutrino (see complete reviews on Refs.~\cite{werner,vergados,pas}). The antineutrino emitted in the transformation of a first neutron can be absorbed by the second neutron only as a neutrino (which requires they coincide, see Fig. \ref{doblebeta} (right)). In addition, also chirality has to be adjusted; this implies the existence of a mass term and/or the existence of non-standard V+A interactions in the weak hamiltonian. In any case, it has been shown that whatever the mechanism is, the existence of neutrinoless DBD implies the existence of a Majorana mass term for the neutrino \cite{valle}. These are the reasons why the searching for the neutrinoless DBD allows to explore the Physics beyond the Standard Model. The particular mechanism occurring in neutrinoless DBD could be identified by measuring the angular distribution of the emitted electrons or the spectrum of each of these electrons.

Historically, the DBD with emission of neutrinos was already proposed in 1935 by M.~Goeppert-Mayer \cite{goeppert}, just some years after the proposal of the existence of neutrinos and the development of the Fermi theory of the beta decay. In 1939, W.H.~Furry \cite{farry} proposed neutrinoless DBD following the ideas developed by G.~Racah \cite{racah} on the symmetry between particle and antiparticle. In the eighties, a third channel was proposed considering the emission together with the electrons of one or more neutral bosons, like a Majoron $M$, a Goldstone boson coupled to the exchanged virtual neutrino, which appears in the spontaneous breaking of the symmetry generating the neutrino mass:
\be
    (2\beta^{-})_{M}: (A,Z)\rightarrow (A,Z+2) + 2e^{-} + M \;. \label{eqdbdm}
\ee
In this case, having four bodies in the final state, the energy spectrum of electrons is again continuous as in the standard channel, but with the maximum shifted to the right.

DBD processes in which electrons are emitted, as those presented in Eqs.~\ref{eqdbdnu}-\ref{eqdbdm}, are the most commonly studied; but, as in ordinary beta decay with the emission of electrons ($\beta^{-}$), positrons ($\beta^{+}$) or electron capture ($EC$), DBD can also happen in other ways where the nuclear charge is diminished by two units provided the atomic masses of the nuclei $M(A,Z)$ satisfy the required conditions (taking into account the electron mass and binding energy):
\bea
(2\beta^{+})_{2\nu}: (A,Z)\rightarrow (A,Z-2) + 2e^{+} + 2\nu_{e}  \\
M(A,Z)>M(A,Z-2)+4m_{e} \;,
\eea
\bea
(EC\beta^{+})_{2\nu}: (A,Z)+e^{-} \rightarrow (A,Z-2) + e^{+} + 2\nu_{e} \\
M(A,Z)>M(A,Z-2)+2m_{e} +E_{{\mathrm{binding~e^{-}}}} \;,
\eea
\bea
(2EC)_{2\nu}: (A,Z)+2e^{-} \rightarrow(A,Z-2) + 2\nu_{e} \\
M(A,Z)>M(A,Z-2)+ 2E_{{\mathrm{binding~e^{-}}}} \;.
\eea
These processes having two neutrino emitted observe electronic lepton number conservation ($\Delta L=0$) and they are allowed in the standard weak theory with very long lifetimes. The first observation of two-neutrino double electron capture in $^{124}$Xe was recently reported by the XENON1T experiment \cite{xenon1tECEC}. The corresponding processes without neutrino emission could also be possible beyond the Standard Model \cite{neutrinolessECEC}. It is worth noting that the double electron capture is strongly suppressed respect to the other processes. Limits have been set for the neutrinoless double electron capture of $^{36}$Ar by the GERDA experiment \cite{gerdaECAr} and for various isotopes and processes from germanium gamma spectroscopy of different materials \cite{tgv,jeskov,belliECEC1,belliECEC2,belliECEC3,belliECEC4}.

Typically, DBD occurs between ground states; as in even-even nuclei these states have spin and parity $0^{+}$, the transition $0^{+}\rightarrow0^{+}$ is always possible, but transitions to low-lying excited states can also occur if they are kinematically allowed. The transitions to $2^{+}$ states have the distinctive feature that they can only be engendered by right-handed currents. The disadvantage to detect transitions to excited states is that rates are even lower, but on the other hand, they can be better identified thanks to the gamma emissions in coincidence. This kind of processes has been observed for $^{100}$Mo and $^{150}$Nd for DBD with two-neutrino emission to the first $0^{+}$ excited state of the daughters nuclei. Reference~\cite{barabashexc} presents a review of DBD to excited states.

\subsection{Decay rates and relationship with neutrino mass}

The theoretical calculation of decay rates of DBD emitters is in some way analogous to that of ordinary beta emitters, based on the Fermi Golden Rule \cite{vogelnu,kimnu,suhonen98}. The transition probability, inversely proportional to the half-life $ T_{1/2}$, is typically written for the DBD with emission of neutrinos as:
\be
    (T_{1/2}^{2\nu})^{-1}=G_{2\nu}|M^{2\nu}_{GT}|^{2},  \label{t2nu}
\ee
being $G_{2\nu}$ the exactly calculable phase space integral containing all the relevant constants and $M^{2\nu}_{GT}$ the Gamow-Teller nuclear matrix element. For the neutrinoless channel, which has a larger phase space, the corresponding expression (considering only the mass term) is:
\be
    (T_{1/2}^{0\nu})^{-1}=G_{0\nu}|M^{0\nu}|^{2} m_{\beta\beta}^{2},  \label{t0nu}
\ee
with $G_{0\nu}|M^{0\nu}|^{2}=F_{N}$ the so-called nuclear factor-of-merit, being $G_{0\nu}$ the integral of the phase space for this process and

\be
M^{0\nu}=M^{0\nu}_{GT} - (g_{V}/g_{A})^{2} M^{0\nu}_{F}, \label{M0nu}
\ee
with $M^{0\nu}_{GT}$ and $M^{0\nu}_{F}$ the Gamow-Teller and Fermi nuclear matrix elements, respectively, and $g_{V}$ and $g_{A}$ the vectorial and axial-vectorial coupling constants. The factor $m_{e}$ is the electron mass and
\be
    m_{\beta\beta}=|\sum_{j=1}^{3} U_{ej}^{2} m_{j}|
\ee
is the neutrino effective mass, where $m_{j}$ is the mass eigenvalue of state $j$ and $U_{ej}$ are the elements of the matrix describing the mixing of neutrinos $j$ with the electronic neutrino. This neutrino mass parameter can be expressed in terms of the PNMS matrix elements given in Eq.~\ref{eqPMNS} as:
\be
m_{\beta\beta} =
c_{12}^2 c_{13}^2 m_1+s_{12}^2 c_{13}^2 e^{i\alpha_1} m_2+s_{13}^2 e^{i\alpha_2}m_3 \;.
\ee\label{eqmnu}

It must be noted that in other kind of experiments trying also to measure the absolute value of the neutrino mass (based on the Kurie plot of single beta decay or on Cosmology, as mentioned in the beginning of this Sec.~\ref{secdbd}) the observable is a different combination of neutrino mass eigenvalues. These three observables are correlated among each other and bounded by the oscillation results within well defined regions.

Equation~\ref{t0nu} shows how from the decay rate of neutrinoless DBD is possible to derive a value for the neutrino effective mass:
\be
    m_{\beta\beta}=(F_{N}T_{1/2}^{0\nu})^{-1/2} \;.
\ee

Because of the uncertainties in the nuclear factor of merit, a restrictive limit to the half-life of a DBD emitter does not guarantee for all the nuclei a restrictive bound to the effective mass. Due to these uncertainties, it would be advisable to extend the searches for neutrinoless DBD decay to different emitters. Present limits on the neutrino effective mass from the neutrinoless DBD are in the range from 60 to 600~meV\footnote{The individual results of each experiment are strongly dependent on the particular nuclear matrix element considered.}; future experiments plan to reach sensitivities down to 10~meV, covering the allowed parameter space for inverted neutrino mass ordering.

Precise calculations of the kinematical factors $G_{0\nu}$ are not problematic in principle (results can be found in Refs.~\cite{calg1,calg2,calg3}), but those of the nuclear matrix elements quantifying the transition probability are complex and depend strongly on the nuclear model considered. The theoretical aspects of neutrino–nuclear responses for DBD have been reviewed in \cite{EjiriPR2019}. Calculations of the nuclear matrix elements quantifying transition probabilities have to consider many states of open-shell nuclei and are based on different approaches for describing their complicated nuclear structure \cite{povesilias,ejiri10,vogel12}: the shell model \cite{shell}, which can now be used for almost all DBD emitters, the ``Quasiparticle Random Phase Approximation'' (QRPA) \cite{faessler18} or the ``Interacting Boson Model'' (IBM) \cite{ibm}. Shell model predictions for the matrix elements are typically lower than those from QRPA or IBM. It has been discussed the tuning of some parameters for the neutrinoless channel by means of the agreement of model predictions with experimental results for the channel with neutrino emission and many different effects, as  deformation of nuclei, are increasingly taken into account. The possible ``quenching'' of the weak axial-vector coupling constant $g_{A}$ (following the results for single beta decay matrix elements), affecting the calculations for DBD and therefore the sensitivity of experiments, has been deeply analyzed \cite{suhonengA}; the impact for neutrinoless DBD of the very recent results of the most advanced nuclear theory calculations reproducing experimental data for beta decay without any ``quenching'' \cite{naturegA} is presently under consideration. Overall, an encouraging convergence between different calculations is taking place in the last years and specific measurements are made to help nuclear calculations \cite{frekers,EjiriPR2019,EjiriJPG2019}; Gamow-Teller nuclear matrix elements with low momentum are relevant to DBD with two neutrino emission. The comparison and status of different nuclear matrix calculations for DBD processes is thoroughly discussed for instance in Refs.~\cite{reviewgiuliani,vergados,bilenky15,menendez17}.

DBD experiments can inform not only on the absolute neutrino mass but also on the mass hierarchy (normal or inverted, according to the relationship between the mass eigenvalues $m_{j}$, $j=$1-3) and even on the CP violation \cite{bac04,bil05,pas06,fogli08,vallerep}. Taking into account the dependence of the effective neutrino mass on the lightest mass eigenvalue for the different hierarchies and the expected sensitivities of DBD projects ongoing, if the masses of the eigenstates are degenerated or following the inverted hierarchy, they can be explored in the proposed experiments; however, to reach sensitivity to masses for normal hierarchy seems harder for the time being.

If neutrinoless DBD was not produced by the exchange of a light Majorana neutrino, it would be possible to obtain information on parameters other than the neutrino mass, like V+A couplings in weak interactions
\be
\langle \lambda \rangle = \lambda \sum_{i} U_{ei} V_{ei}
\ee
and
\be
\langle \eta \rangle= \eta \sum_{i} U_{ei} V_{ei},
\ee
with $\lambda$ and $\eta$ the interaction coefficients describing respectively the coupling between right-handed leptonic currents and right-handed or left-handed quark currents in the weak Hamiltonian and V the mixing matrix analogous to U but for right-handed neutrinos. $\lambda$ and $\eta$ are given by the mixing angle $\theta_{LR}$ between the left and right weak bosons and by their masses $M_{L}$ and $M_{R}$ as
$\lambda\approx(M_{L}/M_{R})^{2}$ and $\eta\approx-\tan\theta_{LR}$.

\subsection{Detection of Double Beta Decay}
\label{det}

Two different approaches have been followed to try to detect the DBD of nuclei since the proposal of its existence:
\begin{itemize}
\item On one hand, indirect searches trying to identify an abnormal concentration of the daughter nuclei. In {\it geochemical experiments}, by means of isotopical analysis, the excess of the daughter nuclei (A,Z+2) accumulated in very large times is searched for in rocks containing nuclei (A,Z). This method has been used to see DBD in $^{82}$Se, $^{128}$Te, $^{130}$Te, $^{96}$Zr and, more recently, in $^{100}$Mo. In fact, this kind of experiments gave the first evidence of DBD in the fifties. In {\it radiochemical experiments}, the daughter nuclei of a DBD emitter are accumulated, extracted and counted. These daughter nuclei must therefore be radioactive and produce a distinctive signal. This method has been used for $^{238}$U.
\item On the other hand, in {\it direct counting experiments}, the energy spectrum of electrons emitted by DBD has to be registered and analyzed. This method has the advantage of discriminating the DBD channel. Proposed in the sixties \cite{fiorini67}, the first direct evidence of DBD with the emission of neutrinos came in 1987, for $^{82}$Se using a TPC. Most of the experimental efforts are presently devoted to this direct search. Different kinds of detectors have been used in counting experiments in the past and are being considered for present and future projects: semiconductors, scintillators, chambers and cryogenic detectors; pros and cons of each experimental technique have been discussed in specialized reviews \cite{reviewelliot,reviewgiuliani,reviewgomez,reviewcremonesi}. Tracking detectors allow to measure not only the energy deposit but also the electron tracks to discriminate the signal from the background.
\end{itemize}

Due to the extremely low expected rates of DBD, very sensitive and special detection systems are mandatory in counting experiments, demanding conditions that cannot be satisfied simultaneously in a single experiment. The following requirements must be considered for a DBD search:
\begin{itemize}
\item Large masses of DBD emitters should be accumulated. Following the scale of a few kg used in past experiments, masses at the tonne scale are already being used and envisaged in future projects. If the natural isotopic abundance of the DBD emitter is low (see Table~\ref{dbdinfo} for the particular values of the commonly investigated nuclei), enrichment techniques must be applied, but they are usually complicated and expensive. Isotopic enrichment is necessary for almost all DBD projects.
\item Long data taking periods are mandatory. Detector technologies offering an easy and stable operation allowing to reach high live times are advantageous.
\item A good energy resolution is also necessary to identify the neutrinoless DBD signal, in order to avoid the interference of the final part of the signal of DBD with emission of two neutrinos (see Fig.~\ref{espbeta}). Energy resolutions better than 1-2\% (Full Width Half Maximum, FWHM) are particularly convenient.
\item The approach detector$=$source, being the DBD emitters located inside the detector, is specially interesting for maximizing the detection efficiency to the signal. The approach can be followed quite easily for some isotopes, like $^{76}$Ge in germanium detectors, $^{136}$Xe in gas and liquid TPCs and $^{130}$Te in bolometers. Self-absorption in DBD sources external to the detector makes difficult to accumulate large masses of the DBD emitter.
\item All DBD experiments, as other rare event searches, have a common challenge: to reduce the detector background to achieve enough sensitivity to the signal. Underground operation, under hundreds of meters of rock to shield cosmic rays, is a must. But even deep underground, primordial or cosmogenically induced activities in materials (in bulk or also on surfaces), radon from the air, environmental neutrons or residual muons can induce events indistinguishable from the signal. Solar neutrinos could also become a background limitation for some types of DBD experiments \cite{barros}. To minimize the effect of all these background sources, different strategies are necessary \cite{heusser,formaggio,cebriancosmogenic}: passive shieldings combining heavy materials like copper or lead against the gamma background with hydrogenous material as neutron moderator; active veto systems using additional detectors to identify coincidence events; and extreme control of the radioactivity of the materials in the set-up to select only those fulfilling the stringent radiopurity requirements.
\item In addition, sophisticated methods to discriminate the DBD signal and the irreducible backgrounds are typically developed, profiting from the particular features of each detector technology.
\item Concerning the type of emitters, in addition to the natural isotopic abundance, other relevant features must be considered:
\begin{itemize}
\item One is the transition energy $Q$; Table~\ref{dbdinfo} presents the values for DBD emitters having $Q>2$~MeV. The phase space in neutrinoless DBD grows as $Q^{5}$. In addition, as the natural radioactive background decrease with energy, almost vanishing beyond $\sim$3~MeV, the higher $Q$, the lower background will entangle the region of interest where the signal is expected to appear.
\item Another one is the nuclear matrix element, contributing to the transition probability of the DBD, as shown in Eqs.~\ref{t2nu} and \ref{t0nu}. A slow decay rate for the DBD channel with the emission of two neutrinos helps to reduce the overlap of the final part of its signal with the neutrinoless DBD peak.
\end{itemize}
\end{itemize}

A detector factor-of-merit $F_{D}$, giving an estimate of the limit (at 1$\sigma$ confidence level) to $T_{1/2}$ achievable for the neutrinoless DBD, has been defined and  commonly used \cite{avignone05} in an attempt to quantify the goodness of an experiment concerning most of the conditions and requirements which have been discussed above. In the detector$=$source approach, $F_{D}$ expressed in years can be written as:
\begin{equation}
F_{D}=\ln 2 \times N_{A} \times 10^{3} \times \frac{a}{A}\sqrt{\frac{M t}{b
\Gamma}} \times \epsilon, \label{fd}
\end{equation}
being $N_{A}$ the Avogadro number, $A$ the atomic mass, $a$ the isotopic abundance, $M$ the detector mass in kg, $t$ the time of measurement in years, $b$ the background as counts keV$^{-1}$ kg$^{-1}$ y$^{-1}$, $\Gamma$ the energy resolution (FWHM, in keV) and $\epsilon$ the detection efficiency. $F_{D}$ gives the half-life corresponding to the maximum signal which can be hidden by the background fluctuations. Equation~\ref{fd} is derived assuming Poisson statistics and obtaining the expected total number of background events by integrating over an energy window given by the FWHM resolution of the detector. It must be noted that other formulae to evaluate experimental sensitivity have been proposed and used, based on a background rate normalized to the mass of the DBD isotopes, or derived specifically for zero background conditions. Therefore, comparisons between sensitivity values for different experiments must be done cautiously as different parameters or confidence levels may have been used. The discovery probability of next-generation neutrinoless DBD experiments has been studied following Bayesian methods in \cite{benato}.

The goal of DBD direct experiments is to register the properties of the two emitted electrons. Together with the collection of the sum energy spectrum of the electrons and the registration of time correlations, other additional pieces of information can be recorded in some cases, such as the single-electron energy, the angular correlations, the event topology and even the identification and/or counting of the species of the daughter nucleus. This additional information is extremely helpful in identifying background contributions and it could be used to constrain theoretical models. The identification of the daughter nucleus could result in a large improvement of the signal to background ratio of an experiment. But it must be kept in mind that exploiting these complementary signatures is not simple at all in many cases.

In the rest of this section, the main detection techniques used to directly detect DBD are described. A complete review of this topic has been made in \cite{reviewcremonesi}. The relevant results of the experiments made in the past or presently underway will be mostly presented in next sections, together with the discussion of the experiments carried out at the Canfranc Underground Laboratory.

\subsubsection{Semiconductor detectors}
\label{germa}

Different types of semiconductor radiation detectors are being used in the search for DBD, but germanium diodes are the ones largely used. Experiments using germanium detectors investigating the DBD of $^{76}$Ge (having $Q$=2.039~MeV and 7.4\% of natural isotopic abundance, as indicated in Table~\ref{dbdinfo}) have been in operation for about five decades \cite{morales03}. Exploiting the detector$=$source method, these experiments have very high detection efficiency for signal and, as semiconductor detectors, they enjoy excellent energy resolution and are based on well established technologies. In addition, germanium is a very radiopure material. The transition energy, precisely measured as $Q=$(2039.061$\pm$0.007)~keV \cite{qge}, is acceptably high and the nuclear matrix element is favorable. Background discrimination techniques have been developed analyzing electric pulse shapes from the detector; on an event-by-event basis, it can be deduced if energy has been deposited only at one point, as expected for DBD, or at different places, as it is the case for many kinds of background \cite{igexpsd,psdradial1,klapdortracks2}. The main disadvantage of germanium DBD experiments is the need of enrichment of $^{76}$Ge, due to its low abundance. The study of the half-life of this isotope for neutrinoless DBD using HPGe detectors in different experiments gave the best bound to the effective neutrino mass for years; the ``International Germanium EXperiment'' (IGEX), based at the Canfranc Underground Laboratory, was one of those experiments. IGEX will be presented in detail in Sec.~\ref{igex} and the main results and the status of DBD searches using semiconductor detectors worldwide will be reviewed in particular in Sec.~\ref{igexww}, including the past Heidelberg-Moscow experiment, the present GERDA and MAJORANA, and the future LEGEND project.

\subsubsection{Bolometric detectors} \label{bolo}

Cryogenic detectors were proposed for DBD searches in 1984 \cite{fiorini84}. They are calorimeters made of suitable materials to measure the deposited energy through the consequent temperature rise; working at very low temperatures, typically below 100~mK, the heat capacity is small enough to generate measurable signals. As many different absorber materials can be used in this type of detectors, different DBD emitters can be studied by this technique. Experiments can be based on the detector$=$source approach optimizing efficiency. Bolometers have in addition excellent energy resolution, with values from 5 to 10~keV at the transition energy. Surface contaminations are a particularly relevant background in these detectors and effective cleaning techniques have been developed to reduce them; hybrid detectors based on bolometers built with scintillating materials have been designed to identify the type of interacting particles and to reject surface events, with very good results.

The group from Milan is leading since many years the direct detection of DBD following this technique for investigating several DBD emitters. $^{130}$Te has been largely studied using bolometers \cite{cremonesi18,brofferio18,brofferio19}, from the MIBETA to the CUORE experiments. Its main advantage is that, due to an exceptionally natural isotopic abundance (34\%, as shown in Table~\ref{dbdinfo}), the use of isotopic enrichment is not indispensable, and therefore detectors of tellurium oxide are not expensive and large masses can be accumulated. Crystals of tellurium compounds are quite free of radioactive impurities, although as bolometers they must operate at very low temperatures, which makes difficult stable operation for long time. As presented in Table~\ref{dbdinfo}, the  transition energy is quite high (with precisely measured values around 2527~keV \cite{qte1,qte2,qte3}, which gives a reduced background in the region of interest. Bolometers of tellurite have reached very good energy resolutions.

Alpha particles are a dominant background source for DBD emitters with a large transition energy. For tellurite crystals, an efficient alpha particle discrimination is possible based on the simultaneous measurement of the tiny light signal dominated by Cherenkov radiation induced by gamma and beta radiation in the TeO$_{2}$ detector \cite{berge}, as proposed in \cite{tabarelli}. On the other hand, hybrid scintillating bolometers, with simultaneous detection of heat and light, allow to identify and reject alpha particles very efficiently thanks to the different ratio between light and phonon yield for different types of particles. Following the available background models developed, background levels of the order of 10$^{-4}$~counts keV$^{-1}$ kg$^{-1}$ y$^{-1}$ are expected for scintillating bolometers thanks to the suppression of the alpha events \cite{cupidbkgmodel}. Different scintillating compounds have been studied for several DBD emitters; many efforts are focused on $^{82}$Se and $^{100}$Mo. The advantages of a multi-isotope bolometric experiment for neutrinoless DBD searches have been highlighted in \cite{giuliani18}. In particular, the CROSS project, being commissioned in the Canfranc Underground Laboratory, plans to use arrays of TeO$_{2}$ and Li$_{2}$MoO$_{4}$ bolometers enriched in the DBD isotopes $^{130}$Te and $^{100}$Mo, respectively, developing effective discrimination techniques on pulse shapes by exploiting Solid-State-Physics phenomena in superconductors, in order to reject events from alpha and beta surface radioactive impurities. CROSS will be described in Sec.~\ref{crosssec} and the status of DBD searches using bolometric detectors worldwide, including CUORE, CUPID or AMORE, will be reviewed in particular in Sec.~\ref{boloww}.

\subsubsection{TPCs and Tracking detectors}
\label{trackdet}

The different DBD experiments following the approaches presented in Secs.~\ref{germa} and \ref{bolo} are based on the measurement of the energy deposited in the detector by the electrons emitted in the decay. The possibility to obtain additional information of the events, like the track of the detected particles, is an asset for a DBD search, pursued by other experiments using different technologies. An event due to a DBD consists of two short tracks coming from the same point and can be distinguished from events with different topology, like multi-site energy depositions from gamma and beta emissions or much longer tracks due to muons.

In some cases, the source of DBD emitters is outside the detector, in very thin sheets. The detection system is made of gaseous detectors to register tracks, surrounded by scintillators acting as calorimeters, using a moderate magnetic field. The main advantage of this ``tracko-calo'' approach is the possibility of identifying the kind of particle interacting and registering the tracks of electrons from DBD; this allows to reject background and even to have angular information of events. In addition, different emitters can be studied at the same set-up. As disadvantages, detection efficiency is quite low and energy resolution limited. The ``Neutrino Ettore Majorana Observatory'' (NEMO) experiment operated following this approach at the Modane Underground Laboratory (France). The project SuperNEMO keeps the same approach of the NEMO experiment, increasing the mass of the sources, which requires a change in the detector geometry with a planar design. Control of the required radiopurity in the DBD source has been a real challenge in this project and a specific detector based on the identification of BiPo events from the natural radioactive chains has been developed and operated at the Canfranc Underground Laboratory \cite{bipo}; the concept of this detector and the obtained results will be presented in Sec.~\ref{biposec}.

Other types of DBD tracking detectors use TPCs; particularly relevant is the case of experiments investigating $^{136}$Xe, contained in the filling xenon liquid or gas, as described in \cite{reviewmichel}. The charge produced by ionization is used to get the tracks,  while the energy is obtained from the scintillation signal also produced in xenon. In Ref. \cite{revHPXe}, the fundamentals of the technology of high pressure gas TPCs for DBD searches and the historical development of the field are reviewed. As it will be discussed in detail in Sec.~\ref{nextsec}, several reasons make $^{136}$Xe a very attractive DBD emitter and various experiments are searching for its decay; in the case of the NEXT experiment, operating in the Canfranc Underground Laboratory, a gas xenon TPC is used. NEXT will be presented in Sec.~\ref{nextsec}, describing its main features and the obtained results; the status of other DBD searches worldwide using tracking detectors, like NEMO-3 and SuperNEMO, and TPCs, like EXO and nEXO, will be reviewed in particular in Sec.~\ref{trackdetww}.

\subsubsection{Scintillators}
\label{scint}

Large liquid-scintillator or water-Cherenkov detectors used in neutrino oscillation studies have been reoriented to investigate also DBD, giving rise to the KamLAND-ZEN  and SNO+ experiments. New techniques for loading a large amount of nuclei into liquid scintillator enable sensitive searches for DBD with various nuclei. Although liquid scintillators suffer from poor energy resolution, making relevant the contribution of the two neutrino DBD channel at the peak expected for the neutrinoless signal, the definition of fiducial volumes and the capability of event reconstruction are very useful for the background suppression in this approach. Reference~\cite{reviewchen} reviews this type of experiments, highlighting the results achieved and describing the techniques involved in loading the liquid scintillator and in the purification to very low background levels. No DBD experiment based on this technology has been carried out at the Canfranc Underground Laboratory up to now; the status of DBD searches worldwide using scintillators will be presented in the rest of this subsection.

For the KamLAND-ZEN experiment, a transparent spherical nylon balloon, with a $\sim$3~m diameter, filled with 13~tons of a Xe-loaded liquid scintillator (80\% dodecane, 20\% pseudocumene plus PPO) was added in the core of the KamLAND detector, located in the site of the earlier Kamiokande experiment in Japan. The scintillation light generated in the detector is recorded by an array of photomultiplier tubes (PMTs) surrounding it. To search for the DBD of $^{136}$Xe, this isotope was dispersed in the liquid scintillator acting as a passive source. Around 320~kg of xenon gas with an isotopic abundance of 90.9\% for $^{136}$Xe was added to the liquid scintillator. An improved limit to the half-life for the neutrinoless DBD was derived after significant reduction of the $^{110m}$Ag identified component thanks to the liquid scintillator purification: $T_{1/2}^{0\nu}>1.07 \times 10^{26}$~y (90\% C.L.), giving upper limits on the effective Majorana neutrino mass in the range 61-165~meV using commonly adopted nuclear matrix element calculations \cite{kamlandzen0nu}. Relevant results have also been obtained for the half-life of the DBD mode with two neutrinos emission \cite{kamlandzen2nu}, for decays to excited states of the daughter nuclei \cite{kamlandzenexc} and for Majoron-emitting DBD decays \cite{kamlandzenmaj}. A new phase is ongoing, using a larger balloon with 745~kg of xenon.

After the conclusion of the Sudbury Neutrino Observatory (SNO) experiment devoted to solar neutrino measurements in the SNOlab site, the heavy water filling an acrylic sphere has been replaced by 780~tonnes of liquid scintillator (linear alkylbenzene with PPO) to be loaded with a DBD emitter for operation of the SNO+ experiment, conceived as a multi-purpose neutrino experiment. Although $^{150}$Nd was firstly considered, natural tellurium with 1.3~tonnes of $^{130}$Te (0.5\% by mass) has been  finally chosen for the first phase. SNO+ aims at exploring the Majorana neutrino mass parameter space down to the inverted mass hierarchy region. Data taking with a detector fully filled with ultrapure water has been carried out before filling with liquid scintillator and loading with tellurium \cite{sno}.

Other DBD experiments based on scintillators were proposed or are presently at different stages of development, like MOON (``Mo Observatory Of Neutrinos'') for $^{100}$Mo in Japan \cite{moon} or CARVEL (``CAlcium Research for VEry Low neutrino mass'') for $^{48}$Ca in Ucraine \cite{carvel}. The CANDLES (``CAlcium fluoride for the study of Neutrinos and Dark matters by Low Energy Spectrometer'') project \cite{candles} uses also $^{48}$Ca isotope, which has the longest transition energy (4.26~MeV) among all possible DBD emitters, as shown in Table~\ref{dbdinfo}; the CANDLES III detector is running at the Kamioka Underground Observatory with 305~kg of calcium fluoride (CaF$_{2}$) crystals immersed in a liquid scintillator acting as an active shield and background studies are underway \cite{candlesbkg}. Different tungstate compounds are being used in crystal scintillators to study the DBD of $^{106}$Cd and $^{116}$Cd \cite{bellicd106}.

\section{The Canfranc Underground Laboratory} \label{seclsc}

The Canfranc Underground Laboratory (``Laboratorio Subterr\'{a}neo de Canfranc", LSC)\footnote{Website: https://lsc-canfranc.es} is presently an international multidisciplinary research facility equipped with different services for underground science \cite{julio2004,bettini2012,ianni2016,ianni2016b}. It is managed by a consortium between the Spanish Ministry of Science, the Government of Arag\'{o}n and the University of Zaragoza and it is recognized as Singular Scientific and Technological Infrastructure (``Instalaci\'{o}n Cient\'{i}fica T\'{e}cnica Singular'', ICTS) by the Spanish Government. It is the second largest underground laboratory in Europe after the
``Laboratori Nazionali del Gran Sasso'' (LNGS) in Italy; a complete comparison of the features of LSC with other underground laboratories all over the world can be found in \cite{aldoijmpa2017}. The research activity at LSC is focused on the direct detection of dark matter and the investigation of neutrino properties looking for neutrinoless DBD, but includes also studies on Geophysics and Biology. In this section, a brief description of the main features of the laboratory, including the environmental backgrounds, will be presented and its scientific program will be outlined.

\subsection{Features}
\label{seclscfea}
\begin{figure}[tb]
\centerline{\includegraphics[width=14cm]{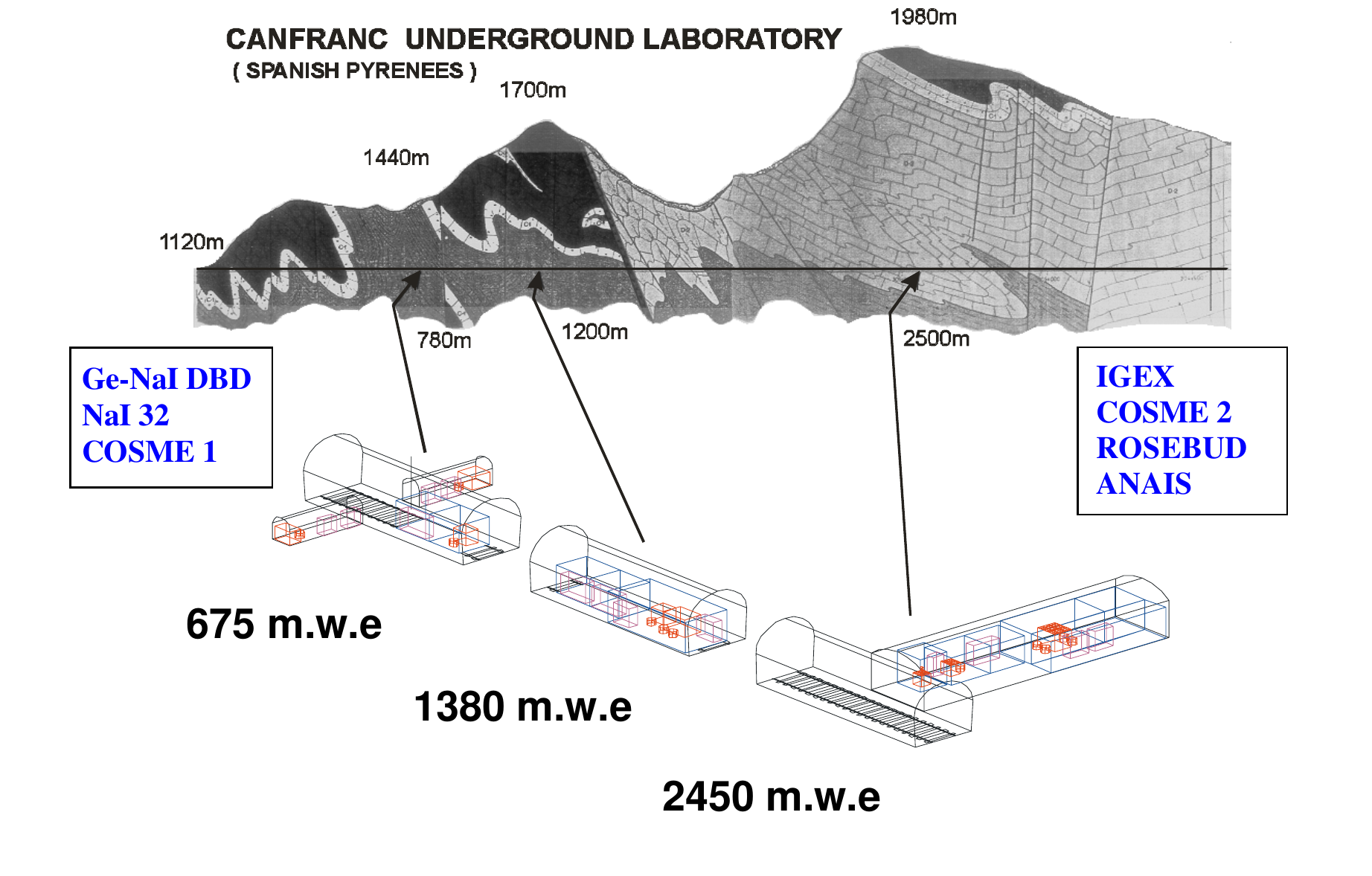}}
\begin{center}
\begin{minipage}[t]{16.5 cm}
\caption{Mountain profile and location along the railway tunnel of the first facilities of the LSC, used before the construction of the present LAB2400. The corresponding rock overburden and the main experiments carried out there are also indicated. The facilities at 780~m and 2500~m from the Spanish entrance of the train tunnel (now referred as LAB780 and LAB2500, respectively, see also Fig.~\ref{layout}) are still in use. \label{oldlabs}}
\end{minipage}
\end{center}
\end{figure}

\begin{figure}[tb]
\centerline{\includegraphics[width=18cm]{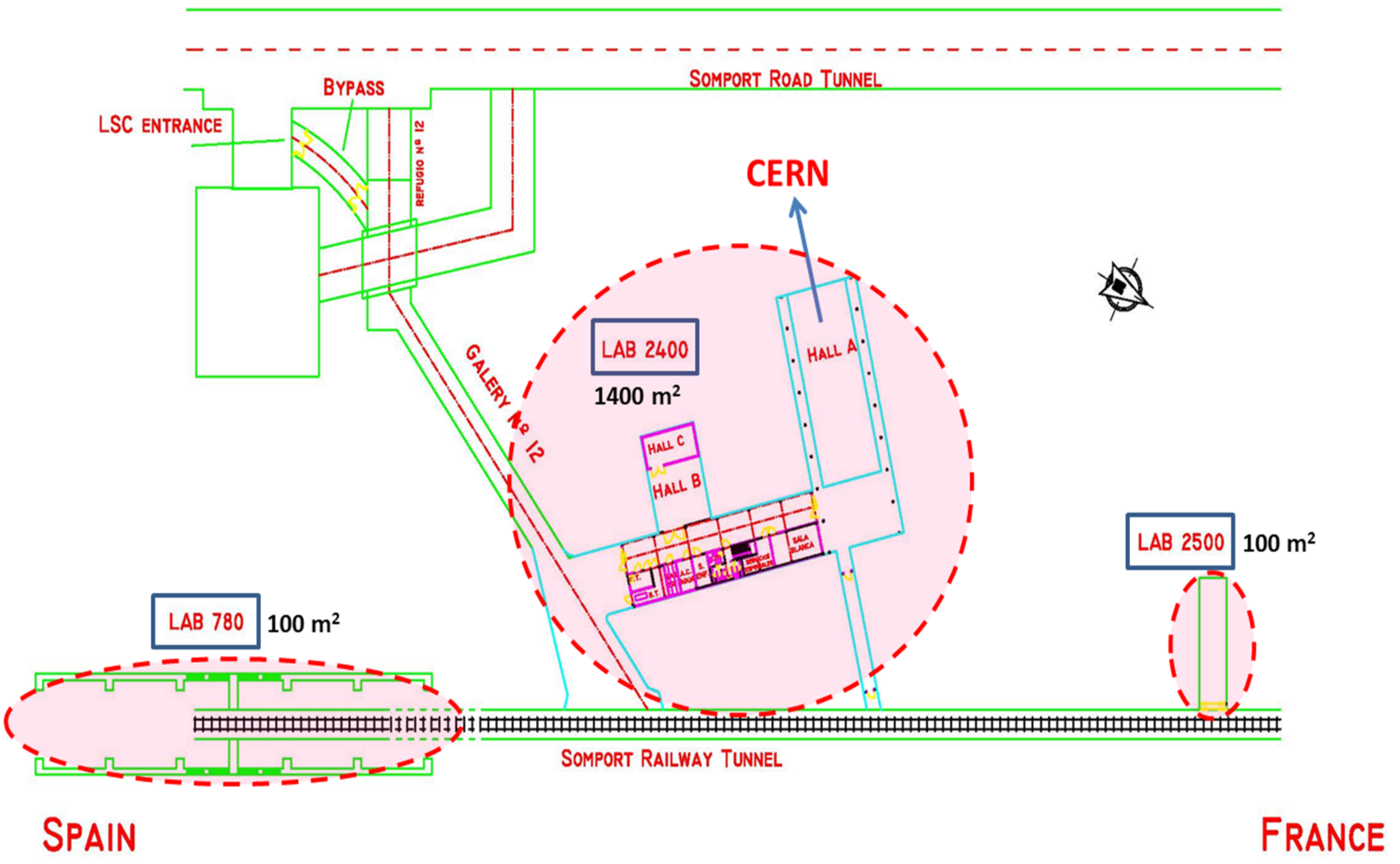}}
\begin{center}
\begin{minipage}[t]{16.5 cm}
\caption{Layout of the present underground facilities at LSC. The main infrastructures, LAB2400 with halls A and B and LAB2500, are located between the railway and road tunnels (courtesy of LSC).\label{layout}}
\end{minipage}
\end{center}
\end{figure}

The history of the laboratory dates back to the mid-eighties \cite{julio2004,bettini2012}, when a team from the University of Zaragoza led by Prof. Angel Morales started using the abandoned train tunnel connecting Spain and France to carry out the first experiments using two old service cavities of 12~m$^{2}$ each placed at a depth of 675~m.w.e.\footnote{Meters of water equivalent.}, presently refurbished and known as LAB780\footnote{The present LSC underground facilities are identified taking into account the distance in meters to the Spanish entrance of the train tunnel.}. Later, also two prefabricated mobile cabins with the convenient conditioning (electrical power installation, telephone, ventilation) were installed over the railway at 1200~m from the Spanish entrance below an overburden of 1380 m.w.e. In the nineties, a new 8-km long road tunnel (named Somport tunnel) was excavated and then an experimental hall of 118~m$^{2}$ was built at the place with the maximum rock overburden of 2450~m.w.e., $\sim$2500~m away from the Spanish entrance (now referred as LAB2500). This hall, managed by the University of Zaragoza, hosted since 1995 a number of experiments, searching not only for neutrinoless DBD but also for Weakly Interacting Massive Particles (WIMPs) which can make part of the galactic dark matter and for solar axions through Bragg-scattering; in these searches different detection techniques were used: germanium detectors (as in the Cosme detector \cite{cosme1,cosme2} and in the IGEX-DM experiment \cite{igexdm}), sodium iodide scintillators (the NaI32 experiment carried out the first annual modulation studies for the dark matter signal \cite{nai32}) and bolometers (in the ROSEBUD experiment \cite{rosebud}). The mountain profile and the location along the train tunnel of all these facilities and experiments are depicted in Fig.~\ref{oldlabs}. LAB780 and LAB2500 are still in use (see also Fig.~\ref{layout}).

\begin{figure}[tb]
\centerline{\includegraphics[width=11cm]{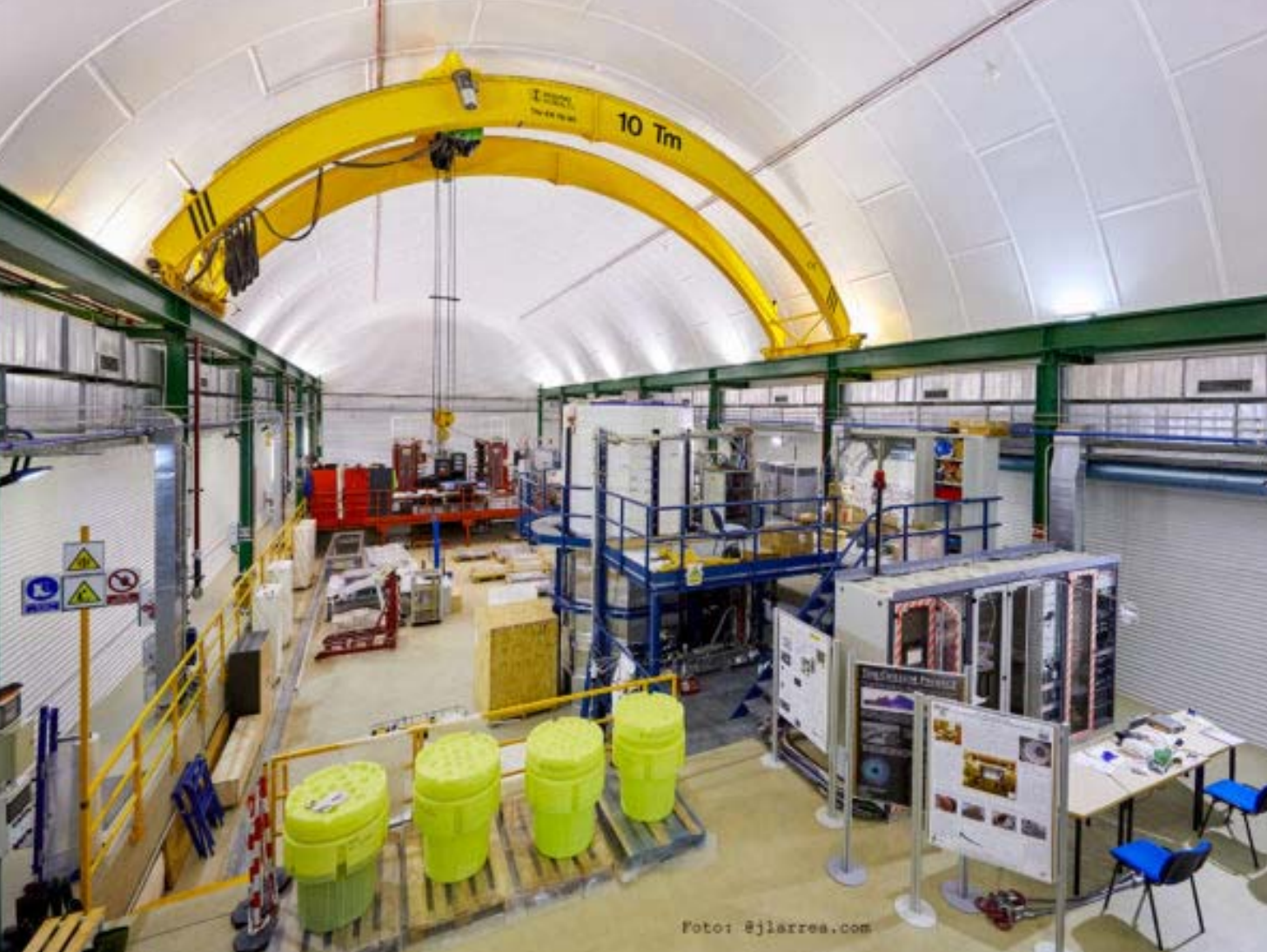}}
\begin{center}
\begin{minipage}[t]{16.5 cm}
\caption{Picture of the hall A of the LSC (courtesy of LSC).\label{hallA}}
\end{minipage}
\end{center}
\end{figure}

The good results obtained in the first LSC facilities fostered a larger excavation in the rock under the Mount Tobazo, aiming at converting Canfranc into an international underground laboratory for Astroparticle Physics. This new facility, known as LAB2400, is divided into two halls, named A and B, with a volume of 40$\times$15$\times$12~(height)~m$^{3}$ and 15$\times$10$\times$7~(height)~m$^{3}$, respectively. Both the old railway tunnel and the new road tunnel are used as access routes to the laboratory area. The LSC was in addition recognized as a legal entity having its own staff. It officially opened in 2006, but the activity in the new facilities started later, as signs of rock instabilities appeared and the lab was closed for safety issues for a long time to carry out a complete revision of the rock support structures. Now, the total underground space available for research activities is about 10000~m$^{3}$, with a surface of 1250~m$^{2}$. The layout of the underground laboratory is depicted in Fig.~\ref{layout} and a picture of the hall A is shown in Fig.~\ref{hallA}. The coordinates of the LSC are known with accuracy of $\pm$5~cm. The altitude of the floor of the Hall A is 1204~m above the sea level, the longitude is 0$^{o}$ 31' 44.85570'' W, and the latitude is 42$^{o}$ 45' 16.95821'' N. Being placed below the Spanish Pyrenees, LSC has about 800~m rock overburden, which corresponds to 2450~m.w.e. depth. The composition of the rock is limestone, made of mainly calcium carbonate.

The LSC is equipped with several underground service facilities, like a mechanical workshop, a gas storage room and other general utilities located in the access corridor to the main halls. There is a clean room of 45~m$^{2}$ type ISO7 (ISO6 in a sector) and a radon abatement system is in operation in the hall A. This facility, based on absorption on charcoal filters, can deliver 220~m$^{3}$ h$^{-1}$ of radon-free air at the level of few mBq m$^{-3}$ of radon \cite{lscari2017}, which means a reduction factor in the radon activity of about 10000. A custom radon monitor has been developed in collaboration with the Jagiellonian University in Krakow with sensitivity to measure the radon level in the radon-free air.

\begin{figure}[tb]
\centerline{\includegraphics[width=10cm]{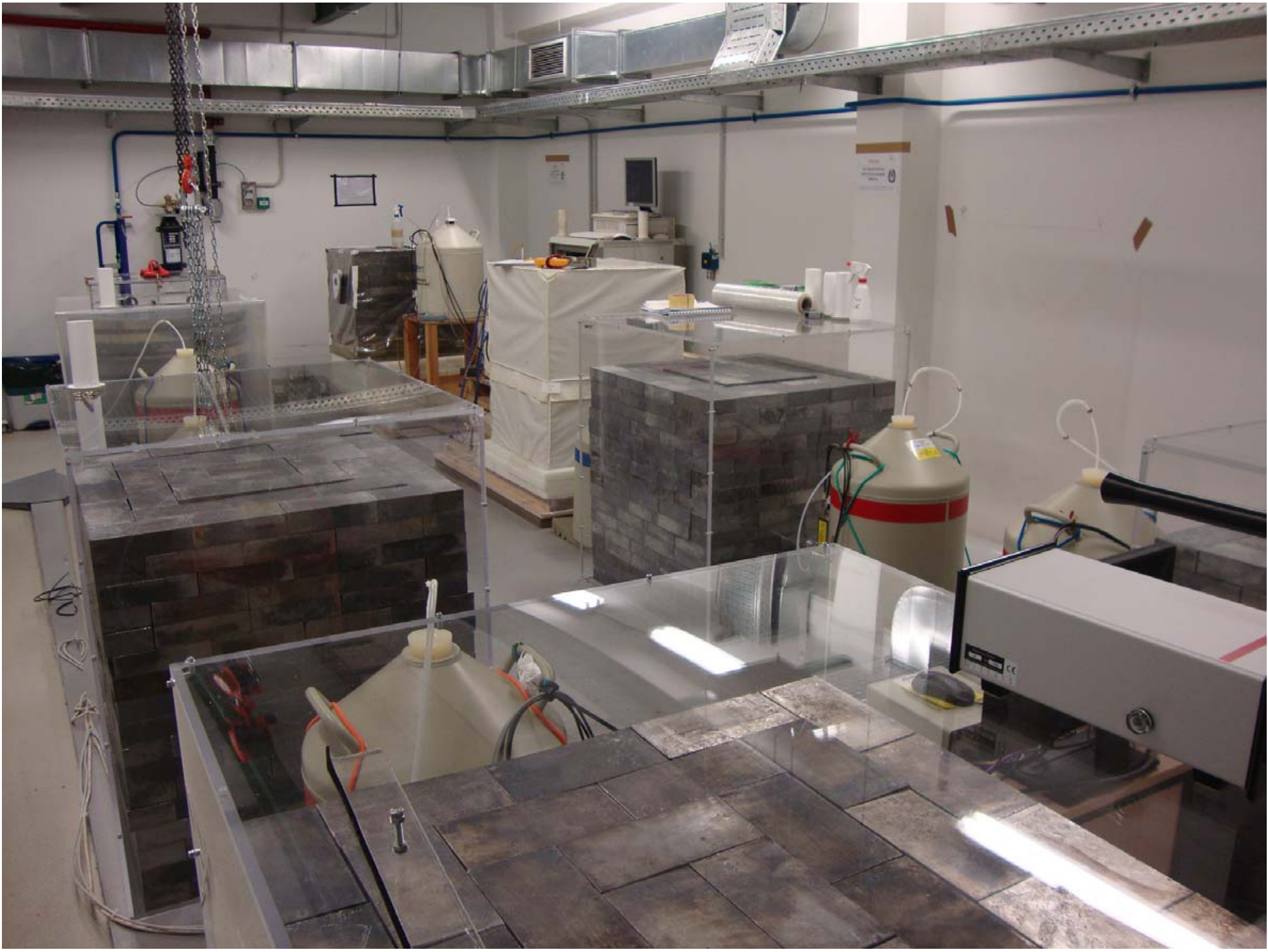}}
\begin{center}
\begin{minipage}[t]{16.5 cm}
\caption{Picture of the room next to the Hall B of the LSC where several HPGe detectors inside their shieldings are operated for material radiopurity screening (courtesy of LSC).\label{gelsc}}
\end{minipage}
\end{center}
\end{figure}

The lab offers an Ultra-Low Background Service with a screening facility with several HPGe for gamma spectroscopy \cite{lscari2017} and an Inductively Coupled Plasma Mass Spectrometry (ICPMS) facility. The ICPMS facility is based on a Thermo Scientific iCAP RQ ICP-MS apparatus and it is in operation since 2018. Its mass range is 2 to 290~amu and the detection limits are 1 ppt for thorium and 0.1~ppt for uranium. Presently, there are several close-end coaxial p-type HPGe detectors ($\sim$100-110\% relative efficiency, with masses of $\sim$2.2~kg and volume 400~cm$^{3}$), made by Canberra. Other new different units like a SAGe well detector are being commissioned. The seven large detectors are named as Asterix, Obelix, GeAnayet, GeAspe, GeLatuca, GeOroel and GeTobazo. They have aluminium or copper cryostats and are operated inside 10-cm-thick Oxygen Free copper and 20-cm-thick low activity lead shieldings, inside a methacrylate box (see Fig.~\ref{gelsc}). The box was flushed with nitrogen gas to avoid airborne radon intrusion, but recently, the detectors have been connected to use radon-free air from the LSC facility. Data acquisition is
based on Canberra DSA 1000 modules.
The sensitivity of the detectors can reach levels of tenths of mBq/kg for the lower part of the $^{238}$U and $^{232}$Th chains and $^{40}$K. A Monte Carlo simulation program (GEBIC) developed using Geant4 and adapted to the specific characteristics of each detector is used to calculate the detection efficiency for each sample. Several hundreds of samples have been analyzed since 2011 for different experiments connected to LSC like NEXT, SuperK-Gd, ArDM, BiPo, ANAIS, TREX-DM, DarkSide, KSTAR and CUNA. Besides the detectors of the Ultra-Low Background Service of LSC, two smaller HPGe detectors (with volume 190~cm$^{3}$ and mass $\sim$1~kg) are operated in LAB2500 by the University of Zaragoza; one of them is underground since 1988.

In addition to the underground facilities, there are two external buildings located in the village of Canfranc Estaci\'{o}n with 23 offices for scientific users and staff, specialized laboratories (chemistry, electronics, \dots), a mechanical workshop, meeting, conference and exhibition rooms and two apartments. At surface, in the Chemistry Laboratory, there is also a Copper Electroforming Service, developed in collaboration with the University of Zaragoza, to obtain high-purity copper pieces for the experiments \cite{lscari2017}; a new electroforming set-up inside the LSC underground clean room is planned. Electroforming is a metal forming process producing parts through electrodeposition of a metal onto a mandrel, which is subsequently removed; copper electroforming is known to be an effective way to obtain high-purity copper free of induced cosmogenic activation required in rare event search experiments. In the LSC, the used technique is direct fixed-current density electroplating. All these ancillary infrastructures  play a very important role during the design and installation of the set-ups of the experiments in Canfranc.

\subsection{Background characterization}
\label{lscbkg}

A proper design, planning and operation of experiments in underground facilities requires a good knowledge of the backgrounds at the laboratory where they operate. Different campaigns for radon and environmental measurements have been accomplished in order to characterize the different background components at LSC \cite{lscari2017}.
\begin{itemize}
\item The radon concentration in air is continuously monitored in the lab using several AlphaGUARD P30 detectors; four monitoring stations are located inside the underground laboratory at the different experimental halls. At LSC the ventilation system supplies around 11000 m$^{3}$ h$^{-1}$ of fresh air from the outside (through a 250~m long vertical pit taking air from the mountain) to the main experimental areas. In the underground laboratory LAB2400 the radon level varies between 50 and 80~Bq m$^{-3}$ \cite{bettini2012}; for instance, the weighted average radon concentration in air at Hall A was measured as (63$\pm$1)~Bq m$^{-3}$.

\item The muon flux integrated over the angles was determined to be between 2 and 4$\times$10$^{-3}$~m$^{-2}$s$^{-1}$, depending on the location \cite{bettini2012}, following the first results obtained in 2006 using large plastic scintillators \cite{luzonidm2006} and new measurements carried out afterwards. More recently, the residual flux and angular distribution of high-energy cosmic muons have been measured in two underground locations at the LSC using a dedicated Muon Monitor, consisting of three layers of fast scintillation detector modules operating as 352 independent pixels. For each site the data were collected over nearly 600~days.
The measured integrated muon flux is (5.26$\pm$0.21)$\times 10^{-3}$~m$^{-2}$s$^{-1}$ in the Hall A of the LAB2400 and (4.29$\pm$0.17)$\times 10^{-3}$~m$^{-2}$s$^{-1}$ in LAB2500 \cite{mulsc1}. These results supersede previous values \cite{mulsc2}. The angular dependence is consistent with the known profile and rock density of the surrounding mountains. In particular, there is a clear maximum in the flux coming from the direction of the Rioseta valley; as a result, the integrated muon flux is larger than expected from the thickness of the overburden directly above the site. Figure~\ref{muonflux} presents the decrease of the muon flux with the depth, showing the values for different underground facilities all over the world including Canfranc.

\begin{figure}[tb]
\centerline{\includegraphics[width=12cm]{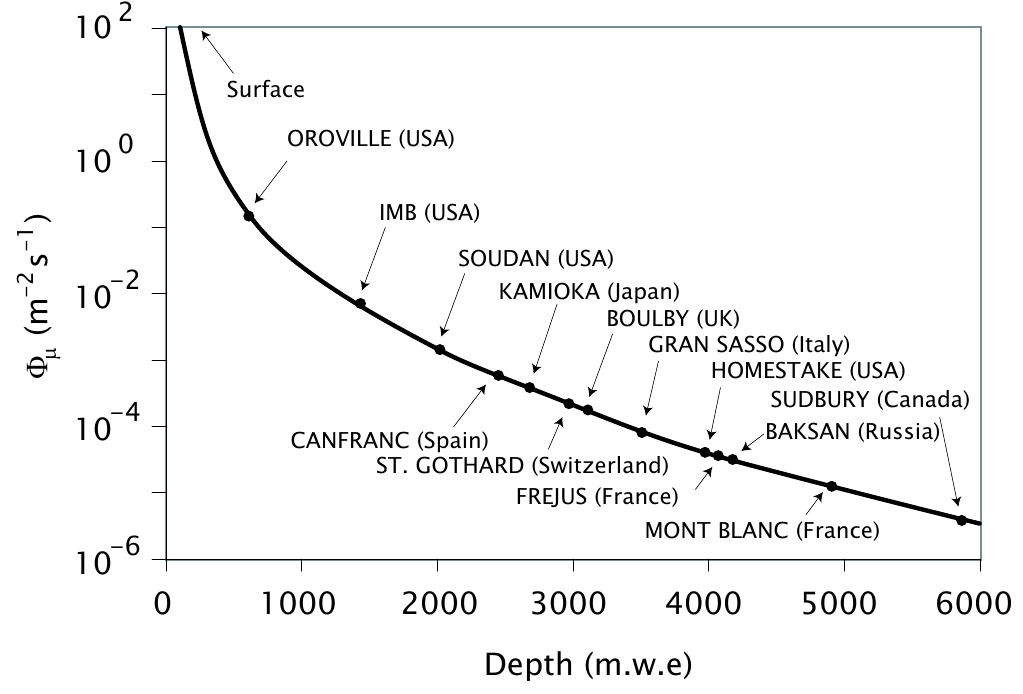}}
\begin{center}
\begin{minipage}[t]{16.5 cm}
\caption{Dependence of the muon flux on the depth expressed in m.w.e. Positions of different underground facilities all over the world including the Canfranc Underground Laboratory, at 2450~m.w.e., are marked. \label{muonflux}}
\end{minipage}
\end{center}
\end{figure}

\item The first determination of the neutron flux at the LAB2500 of LSC was derived from the data of the IGEX experiment taken in different shielding conditions concerning the neutron moderator \cite{Carmona}; afterwards, the integral neutron background flux and its energy distribution in the range between 1 eV and 10~MeV were studied in Hall A using a novel approach based on the combination of the information obtained with six large high-pressure $^{3}$He proportional counters embedded in individual polyethylene blocks of different size \cite{Jordan}. The total neutron flux measured in that energy range is (3.44$\pm$0.35)~$\times$10$^{-6}$~cm$^{-2}$s$^{-1}$, which is about four orders of magnitude smaller that the neutron background at sea level. In addition, measurements using a CLYC scintillator have been made in the Hall A and studies in collaboration with CIEMAT are underway to develop a CLYC detector as a monitor for radiogenic neutrons in underground facilities (CLYC-N project) \cite{clyc}.

\item The gamma-ray flux at Hall B was firstly determined from measurements using a 3''$\times$3'' NaI(Tl) detector \cite{luzonidm2006}; new measurements were carried out later also at Hall A  \cite{bettini2012}, pointing to a total flux of 1-2~cm$^{-2}$s$^{-1}$ considering emissions coming from $^{238}$U and $^{232}$Th chains and $^{40}$K in the whole relevant energy range up to $\sim$3~MeV.
\end{itemize}

In addition, a series of other monitoring measurements are regularly performed at LSC, related to the rock stability underground (through a dedicated structure of optical fibers to monitor continuously the rocks) and water quality.

\subsection{Scientific program}
The main research activity at LSC is about Astroparticle Physics, dark matter searches and neutrino Physics; but also activities in Nuclear Astrophysics, Geophysics and Biology are carried out. In addition, support activities for projects performed in other laboratories are developed by means of all the facilities described before. This article is devoted to the DBD searches performed at LSC; but in this section a brief description of all the other present projects is provided.

Presently, LSC hosts three projects, at different stages of development, devoted to the direct detection of WIMPs which could be pervading the galactic halo:
\begin{itemize}
\item The ANAIS (``Annual modulation with NaI(Tl) Scintillators'') experiment aims at the confirmation or refutation of the positive annual modulation signal in the low energy detection rate observed by the DAMA/LIBRA experiment, using the same target and technique. DAMA/LIBRA, operated at LNGS, has reported the effect over 20~years with a 12.9~$\sigma$ confidence level \cite{dama}. The annual modulation in the dark matter signal is expected by the revolution of the Earth around the Sun, which distorts the particle velocity distribution function as seen by the detector. ANAIS-112, consisting of nine 12.5~kg NaI(Tl) modules produced by Alpha Spectra Inc., is taking data smoothly in ``dark matter search'' mode since August, 2017, after a commissioning phase and operation of the first detectors during previous years in various set-ups in Canfranc. The cylindrical crystals are housed in Oxygen Free Electronic copper and coupled to highly efficient Hamamatsu PMTs. Before unblinding the data, the whole analysis procedure was fixed \cite{anaisanalisis}, the background of the experiment thoroughly studied \cite{anaisbkg} and the expected sensitivity evaluated \cite{ivan}. Triggering at 1~keV (electron equivalent energy) has been reached thanks to the outstanding light collection measured for the nine modules, at the level of 15~phe/keV. ANAIS-112 could detect the annual modulation observed by DAMA/LIBRA or, otherwise, exclude the 3$\sigma$ region singled out by DAMA/LIBRA, in five years of data taking. The first published results, corresponding to 1.5~y \cite{anaisprl}, have been updated using two years of data (220.69~kg$\cdot$y) \cite{anaistaup2019}; the null hypothesis is well supported and the best fits for the modulation amplitude are incompatible with DAMA/LIBRA results at 2.6$\sigma$. As ANAIS-112, the COSINE-100 experiment has released also their first annual modulation results from 97.79~kg$\cdot$y \cite{cosineprl}.

\item ArDM (``Argon Dark Matter'') is a two-phase TPC using 2~tonnes of liquid argon to search for WIMPs recoils. The detector consists of a cylindrical vessel with an array of PMTs on top and at the bottom. An electric field is present inside the vessel to drift electrons toward the upper side of the vessel. The scintillation in the liquid argon makes the prompt signal (S1). Close to the array of PMTs at the top of the vessel, a small argon gas layer makes the delayed signal (S2). The different characteristics of S1 and S2 can be analyzed to disentangle nuclear recoils due to WIMPs from electron recoils due to background sources. Moreover, in liquid argon the scintillation light emitted by nuclear recoils has a characteristic timing very different than that from electron recoils, allowing to perform pulse shape analysis. Results from operation in single phase were presented \cite{ardm} and the data taking in the two-phase mode was completed in 2019. In the next future, the ArDM infrastructure (after a minimal modification to install in the middle of the ArDM vessel a radiopure copper single-phase liquid argon detector of about 1~l volume) will be used by the DArT experiment to carry out precise measurements of the $^{39}$Ar activity in underground argon, promoted by the DarkSide-20k and the Global Argon Dark Matter Collaboration \cite{dart}. The underground argon will be extracted from the Urania facility and purified in the Aria plant, being developed in Colorado, US and Sardinia, Italy, respectively; a reduction factor of 1400 for $^{39}$Ar activity relative to atmospheric argon has been measured by the DarkSide-50 experiment in argon from the deep underground sources in Colorado \cite{agnesar}.

\item TREX-DM (``TPC for Rare Event eXperiments-Dark Matter'') has been conceived to look specifically for low-mass WIMPs (below 10~GeV/c$^{2}$) using a gas TPC equipped with Micromegas readout planes  \cite{iguaz16,trexdmjcap}. The detector can hold in the active volume about 20~l of pressurized gas at a pressure of 10~bar inside a copper vessel; this corresponds to 0.30~kg of Ar or 0.16~kg of Ne. The Microbulk Micromegas being used \cite{adriamonje}, manufactured at CERN, have the largest surface ever made with this technology (25$\times$25~cm$^{2}$). The detector has been built with low background specifications, and although not focused on directionality, topological information of the events can be obtained to further discriminate backgrounds from the expected signal by dark matter \cite{trexdmbkg}. The Micromegas are read with a self-triggered acquisition, allowing for thresholds below 0.4~keV (electron equivalent energy). All these features make TREX-DM competitive in the search for low-mass WIMPs. At the end of 2019, the commissioning phase is finishing and the data taking is expected to start soon.
\end{itemize}

SuperKamiokande doped with Gadolinium is a very important project in the field of neutrino Physics. The Ultra-Low Background Service of LSC is being used to select low radioactivity Gadolinium salts, such as Gd$_{2}$(SO$_{4}$)$_{3}$, to be used in the project and a huge number of measurements have been already carried out in the framework of the SuperKGd project using HPGe detectors in Canfranc \cite{tesisjavi,egads}.

LSC is equipped with an underground and surface geodynamic facility named GEODYN aimed at studying both local and global phenomena (low-frequency seismic waves, Earth free oscillations, possible local aseismic stress release, \dots). The facility consists of a broadband seismometer and accelerometer, two 70~m long laser strainmeters and two GPS stations on surface in the surroundings of the underground site. The two lasers are installed at two almost perpendicular directions in LAB780 and in one by-pass gallery between the road and train tunnels. As the LSC site is exceptionally low noise, the strainmeters set-up has been able to analyze for instance non-linear ocean load tides more than 100~km away or hydrological signatures due to the load of rain water. In addition, in the framework of the Einstein Telescope project, several seismic sensors have been installed in the tunnel galleries and taken data to study noise for the underground detection of gravitational waves (ETSEC project).

Research concerning life in extreme environments is also performed at LSC with the GOLLUM project. The Canfranc railway tunnel offers a unique environment to study microorganism communities. The GOLLUM project aims to characterize subterranean microbial communities by extraction of DNA in rock samples. Drilling tests along the train tunnel are being made to collect samples and finalize the protocol for DNA characterization.

\section{First Double Beta Decay searches} \label{firstdbd}

The very first DBD searches at LSC were performed in the late eighties and early nineties. A fruitful collaboration between the University of Zaragoza, the University of South Carolina (USC) and the Pacific Northwest National Laboratory (PNNL) started in 1988, when the first ultralow background detector, a HPGe detector, was taken to Canfranc. Preliminary measurements of materials (lead, copper, polyethylene, \dots) were carried out with this detector and at the end of 1989, the first experiment started in Canfranc, using a multidetector system.

An experiment to investigate the neutrinoless DBD of $^{76}$Ge to excited states of $^{76}$Se was carried out by the University of Zaragoza group looking for coincidences between the DBD signal and gamma emissions using a natural germanium detector surrounded by 14~NaI(Tl) hexagonal scintillators at the facility at 675 m.w.e. (presently LAB780) \cite{morales1991}. The detector was a 208~cm$^{3}$ HPGe detector made by Princeton Gamma Tech (PGT, in US) in collaboration with PNNL/USC, incorporating the latest well-known achievements regarding the ultralow background cryostats and detector components. A complete passive shielding composed of 5~cm of 2000~y old lead, in the inner layer, 20~cm of low activity lead, cadmium sheets and borated polyethylene was used. The first excited level of $^{76}$Se has an energy of 559.1~keV with J$^{P}$=2$^{+}$; therefore, the transition $0^{+} \longrightarrow 2^{+}$ should show a signal shifted to the left by this energy (in comparison to the transition to the ground state of the daughter) in the germanium two-electrons sum energy spectrum. This experiment in Canfranc was intended to further investigate a small, unexplained coincident effect close to this signal region found in a a previous search carried out at the Frejus tunnel in France \cite{morales1988}. This experiment consisted of four germanium detectors (having a total mass of 2.16~kg) surrounded by 19~NaI detectors (including those later used in Canfranc); the intriguing peak found at the level of 2.5$\sigma$ was close to the expected energy for the neutrinoless DBD signal, disappeared when the NaI window was moved away from the relevant region and it could not be attributed to any known background source. The limit set on the half-life for the neutrinoless DBD of $^{76}$Ge to the first excited state 2$^{+}$ of the daughter was $T_{1/2}>0.6 \times 10^{23}$~y. From 6062.5~h of data collected in the experiment in Canfranc, no accumulation of counts in the region of interest was observed for the coincidence events having energy deposits of 559.1~keV in the NaI detectors. These results allowed to reject as a neutrinoless DBD effect at the confidence level of 95\% the peak found in Frejus; in Canfranc the achieved background level in the region of interest was three times better that in the previous experiment.

The DBD of $^{78}$Kr to $^{78}$Se was studied investigating double positron decay and electron-positron conversion processes, with and without neutrino emission, in a collaboration between the University of Zaragoza and the Institute of Nuclear Research (INR) of Moscow \cite{nimakr}. This isotope has a transition energy of 2881~keV, the largest value among the potential double positron decay emitters; the double electron capture is also possible and has been theoretically analyzed \cite{kryptonth}. A high pressure, high resolution ionization chamber filled with krypton enriched to 94\% of $^{78}$Kr in a coincidence set-up was used. The cylindrical chamber, 14~cm long and 10.6~cm in diameter, was filled with 35~l of krypton gas in the fiducial volume and placed inside a hexagonal array of six large NaI scintillators produced by Bicron. The 511~keV annihilation gammas were tagged thanks to the NaI array. This experiment was performed at the facility at 675 m.w.e. (presently LAB780).  After 4434~h of counting time, the following half-life lower limits were set at 68\% C.L. \cite{krypton}:

\be
T_{1/2}(EC \beta^{+})_{0\nu}\geq 5.1 \times 10^{21}~\mathrm{y},
\ee
\be
T_{1/2}(EC \beta^{+})_{2\nu}>1.1 \times 10^{20}~\mathrm{y},
\ee
\be
T_{1/2}(2\beta^{+})_{0\nu+2\nu}\geq 2.0 \times 10^{21}~\mathrm{y}.
\ee

These limits represented an improvement of 3-4 orders of magnitude with respect to the ones available at that time for other $2\beta^{+}$ emitters. For $^{78}$Kr the double electron capture with two neutrino emission has been measured, with a recommended value of $T_{1/2}(ECEC)_{2\nu}=1.9^{+1.3}_{-0.8}\times10^{22}$~y \cite{barabash19}; but for the other modes the limits obtained in Canfranc are still referenced \cite{barabash11}. The theoretical half-lives presented in \cite{libroklapdor} are at the level of 10$^{27}$~y for the neutrinoless modes (assuming an effective neutrino mass of 1.0~eV) while for the two neutrino channels are $T_{1/2}(EC \beta^{+})_{2\nu}=5.3\times 10^{22}$~y and $T_{1/2}(2\beta^{+})_{2\nu}=2.3\times 10^{26}$~y.

\section{IGEX}\label{igex}

The ``International Germanium EXperiment'' (IGEX) was developed thanks to the collaboration between several American, Sovietic and Spanish groups: University of South Carolina (Columbia, South Carolina), Pacific Northwest National Laboratory (Richland, Washington), University of Zaragoza (Zaragoza, Spain), Institute for Theoretical and Experimental Physics (Moscow, Russia), Institute for Nuclear Research, Baksan Neutrino Observatory (Russia), and Yerevan Physical Institute (Yerevan, Armenia). It was devoted to the investigation of the DBD of $^{76}$Ge although other interesting searches were also performed. The very relevant results obtained by the IGEX experiment were important not only for the collaboration but also for the consolidation of Canfranc as a first-level underground facility. In this section, the IGEX detectors and their set-ups in Canfranc will be described; the background studies developed, focused on Pulse Shape Discrimination (PSD) techniques, will be presented; and finally, the results which were obtained will be commented.

\subsection{Detectors and set-up} \label{igdet}
IGEX operated three detectors of $\sim$0.7~kg and three detectors of $\sim$2~kg fiducial mass each, corresponding to a total mass of 8.4~kg. They were produced from germanium isotopically enriched to 86\% in $^{76}$Ge and mounted in ultralow radioactive background cryostats electroformed from purified CuSO$_{4}$ solution. The technical details were published in \cite{igexnpbps99,igexnpbps00,igex99}.

The IGEX detectors were fabricated at Oxford Instruments, Inc., in Oak Ridge, Tennessee. Russian GeO powder, isotopically enriched to 86\% $^{76}$Ge, was purified, reduced to metal, and zone refined to $\sim$10$^{13}$ p-type donor impurities per cubic centimeter by Eagle Picher, Inc., in Quapaw, Oklahoma. The metal was then transported to Oxford Instruments by surface in order to minimize activation by cosmic ray neutrons, where it was further zone refined, grown into crystals, and fabricated into detectors. The three small detectors (named ``Rico''), produced during the first phase of IGEX, were all cut from the same crystal. They had total masses of 1006, 1018, and 896~g, respectively. The total active mass of the three detectors was $\sim$2~kg. The three large enriched detectors (named ``Rico Grande'' and designated RG-I, RG-II, and RG-III) had masses of 2150, 2194, and 2121~g, respectively. The active volumes of all three Rico Grande detectors were measured with a collimated source of $^{152}$Eu in the LSC; the results were in good agreement with the standard Oxford efficiency measurements. All three detectors had active masses of $\sim$2~kg and efficiencies of $\sim$100\% relative to a 3'$\times$3' cylindrical NaI(Tl) detector for 1333~keV gamma rays. The dead layer thickness of RG-I was also measured at various positions on the crystal and resulted in an average thickness of 0.5~mm.

All of the cryostat parts were electroformed onto stainless steel mandrels using a high purity Oxygen Free High Conductivity (OFHC) copper/CuSO$_{4}$/H$_{2}$SO$_{4}$ plating system. The solution was continuously filtered to eliminate copper oxide, which causes porosity in the copper. A Ba(OH)$_{2}$ solution was added to precipitate BaSO$_{4}$, which was also collected on the filter. Radium in the bath exchanges with the barium on the filter, thus minimizing radium contamination in the cryostat parts. The CuSO$_{4}$ crystals were purified of thorium by multiple recrystallization. The cryostats of all the large detectors were fabricated using these improved procedures, following the results obtained for the first three smaller detectors \cite{igex99}.

The ``Rico'' detectors were operated one each in the Homestake gold mine (used now for the present Sanford Underground Research Facility) in South Dakota, US, in the LSC in Spain, and in the Baksan Neutrino Observatory in Russia, in order to evaluate the conditions at these sites. Considering overburden, location, and available space, Canfranc was ultimately chosen as the experimental site. The three``Rico Grande'' detectors were produced one at a time, tested, and eventually operated in Canfranc, while all three small detectors were eventually operated in Baksan. All the available data from Homestake, Baksan and Canfranc were used in the IGEX analysis; the operating periods, exposure, overburden, and shielding conditions for each location are given in Table~\ref{igexinfo}.

\begin{table}
\begin{center}
\begin{minipage}[t]{16.5 cm}
\caption{Summary of the experimental parameters of the three laboratories where the IGEX detectors were operated: operating times, rock overburden, accumulated exposure and shielding conditions.} \vskip 0.5 cm
\label{igexinfo}
\end{minipage}
\begin{tabular}{l|cccc}
\hline
Laboratory & Operating period & Overburden & Exposure & Shielding \\ \hline
Homestake & September 1994-June 1997 & 4000 m.w.e. & 31.13 mol y & passive   \\
Baksan & June 1994-May 1999 & 660 m.w.e. & 33.03 mol y &  passive/active \\
Canfranc & February 1996-June 1999 &2450 m.w.e. & 52.61 mol y & passive/active  \\
\hline
\end{tabular}
\end{center}
\end{table}

The quest for an ultralow background started in IGEX with a thorough radiopurity screening of the materials to be used in the detectors and in the inner components of the shielding. As already mentioned, the copper parts of the cryostats were produced by special techniques to eliminate Th and Ra impurities. The first stage field-effect transistor (FET) was mounted on a Teflon block a few centimeters apart from the inner contact of the crystal and shielded by 2.6~cm of 500~y old lead to reduce the background. Also the protective cover of the FET and the glass shell of the feedback resistor were removed for such purpose. Further stages of amplification were located 70~cm away from the crystal. All the detectors had preamplifiers modified for pulse shape analysis for background identification.

\begin{figure} 
\centerline{\includegraphics[width=8cm]{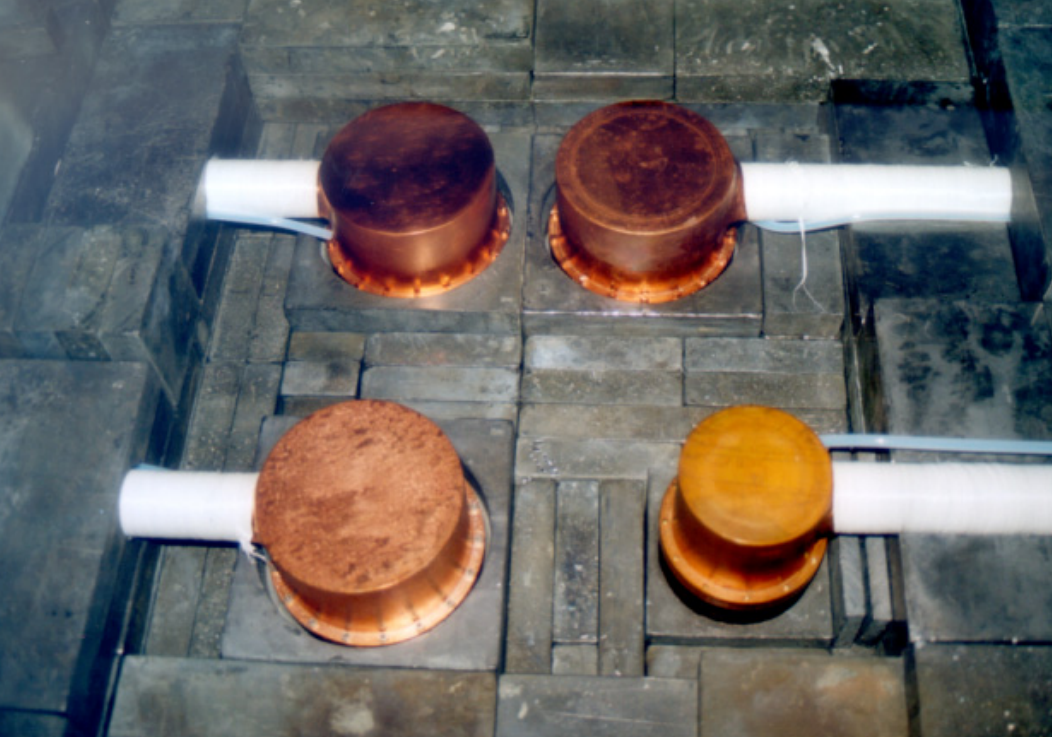}}
\centerline{\includegraphics[width=8cm]{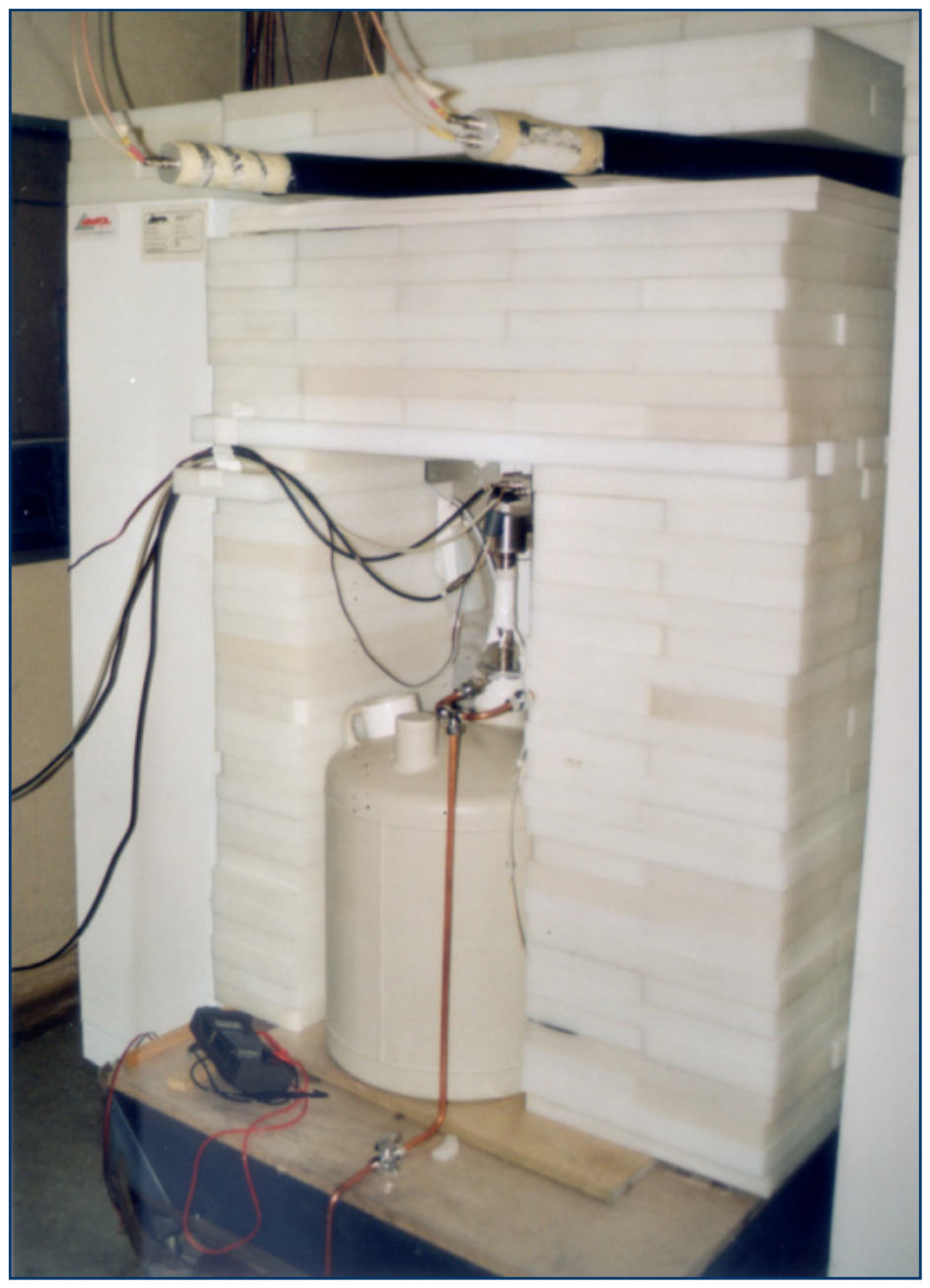}}
\begin{center}
\begin{minipage}[t]{16.5 cm}
\caption{Pictures of the IGEX set-up at LSC: innermost lead shielding with the three large and one small germanium detectors inside their copper cryostats fitted into precision-machined holes (top) and outermost view of the whole shielding, with the polyethylene bricks visible (bottom).\label{igexpictures}}
\end{minipage}
\end{center}
\end{figure}

A heavy shielding enclosing tightly the set of detectors was developed for the IGEX set-up running at LSC. The innermost shielding consisted of 2.5~tons ($\sim$60~cm cube) of archeological lead (2000~y old, from old ships) having a $^{210}$Pb($^{210}$Bi) content of $<$10~mBq/kg \cite{igex99}, where the three large detectors were fitted into precision-machined holes to minimize the empty space around the detectors available to radon (see Fig.~\ref{igexpictures}, top). Nitrogen gas evaporated from liquid nitrogen was injected into the remaining free space to minimize radon intrusion. Surrounding the archeological lead block there was a 20-cm thick layer ($\sim$10~tons) of 70-year-old low activity lead, sealed with plastic and 2-mm-thick cadmium sheets. An active veto rejected muon-induced events, covering the top and sides of the set-up, except where the detector Dewars were located; it consisted of BICRON BC-408 plastic scintillators 5.08~cm $\times$ 50.8~cm $\times$ 101.6~cm with surfaces finished by diamond mill to optimize internal reflection. BC-800 light guides were coupled to Hamamatsu R329 PMTs. The anticoincidence veto signal was obtained from the logical OR of all PMT discriminator outputs. A neutron shielding completed the barrier against external sources of background, made of polyethylene bricks and borated water with a thickness of 20~cm (see Fig.~\ref{igexpictures}, bottom), which was enlarged up to even 80~cm during the last period of the data taking, focused on the direct detection of dark matter particles. The shieldings used in Homestake and Baksan were similar to that used in Canfranc with the exception that their innermost shields were 450-y-old Spanish lead and copper stored underground for more than 20~y, respectively. There was no muon veto in Homestake, placed at 4000~m.w.e.

The data acquisition system had an independent spectroscopy chain for each germanium detector. It was based on standard NIM electronics. For each detector, two Canberra 2020 linear amplifiers having different shaping times were used to filter noise \cite{julio1992} and connected to Canberra 8701 Wilkinson type analog-to-digital converters (ADCs) controlled by computers through parallel interfaces. The preamplifiers were modified to optimize PSD. Each preamplifier fast-pulse output was routed to a LeCroy 9362 or 9360 digital oscilloscope. The sampling rate was 2~ns per point. The threshold was set to 1.5~MeV in plastic scintillators to register low energy deposits of muons and to 100~keV in germanium detectors to avoid unnecessary high count rates. For each germanium event, the following information was recorded: the time elapsed since run started with a precision of 100~$\mu$s, the elapsed time since the last veto signal with a precision of 20~$\mu$s, the ADC channel numbers giving the event energy (range 0-8191, $\sim$1~keV/ch) and the scope trace with pulse shape (500~points, 2~ns point).

Special care was put in maintaining a long-term stability in gain, resolution, noise level and counting rates. The FWHM energy resolutions of the three ``Rico Grande'' IGEX detectors at 1333~keV were 2.16, 2.37, and 2.13~keV, and the energy resolution of the summed data integrated over the time of the experiment was $\sim$4~keV at the transition energy of $^{76}$Ge. Concerning gain stability, the energy shifts at $\sim$1.3~MeV were smaller than 0.5~keV (typically 0.3~keV) over 2~months.

\subsection{Background and Pulse Shape Discrimination}

\begin{figure} 
\centerline{\includegraphics[width=11cm]{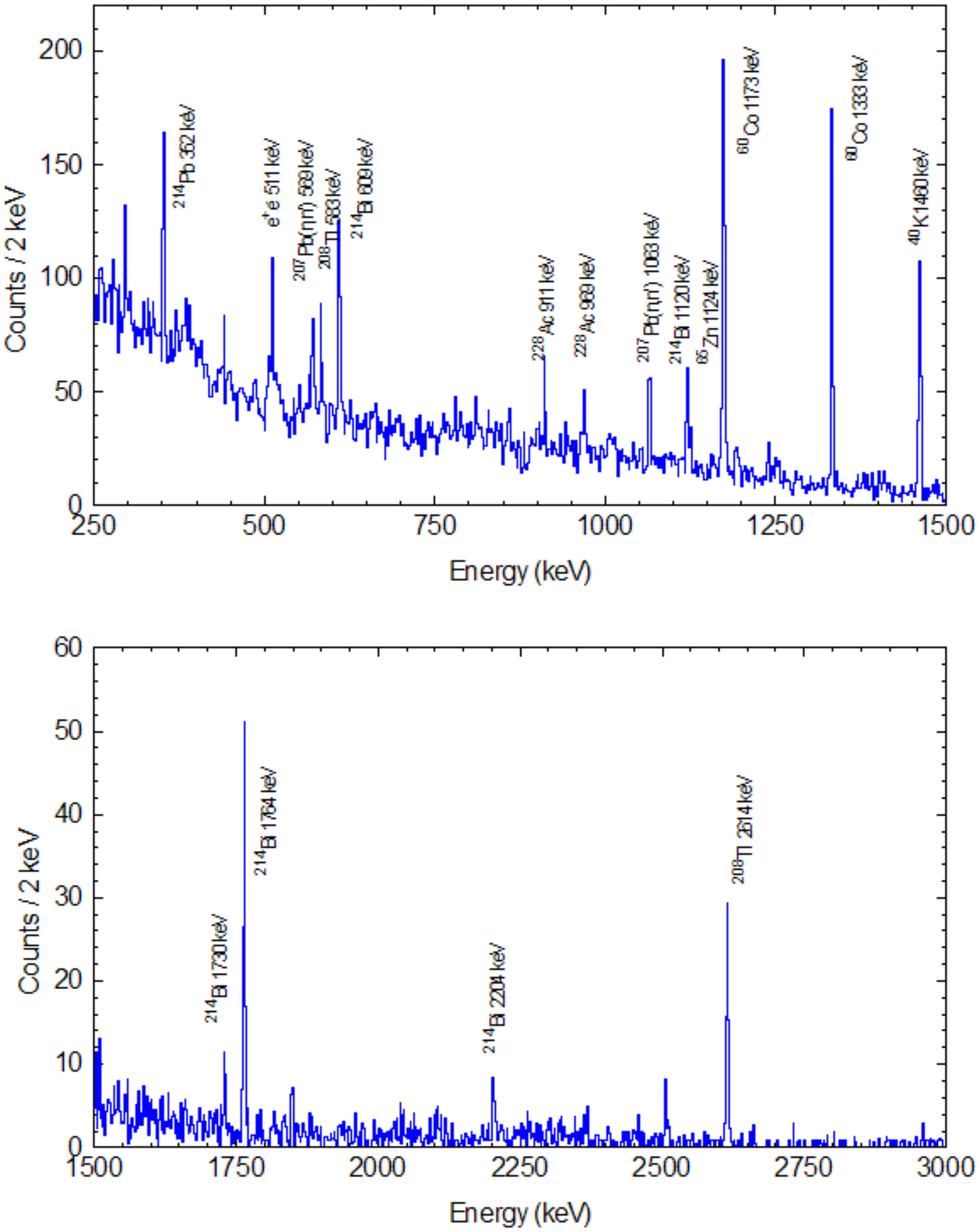}}
\begin{center}
\begin{minipage}[t]{16.5 cm}
\caption{Energy spectrum measured by one of the large IGEX detector (the one named RG-II) operated in Canfranc for an exposure of 2.75~kg y (as reported in \cite{dbdmorales}). The main identified emissions are labeled indicating energy and the responsible isotope. \label{spcrg}}
\end{minipage}
\end{center}
\end{figure}

A background analysis for IGEX detectors was presented in \cite{igex99}. It was concluded that the majority of the registered background came from cosmic-ray neutron spallation reactions that occurred in the detector and cryostat components while they were above ground. Figure \ref{spcrg} shows as an example the measured energy spectrum of one of the large IGEX detector (the one named RG-II). Most of the background in the region of interest was accounted for by cosmogenic activated nuclei $^{68}$Ge and $^{60}$Co; the production rates of the relevant cosmogenic isotopes for DBD in copper and in both enriched and natural germanium have been computed for instance in \cite{cosmogenicap2010}. The background recorded in the energy region between 2.0 and 2.5~MeV, prior to PSD analysis, was about 0.2~counts keV$^{-1}$ kg$^{-1}$ y$^{-1}$ \cite{igex99}.

In addition to the active and passive shieldings implemented in the IGEX set-ups, as described in Sec.~\ref{igdet}, the final step in the background reduction process was obtained through the event selection via the PSD analysis intended to eliminate multi-site events, characteristic of non-DBD events, as described in detail in \cite{psdnima2003}. The rationale for this PSD is quite simple: in large intrinsic germanium detectors, the electric field increases by a factor 10 or more from the inner to the outer conductors, which are almost 4~cm apart. Electrons and holes take about 300-500~ns to reach their respective conductors. The current pulse contributions from electron and holes are displacement currents, and therefore, dependent on their velocities and radial positions. Accordingly, events occurring at a single site (like DBD events for example) have associated current pulse features that reflect the position in the crystal where the event occurred. More importantly, these single-site events (SSE) have frequently pulse shapes that differ significantly from those due to the most dominant background events that produce electron-hole pairs at several sites by multi-Compton-scattering process, for instance. Consequently, pulse shape analysis can be used to distinguish between these two types of energy depositions.

\begin{figure} 
\begin{center}
\centerline{\includegraphics[width=12cm]{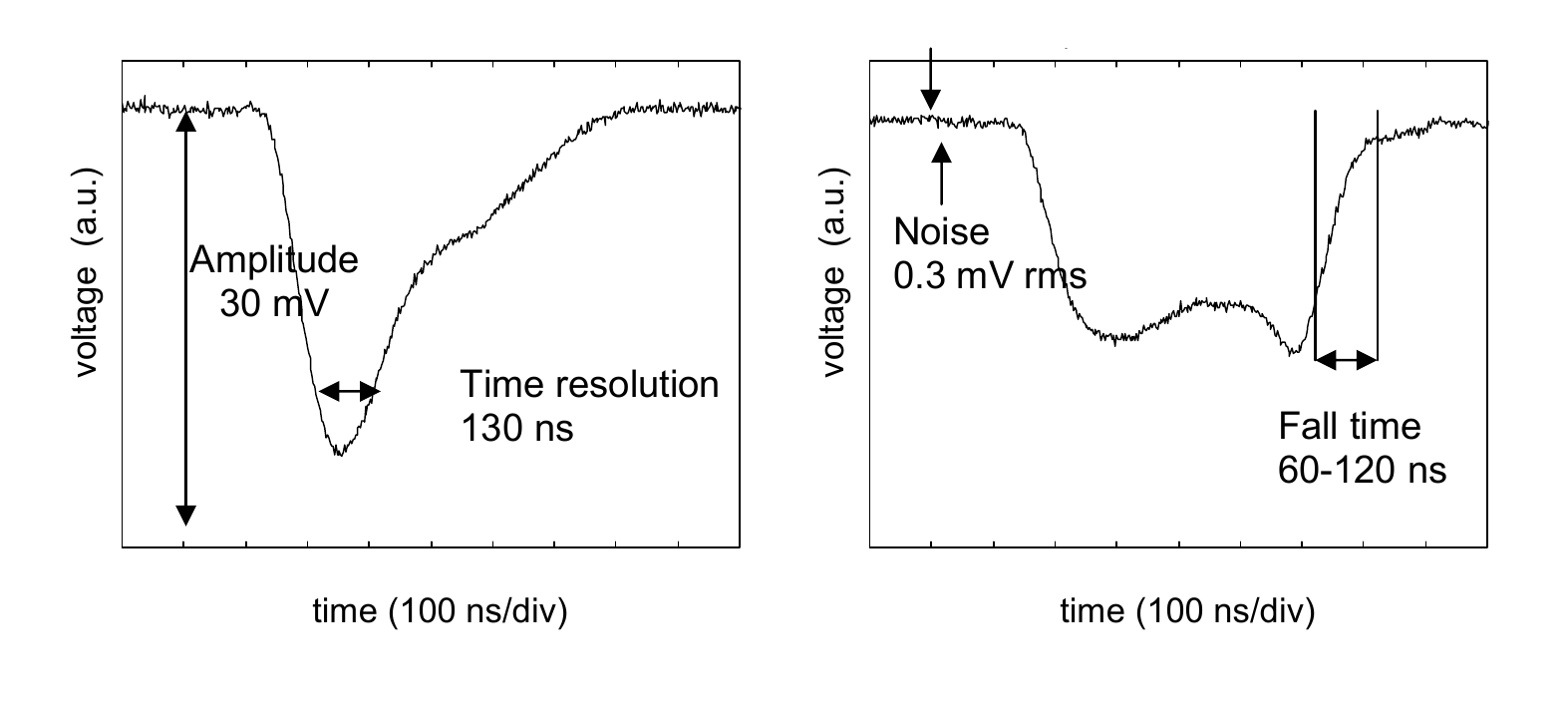}}
\begin{minipage}[t]{16.5 cm}
\caption{Main features of the digitized experimental pulses of the IGEX detectors (amplitude, electronic noise level, fall time and time resolution inferred from the width of the peaked features) and the typical values obtained for each of them from two measured example pulses (reproduced from \cite{psdnima2003}).} \label{igexpulse}
\end{minipage}
\end{center}
\end{figure}

To develop PSD techniques it is helpful to work with a signal as close as possible to the displacement current of the detector, recording the current pulses at a very early stage of preamplification. This allows the development of algorithms that do not depend strongly on the preamplifier electronics in use. Figure~\ref{igexpulse} shows the main features of the digitized pulses for the IGEX detectors. The transfer function of the preamplifier and associated front-end stage was measured for each detector as the response of the preamplifier for a narrow $\delta$-like signal from a signal generator, allowing the reconstruction of the displacement current and easy comparison to computed pulse shapes. The electric field in the crystals was numerically characterized and detailed models of the detectors and associated first stage preamplifier were constructed to simulate pulse shapes from various sources of background \cite{psdnima2003}. DBD events deposit energy at a single site in a detector and most of background events deposit energy at several sites. The models of the structure of the current pulse revealed that SSEs exhibit only one or two features, or ``lobes'', in more than 97\% of the cases. Multiple-site events most often exhibit more than two lobes. The short distance (a few cm) between the gate of the FET  and the detector contact in IGEX detectors resulted in single-site and multi-site pulses of good quality for PSD, similar to those obtained by the Heidelberg-Moscow experiment \cite{psdradial2}.

\begin{figure}
\centerline{\includegraphics[width=13cm]{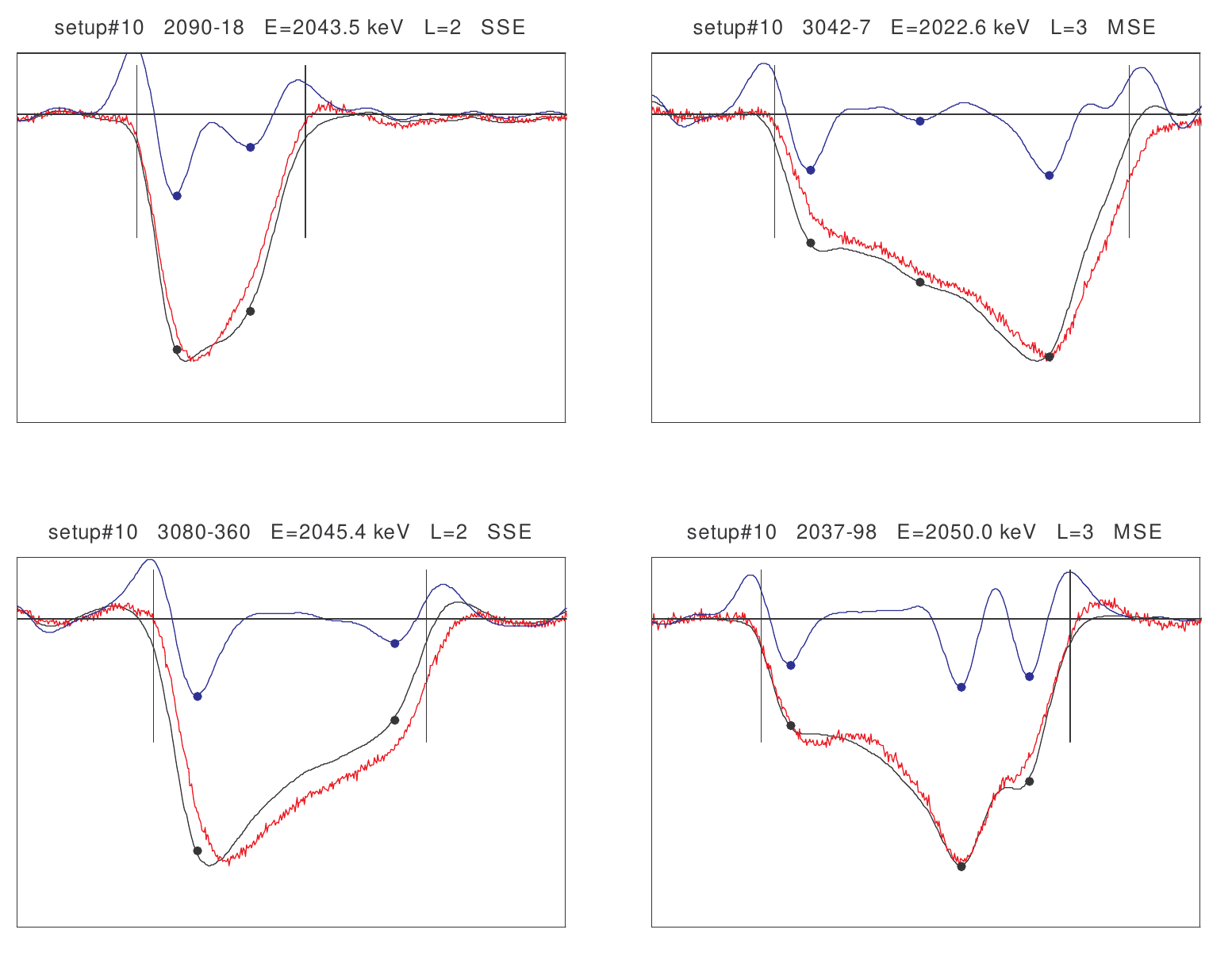}}
\begin{center}
\begin{minipage}[t]{16.5 cm}
\caption{Examples of the effect of applying the ``Mexican-hat'' filter to detect significant lobes (marked by the big points) in the digitized IGEX pulses (reproduced from \cite{psdnima2003}). Four different event pulses with energy in the region of interest are shown: events on the left are accepted (having two lobes), while those on the right are rejected (having three lobes). The raw pulse, the one unfolded and the result after the filter application are depicted (see text).}
\label{lobes}
\end{minipage}
\end{center}
\end{figure}

The PSD method used in IGEX consisted in counting the number of lobes of the pulses and rejecting those events having more than two significant lobes or peaks. A SSE pulse is expected to have at most two lobes, one due to electrons and the other due to holes. Experimental pulses were first unfolded using the transfer function of the preamplifier. Then, to detect lobes, a ``Mexican-hat'' filter (second derivative of a gaussian) of the proper width was applied to the pulse. The filtered signal has a null mean value where there is no lobe in the original signal and a peak where a lobe is present. Therefore, it was straightforward to reject all the events having more than two lobes. The proper width of the filter was chosen to optimize the fraction of misidentified events on a sample of calculated SSE pulses. Figure \ref{lobes} shows the results of applying this method to four different pulses.

The effectiveness of this method was checked by applying it to a $^{22}$Na calibration spectrum and to a set of data taken following a large intrusion of radon in the shielding. It is worth noting that some multi-site events may show only one or two lobes and then they were not rejected by this technique. Table~\ref{psdresults} summarizes the results (exposure, background levels in the region of interest from 2 to 2.5~MeV before and after the PSD and rejection factor) for each detector where the procedure was applied. The use of this PSD method resulted in the rejection of $\sim$60\% of the IGEX background in the energy interval 2.0-2.5~MeV, going down to $\sim$0.09~counts keV$^{-1}$ kg$^{-1}$ y$^{-1}$ \cite{psdnima2003}.

\begin{table}
\begin{center}
\begin{minipage}[t]{16.5 cm}
\centering \caption {Results of applying the PSD in the IGEX experiment: exposure,
background levels B in the 2.0-2.5~MeV region before and after the
discrimination and rejection factors achieved. Results are shown for individual detectors and overall.} \label{psdresults}
\end{minipage}
\vskip 0.5 cm
\begin{tabular}{c|cccc}
\hline
   & exposure  & B before  &  B after  & rejection factor \\
& (kg y) & (counts keV$^{-1}$ kg$^{-1}$ y$^{-1}$) & (counts keV$^{-1}$ kg$^{-1}$ y$^{-1}$) &  (\%) \\ \hline
RG-II & 2.75 & 0.27 &  0.10 & 62 \\
RG-III & 1.90 & 0.26 &  0.11 & 58 \\  \hline
 total & 4.65 & 0.26 & 0.10 & 60 \\     \hline
\end{tabular}
\end{center}
\end{table}

The background levels finally achieved by IGEX and Heidelberg-Moscow experiments were considered as a stationary limit whose reduction required further and deeper investigation. Besides using larger quantities of the emitter nucleus, the goals of the next extended germanium experiments, presently in operation (see Sec.~\ref{igexww}), were to substantially improve the radiopurity of the detectors and components (both intrinsic and induced) and to suppress, as best as possible, the background originated from external sources. Although after the end of the IGEX experiment, no other investigation of the DBD of $^{76}$Ge has been carried out at LSC, a prospective study was made in \cite{hector2007}; it was intended to assess the attainable background reduction in the energy region where the neutrinoless DBD signal is expected to appear for experiments using germanium detectors, taking into consideration different strategies like the granularity of the detector system, the segmentation of each individual germanium detector and the application of pulse shape analysis techniques to discriminate signal from background events. The required conditions (regarding crystal mass, radiopurity, exposure to cosmic rays, shielding and rejection capabilities) to achieve a background level of 10$^{-3}$ counts keV$^{-1}$ kg$^{-1}$ y$^{-1}$ in the region of interest were evaluated; a background rate even lower has been already achieved in the phase II of the GERDA experiment, as it will be commented in Sec.~\ref{igexww}.

\subsection{Results}

First results from analyzing 74.84~mol y of $^{76}$Ge\footnote{mol y refers to moles of $^{76}$Ge} data were presented in \cite{igex99}, operating the three crystals of $\sim$2~kg of enriched germanium at LSC and one of the smaller ones in Baksan, Russia. A maximum likelihood analysis yielded a lower bound for the half-life for the neutrinoless DBD of $^{76}$Ge:
\be
T_{1/2}^{0\nu}\geq0.8 \times 10^{25}~\mathrm{y} \hskip 0.2 cm (90 \% \mathrm{C.L.}) \label{firstigexlimit}
\ee

\begin{figure}
\centerline{\includegraphics[width=12cm]{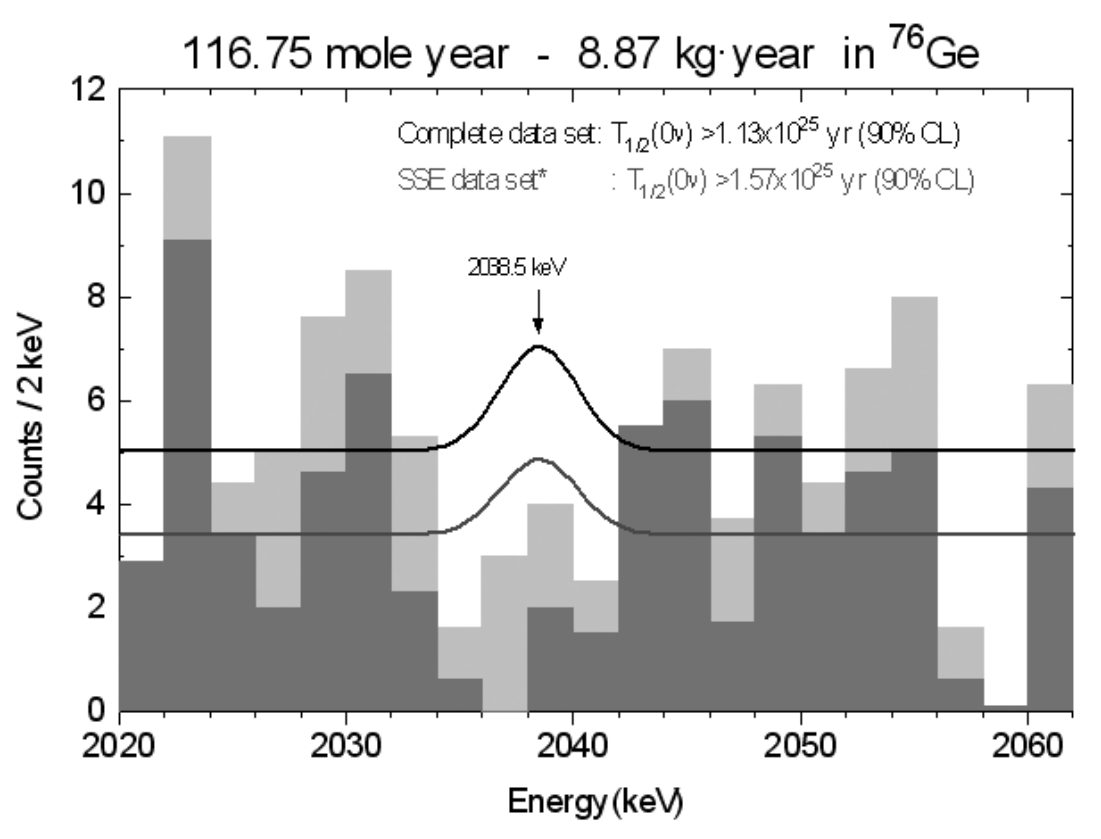}}
\begin{center}
\begin{minipage}[t]{16.5 cm}
\caption{Energy spectrum measured by the IGEX experiment in the region of interest for the neutrinoless DBD of $^{76}$Ge from 117~effective mole years of data (as presented in \cite{dbdmorales}). Energy bins are labeled on the left edge. The dark grey spectrum for the single-site events (SSE) data set results from the application of PSD to $\sim$45\% of the complete data set. The Gaussian curves in the two solid lines represent the 90\% C.L. constraints of $<$3.1(4.3)~events for the SSE (complete) data set, for the neutrinoless DBD peak considering a FWHM of $\sim$4 keV, corresponding to the effective energy resolution of the entire experiment.} \label{igexspc}
\end{minipage}
\end{center}
\end{figure}

For the final results presented in \cite{igexfinal}, the IGEX collaboration analyzed 117~mol y of $^{76}$Ge data (8.89~kg y of $^{76}$Ge, or 10.14~kg y using the total mass including 14\% of $^{74}$Ge) from its isotopically enriched germanium detectors. The PSD analysis (based on the comparison of experimental pulse shapes to computed single-site and multi-site pulses) could only be applied to the most  recent data. Figure~\ref{igexspc} shows the measured energy spectrum in the region of interest where the neutrinoless DBD signal is expected to appear; the darkened spectrum resulted from applying PSD to about 15\% of the first $\sim$75~mol y data set and to the entire additional 41.9~mol y data (45\% of the total data). The number of events for each one of the 2-keV energy bins are presented in Table~\ref{igexdata}, with and without application of PSD. The total number of counts registered in the 40-keV interval from 2020 to 2060~keV prior to PSD corresponded to a rate of 0.24~counts keV$^{-1}$ kg$^{-1}$ y$^{-1}$; after the PSD analysis, the rate in the same region was 0.17~counts keV$^{-1}$ kg$^{-1}$ y$^{-1}$ in the data set. The background computed only from the 45\% of the PSD analyzed data (53~mol y), always in the same 40-keV region, was 0.10~counts keV$^{-1}$ kg$^{-1}$ y$^{-1}$ \cite{igexnpbps00}.
Using standard statistical techniques \cite{bartlett} (the ones recommended at that time by the Particle Data Group), it was derived that there were fewer than 3.1 candidate events at 90\% C.L. under a peak having a FWHM of 4~keV and centered at 2038.56~keV. As no positive signal for the neutrinoless DBD of $^{76}$Ge was identified, the following lower bound on the half-life was set:
\be
T_{1/2}^{0\nu}>1.57 \times 10^{25}~\mathrm{y} \hskip 0.2 cm (90 \% \mathrm{C.L.}) \label{igexlimit}
\ee
It is worth noting that if the complete data set before PSD analysis was considered, the corresponding bound would be $T_{1/2}^{0\nu}>1.13 \times 10^{25}$~y.

The upper limit to the effective neutrino mass presented in \cite{igexfinal} using the results with PSD was:
\be
m_{\beta\beta}<(0.33-1.35)~\mathrm{eV},
\ee
depending on the choice of theoretical nuclear matrix elements used in the analysis. Different shell model and QRPA calculations were considered; the smallest effective mass comes from \cite{haxton} (with $F_{N}=1.56\times10^{-13}$~y$^{-1}$) and the largest one from \cite{vogel} (with $F_{N}=9.67\times10^{-15}$~y$^{-1}$). In \cite{igexresponse}, using a more complete and updated list of theoretical calculations for nuclear matrix elements, new values for the upper limits on the effective neutrino mass were derived, ranging from 0.29~eV using results from \cite{simkovic} to 1.17~eV from \cite{stoica}.

\begin{table}
\begin{center}
\begin{minipage}[t]{16.5 cm}
\caption{IGEX data for the bins from 2020 to 2060~keV for 117~mol y (8.89~kg y) in $^{76}$Ge after the partial application of PSD for single-site event (SSE) identification as well as for the complete data set (as reported in \cite{dbdmorales}). The starting energy of each 2-keV bin is given. Counts per bin are presented.}
\label{igexdata}
\end{minipage}
\vskip 0.5 cm
\begin{tabular}{ccc}
\hline E (keV) & SSE data set & complete data set\\ \hline
2020 & 2.9 & 2.9 \\
2022 & 9.1 & 11.1 \\
2024 & 3.4 & 4.4\\
2026 & 2.0 & 5.0 \\
2028 & 4.6 & 7.6 \\
2030 & 6.5 & 8.5 \\
2032 & 2.3 & 5.3 \\
2034 & 0.6 & 1.6 \\
2036 & 0.0 & 3.0 \\
2038 & 2.0 & 4.0 \\
2040 & 1.5 & 2.5 \\
2042 & 5.5 & 5.5 \\
2044 & 6.0 & 7.0 \\
2046 & 1.7 & 3.7 \\
2048 & 5.3 & 6.3 \\
2050 & 3.4 & 4.4 \\
2052 & 4.6 & 6.6 \\
2054 & 5.0 & 8.0 \\
2056 & 0.6 & 1.6 \\
2058 & 0.1 & 0.1 \\
2060 & 4.3 & 6.3 \\
\hline
\end{tabular}
\end{center}
\end{table}

\begin{figure}
\centerline{\includegraphics[width=10cm]{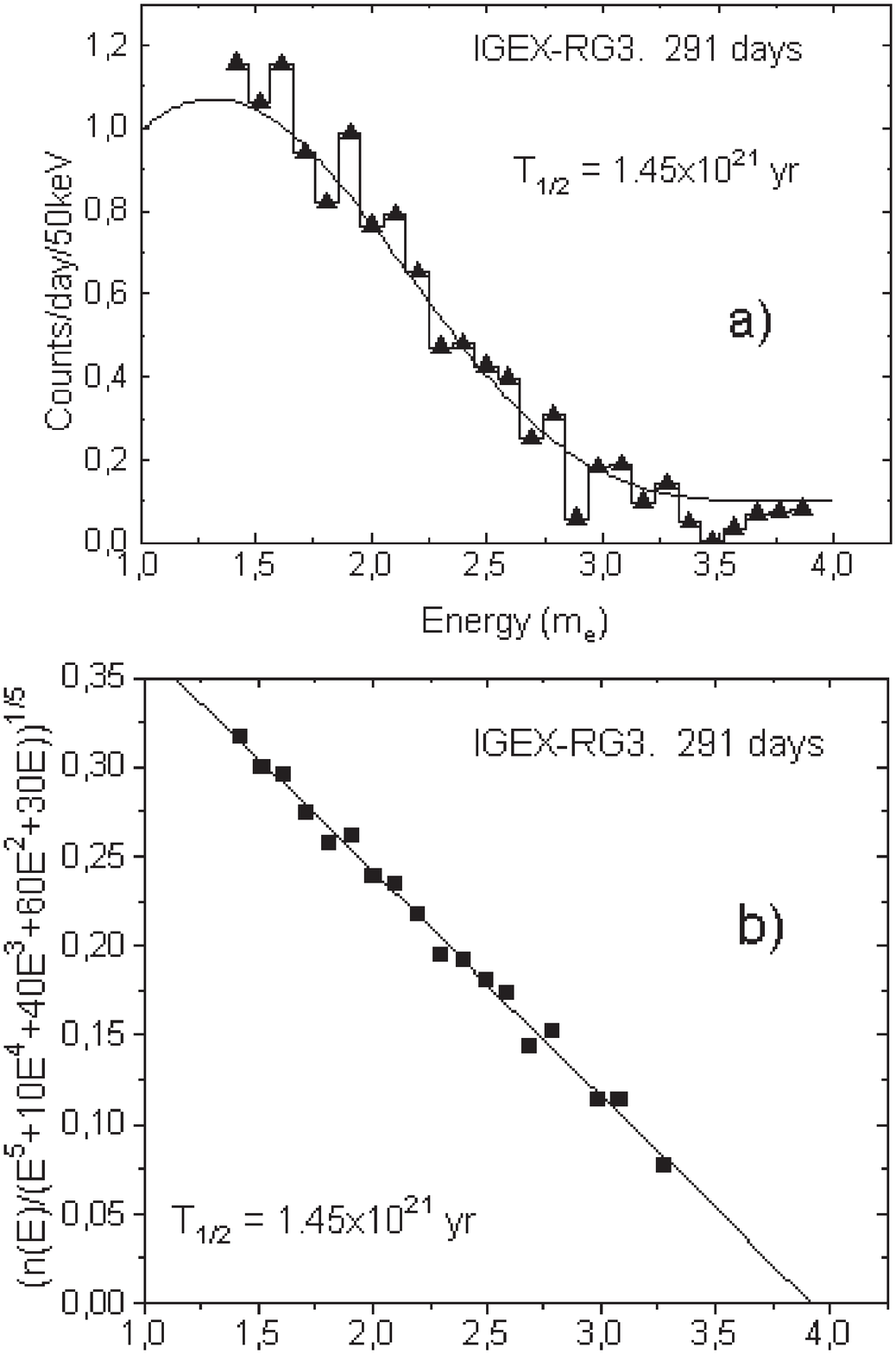}}
\caption{Estimate of the two neutrino DBD half-life of $^{76}$Ge from one of the IGEX detectors (as presented in \cite{dbdmorales}): a) Best fit to the stripped data corresponding to a half-life of $T^{2\nu}_{1/2}=(1.45 \pm 0.20) \times 10^{21}$~y. b) Experimental points fitting the double beta Kurie plot. Note that the energy scale for X-axis is given in units of electron mass.} \label{2nu}
\end{figure}

The channel of DBD with emission of neutrinos was also studied in IGEX \cite{tesisAna}. Data from one of the IGEX detectors (the one named RG-III) corresponding to 291~days, were used to set a value for the two neutrino DBD mode half-life by simply subtracting the Monte Carlo simulated background. In Fig.~\ref{2nu}, the plot (a) presents the best fit to the stripped data and the deduced half-life was:
\be
T^{2\nu}_{1/2}=(1.45 \pm 0.20) \times 10^{21}~\mathrm{y},
\ee
whereas plot (b) shows how the experimental points fit the double beta Kurie plot \cite{dbdmorales,amnu98}.

Together with the analysis of the  DBD of $^{76}$Ge, the IGEX data were used to derive other relevant results dealing with dark matter and background components.
\begin{itemize}
\item An upgrade of the data acquisition system and of the shielding of the IGEX detectors operating in Canfranc made possible to use them from the year 2000 in a search for dark matter particles, through the germanium nuclear recoil produced by the WIMP elastic scattering \cite{igexdm2000} (IGEX-DM experiment). In particular, data from one of the large detectors (named RG-II) reaching an energy threshold of $\sim$4~keV were used. Exclusion plots for WIMP-nucleon spin-independent cross-sections as a function of the WIMP mass were obtained. The results using 40~days of data \cite{igexdm} improved the exclusion limits derived from all the other ionization germanium detectors at that time in the mass region from 20 to 200~GeV, where a WIMP supposedly responsible for the annual modulation effect reported by the DAMA experiment would be located. The IGEX exclusion contour entered, for the first time, the DAMA region singled-out at that moment \cite{dama2000} by using only raw data, with no background discrimination, excluding its upper left part.

\item As the neutron background is particularly relevant for dark matter searches, a quantitative study of the neutron environment at LSC was performed in the context of the IGEX-DM experiment, based on a complete set of simulations tested with IGEX low energy data obtained in different shielding conditions \cite{Carmona}. The results of the study allowed to conclude, with respect to the IGEX-DM background, that the main neutron population, coming from radioactivity from the surrounding rock, was practically eliminated after the implementation of a suitable neutron shielding chosen after successive tests with increasing thickness and with other improvements \cite{igexdm,igexdm2000}. The remaining neutron background (due to muon-induced neutrons in the rock and shielding) was substantially below the background level thanks to the muon veto system used. In addition, this study gave a further insight on the effect of neutrons for other experiments in the LSC. The comparison of simulations with the body of data available allowed to set the flux of neutrons from radioactivity of the Canfranc rock as (3.82$\pm$0.44)$\times 10^{-6}$~cm$^{-2}$s$^{-1}$, as well as the flux of muon-induced neutrons in the rock ((1.73$\pm$0.22(stat)$\pm$0.69(syst))$\times 10^{-9}$~cm$^{-2}$s$^{-1}$) or the rate of neutron production by muons in the shielding lead ((4.8$\pm$0.6(stat)$\pm$1.9(syst))$\times 10^{-9}$~cm$^{-3}$s$^{-1}$). The total neutron flux at LSC measured afterwards using $^{3}$He proportional counters \cite{Jordan}, as mentioned in Sec.~\ref{lscbkg}, is in very good agreement with the first estimate from IGEX-DM data.
\end{itemize}

\subsection{Semiconductor Double Beta Decay experiments worldwide}
\label{igexww}

The Heidelberg-Moscow (HM) experiment \cite{GUN97,hm}, operated at LNGS in Italy and devoted also to investigate the DBD of $^{76}$Ge, was contemporary to IGEX. HM used five detectors with a total mass of 10.96~kg of germanium, equally enriched in $^{76}$Ge. From the
analysis of the neutrinoless channel of DBD, the following limits were
presented, slightly better that those derived by IGEX: $T_{1/2}^{0\nu}\geq 1.9 \times 10^{25}$~y and $m_{\beta\beta} \leq 0.35$~eV, at 90 \% C.L. \cite{klapdor01}. Soon after, a few members of the collaboration after a reanalysis of data raised a claim for an evidence \cite{klapdorevidence} opposed by the community
\cite{response,zdesenko,hmotros}. Once finished the data taking, analyzing more
than 13~y of data and after a new analysis \cite{klapdoranalisis}, their final
results pointed to hints of a positive signal \cite{klapdorevidence2,klapdorevidence3,klapdorevidence4},
pending to be confirmed or completely refuted by other experiments. For 71.7~kg y of data, authors
considered a 4.2$\sigma$ confidence level for the signal, corresponding to a
half-life $T_{1/2}^{0\nu}=1.2 \times 10^{25}$~y and an effective neutrino mass
$m_{\beta\beta}=0.44$~eV. The criticism raised by some of the members of this collaboration to the final IGEX results \cite{criticismhm} was answered at \cite{igexresponse}.

These first germanium experiments seemed to have achieved an insurmountable limit in material radiopurity, so new projects developed new strategies to get rid of the radioactive background. Both IGEX and HM experiments used a semi-coaxial detector design but an intense work to search for new germanium detector designs aiming at improving the background suppression compared to the semi-coaxial type has been developed in cooperation with manufacturers. Many efforts were devoted to apply segmented detectors to DBD experiments, applying more sophisticated techniques of background rejection than in conventional germanium detectors \cite{segmented1,segmented2,segmented3,segmented4,segmented5,gomez07,psaseg1,psaseg2,psaseg3,psaseg4,psaseg5}. Development of Broad Energy germanium (BEGe) detectors opened new possibilities for a further improvement of this kind of experiments \cite{broadenergy1,broadenergy2}. For BEGe design, the average mass is typically by a factor 2-3 smaller, but its design has been found to lead to an improved energy resolution and superior background rejection capability by means of pulse shape analysis of the signal waveforms. New discrimination techniques are being explored based on different approaches \cite{psdgerda2019,psdmajorana2019}.

The ``GERmanium Detector Array'' (GERDA) \cite{gerda} is based on the operation in cryogenic liquid of naked germanium diodes with an enriched $^{76}$Ge fraction with the aim to minimize material around the detectors. GERDA has been conceived in two phases operating at LNGS. In the first phase (Phase I) of the experiment, the enriched HPGe detectors used by IGEX and HM experiments ($\sim$18~kg) were refurbished for operation in a bare mode and immersed in liquid argon, after being dismounted from their cryostat. An overall background level of 10$^{-2}$~counts keV$^{-1}$ kg$^{-1}$ y$^{-1}$ was achieved, a factor of ten better than those of the previous experiments, after PSD \cite{gerdaIpsd}. No signal was found and using data collected from 2011 to 2013 (21.6~kg$\cdot$y) a lower limit was set on the half-life for the neutrinoless DBD of $^{76}$Ge $T_{1/2}^{0\nu}>2.1 \times 10^{25}$~y (90\% C.L.) \cite{gerdaphaseI}. A new result for the half-life of the neutrino-accompanied DBD of this isotope with significantly reduced uncertainties was also derived from the Phase I data as $T_{1/2}^{2\nu}=(1.926 \pm 0.094)\times 10^{21}$~y (reported uncertainty combines statistical and systematic ones), together with new limits for the half-lives of DBD with Majoron emission \cite{gerda2nuM}, DBD with two neutrino emission into excited states \cite{gerdaexcited} and the radiative neutrinoless double Electron Capture of $^{36}$Ar \cite{gerdaECAr}. A second phase (Phase II) started at the end of 2015 after a major upgrade with additional 30~BEGe detectors \cite{gerdabe}, which feature an excellent background suppression from the analysis of the time profile of the detector signals. Thanks to the increased detector mass and performance of the enriched germanium diodes and due to the introduction of liquid argon instrumentation techniques in a novel active veto system \cite{gerdaIIupgrade}, it was possible to reduce the background down to an unprecedented level of 5.7$\times$10$^{-4}$~counts keV$^{-1}$ kg$^{-1}$ y$^{-1}$ \cite{gerdanature,gerdaphaseII,gerdascience}. No signal has been observed and a 90\% C.L. lower limit for the half-life of $T_{1/2}^{0\nu}>0.9 \times 10^{26}$~y has been placed when combining with previous data from Phase I \cite{gerdascience}. Phase II will continue for the collection of an exposure of 100~kg$\cdot$y. If no signal is found by then, the GERDA sensitivity will have reached $1.1\times 10^{26}$~y for setting a 90\% C.L. limit and a sensitivity to the effective neutrino mass of 0.07-0.16~eV.

The MAJORANA Collaboration has constructed and operated the \textsc{Majorana Demonstrator}, an ultra-low background, modular, HPGe  detector array with a mass of 44~kg (29.7~kg enriched in $^{76}$Ge) to search also for the neutrinoless DBD in this isotope. The \textsc{Majorana Demonstrator} follows a modular implementation to be easily scalable. The germanium detectors are split between two modules contained in a low background shield at the Sanford Underground Research Facility in Lead, South Dakota, US. Point-contact Ge detectors fabricated from germanium isotopically enriched to 88\% in $^{76}$Ge have been produced following careful processing procedures to minimize the cosmogenic generation of radioactive isotopes and to maximize the yield of detector mass \cite{majprocessing}. The first detector module started low-background data production in early 2016 and the second one was added in August 2016 to begin operation of the entire array. An unprecedented energy resolution of 2.53~keV (FWHM) at the transition energy has been registered and the measured background in the low-background configurations is (11.9$\pm$2.0) counts/(FWHM t y) \cite{maj2019}. The latest results released give a lower limit on the $^{76}$Ge neutrinoless DBD half-life of $T_{1/2}^{0\nu}>2.7 \times 10^{25}$~y (90\% C.L.). Depending on the matrix elements used, a 90\% C.L. upper limit on the effective Majorana neutrino mass in the range of 200-433~meV is obtained. Limits on DBD to excited states have been also deduced \cite{majexc}. The data taken at the \textsc{Majorana Demonstrator}, even during commissioning runs, have been used in addition to derive results for other rare events or new exotic Physics. Limits on bosonic dark matter, solar axions, Pauli Exclusion Principle violation, and electron decay have been obtained from the low-energy spectrum \cite{majbosonic}. A first limit on the direct detection of Lightly Ionizing Particles for electric charge as low as $e/$1000 has been derived with the \textsc{Majorana Demonstrator} \cite{maje}, as well as the first limits for tri-nucleon decay-specific modes and invisible decay modes for germanium isotopes \cite{majnuc}.

The progress achieved by this current generation of germanium DBD experiments relies on superior germanium detector energy resolution and the improved background discrimination of the detectors. Building on the successes of the \textsc{Majorana Demonstrator} and GERDA, the LEGEND collaboration has been formed to pursue a tonne-scale $^{76}$Ge experiment, with discovery potential at a half-life beyond 10$^{28}$ years after ten years of data taking, fully covering the inverted hierarchy region \cite{appec,legend}. The collaboration aims to develop a phased neutrinoless DBD experimental program, starting with a 200~kg measurement using the existing GERDA cryostat at LNGS, and is being supported by the Double Beta
Decay APPEC Committee \cite{appec}.

It is worth noting that other non-germanium DBD searches using different semiconductor detectors are also in place. The COBRA collaboration has run as a demonstrator for a larger scale DBD experiment, consisting of an array of 64 calorimetric CdZnTe semiconductor detectors with a total mass of 380~g at the LNGS in Italy. From the operation of this demonstrator for several years, lower limits on the half-life for the neutrinoless DBD of the nuclides $^{114}$Cd, $^{128}$Te, $^{70}$Zn, $^{130}$Te and $^{116}$Cd have been set \cite{cobra}.

\section{NEXT}
\label{nextsec}

The ``Neutrino Experiment with a Xenon Time-Projection Chamber'' (NEXT)\footnote{Website: http://next.ific.uv.es} is a project developed by a collaboration of around twenty institutions from Spain, Portugal, US, Colombia and Israel.
It is intended to study the nature of the neutrinos (i.e., if the neutrino is its own antiparticle) searching for the neutrinoless DBD of $^{136}$Xe using a TPC filled with high-pressure gaseous xenon in the LSC. It is a recognized experiment at CERN since 2013. In the ``Input for the European Strategy for Particle Physics Update 2020'' document \cite{espp}, it is considered that ``NEXT explores an alternative approach based on gaseous xenon TPC, which has advantageous features both with respect to energy resolution and background suppression''. Moreover, NEXT has been selected as one of the most competitive projects, together with LEGEND and CUPID, by the Double Beta Decay APPEC Committee \cite{appec}. In this section, the concept of the experiment will be outlined and the development of detectors, from the first prototypes to the implementation of the one for NEXT-100 as a preparation of the future ones, will be described. Then, the background studies carried out will be commented and, finally, the obtained results and future plans presented.

\subsection{Concept} \label{nextcon}

As pointed out in section \ref{trackdet}, $^{136}$Xe is an attractive DBD candidate being considered in different projects for various reasons:
\begin{itemize}
\item It has a relatively high transition energy ((2457.83$\pm$0.37)~keV \cite{QXe}), therefore the neutrinoless signal grows in a region that is less affected by radioactive background events.
\item Its mode with neutrino emission is slow (even slower than expected); hence, its contribution in the neutrinoless DBD decay region of interest is less relevant and the experimental requirement for good energy resolution to suppress this particular background is less stringent.
\item It can be quite easily and cheaply enriched (its natural isotopic abundance is 8.86\%). Indeed, there is already more than one tonne of enriched xenon in the world, owned by KamLAND-Zen (800~kg), EXO-200 (200~kg), and NEXT (100~kg).
\item Xenon has no long-lived radioactive isotopes and, being a noble gas, can be easily purified.
\item Xenon can be used for the realization of a homogeneous detector since it provides both scintillation and ionization signals. Pure xenon can be used as detector medium to build a TPC.
\end{itemize}

In natural xenon, the naturally occurring $^{134}$Xe isotope (10.44\% isotopic abundance) could also be a DBD emitter, but with a very low transition energy of 825~keV.

Both liquid and high-pressure gas xenon (HPXe) provide suitable technologies for a TPC. But energy resolution is much better in gas than in liquid, since, in its gaseous phase, xenon is characterized by a Fano factor lower than 1, meaning that the fluctuations in the ionization production are smaller than the ones due to pure Poisson statistics. The application of HPXe TPCs to neutrinoless DBD experiments is specifically reviewed in \cite{revHPXe}.

\begin{figure}
\centerline{\includegraphics[width=12cm]{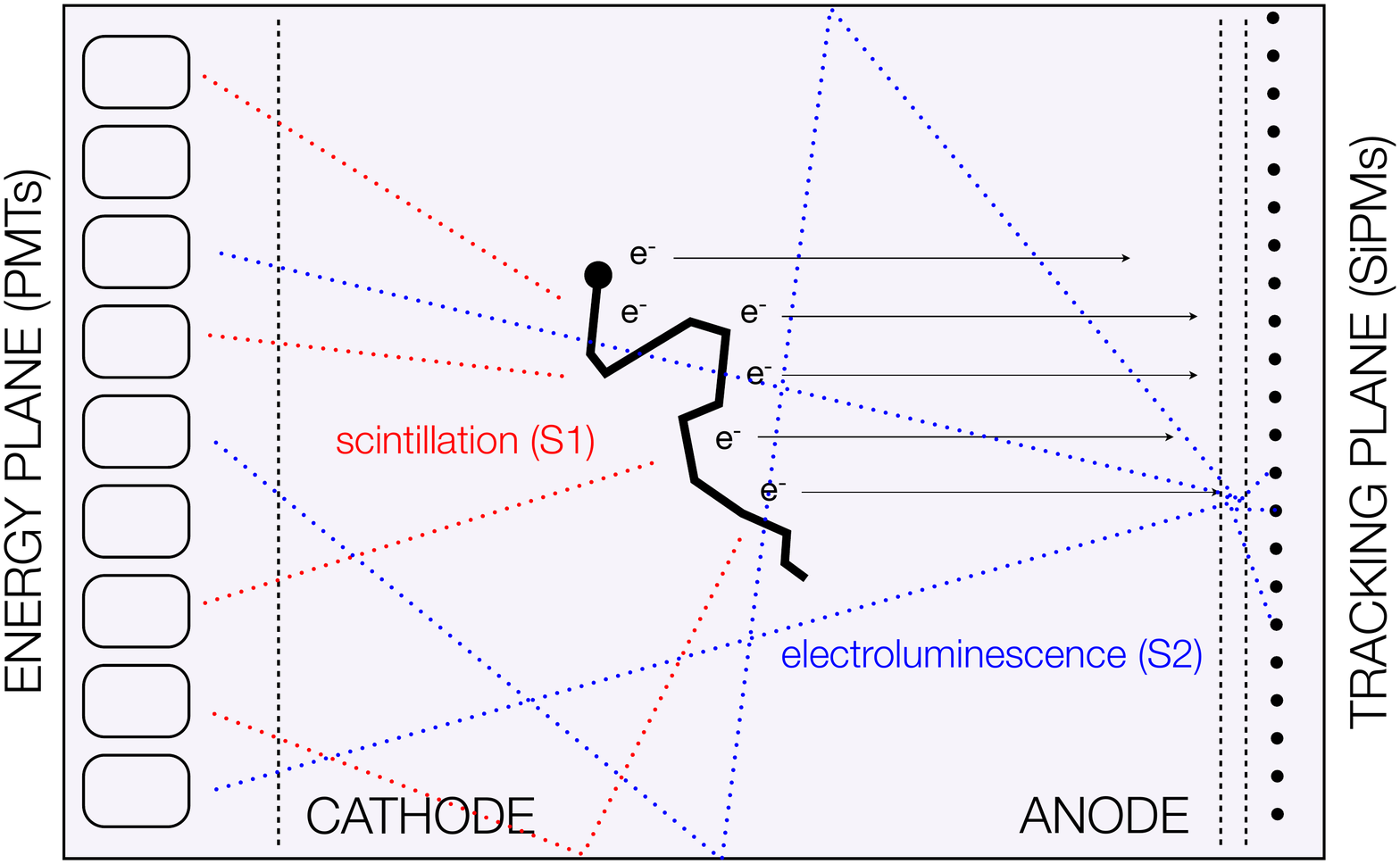}}
\begin{center}
\begin{minipage}[t]{16.5 cm}
\caption{Concept of the NEXT experiment: light from the Xe electroluminescence generated at the anode is recorded both in the photosensor plane right behind it for tracking and in the photosensor plane behind the transparent cathode for a precise energy measurement (courtesy of the NEXT collaboration). Primary scintillation defining the start of the event is also detected by the cathode photosensors.}
\label{nextconcept}
\end{minipage}
\end{center}
\end{figure}

The NEXT program has developed a novel detection concept for investigating the neutrinoless DBD of $^{136}$Xe based on a high-pressure xenon gas TPC with electroluminescent amplification (HPXe-EL) having separated function capabilities for calorimetry and tracking \cite{next2012,next2013,nextahe,next2018,revHPXe}.  As sketched in Fig.~\ref{nextconcept}, light from the Xe electroluminescence generated at the anode is recorded both in the photosensor plane right behind it for tracking and in the plane behind the transparent cathode at the opposite side of the pressure vessel for a precise energy measurement. The electroluminescent-based readout of NEXT gives nearly noiseless amplification for the ionization signal, allowing to achieve optimal energy resolution.

\begin{figure}
\centerline{\includegraphics[width=14cm]{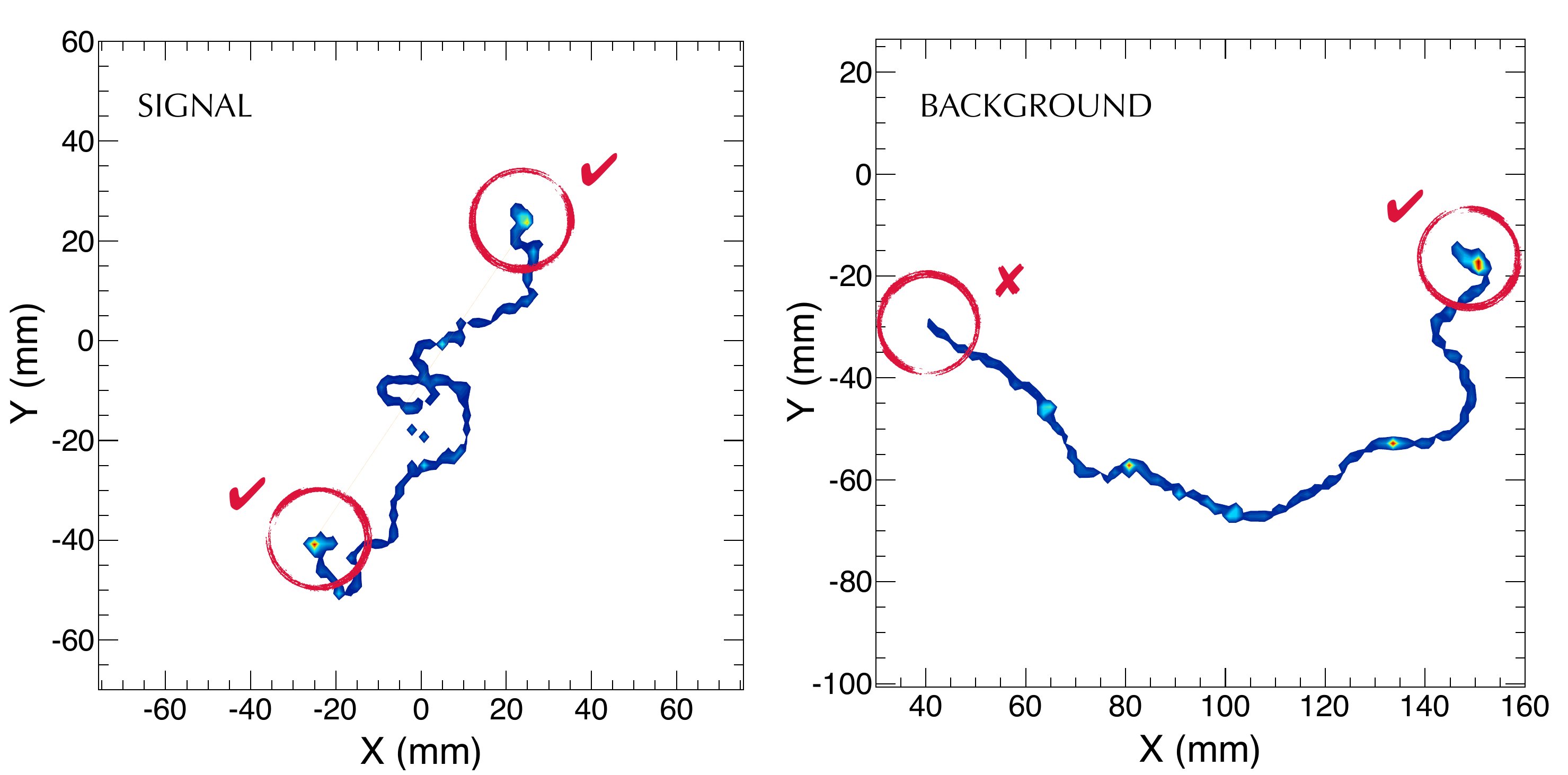}}
\begin{center}
\begin{minipage}[t]{16.5 cm}
\caption{MonteCarlo simulation in xenon gas at 15~bar of a neutrinoless DBD event of $^{136}$Xe (left) and a single electron background event from a 2.44~MeV $^{214}$Bi photon (right); the energy loss per path length is depicted in the X-Y plane (courtesy of the NEXT collaboration). The ionization track is tortuous because of multiple scattering and has larger depositions, referred to as ``blobs'', in both ends for DBD events.}
\label{nexttrack}
\end{minipage}
\end{center}
\end{figure}

Such a detector offers important advantages for the search of neutrinoless DBD, highlighted in the following:
\begin{itemize}
\item Excellent energy resolution, as mentioned before, with an intrinsic limit due to statistical fluctuations of about 0.3\% FWHM at the transition energy of $^{136}$Xe and quite close measured values of 0.5-0.7\% demonstrated by the NEXT prototypes (see Secs.~\ref{nextdetdev} and \ref{nextres}).
\item Tracking capabilities that provide a powerful topological signature to discriminate between signal (two electron tracks with a common vertex) and background (mostly, single electrons). Neutrinoless DBD events have a distinctive signature in gaseous xenon: an ionization track, about 30~cm long at 10~bar, tortuous due to multiple scattering, and with larger energy depositions at both ends, as shown in Fig.~\ref{nexttrack}.
\item A fully active and homogeneous detector, with very radiopure apparatus of large mass. Since 3-dimensional reconstruction is possible, events can be located in a fiducial region away from surfaces, where most of the background arises.
\item Scalability of the technique to larger masses, even to a 1 tonne-scale experiment, thanks to the xenon properties described before.
\end{itemize}

The basics of the detection process in NEXT detectors (see Fig.~\ref{nextconcept}) is as follows. The separate energy and tracking readout planes use different sensors: PMTs are used for calorimetry, and for determining the start time of the event thanks to the detection of the primary scintillation, while Silicon photomultipliers (SiPMs) are used for tracking. Particles interacting in the HPXe transfer their energy to the medium through ionization and excitation. The excitation energy is manifested in the prompt emission of Vacuum Ultra-Violet (VUV) scintillation light (S1) at $\sim$178~nm. The ionization tracks (positive ions and free electrons) left behind by the particle are prevented from recombination by an electric field (0.03-0.05~kV/cm/bar). The ionization electrons drift toward the TPC anode, entering a region, defined by two highly transparent meshes, with an even more intense electric field (3~kV/cm/bar). There, further VUV photons are generated isotropically by electroluminescence (S2). Therefore, both scintillation and ionization produce an optical signal, to be detected with a sparse plane of PMTs (the energy plane) located behind the cathode. The detection of the S1 light constitutes the start-of-event (t$_{0}$), whereas the detection of the proportional EL light S2 provides an energy measurement. Electroluminescent light provides tracking as well, since it is detected also a few millimeters away from production at the anode plane, via an array of $\sim$1~mm$^{2}$ SiPMs, $\sim$1~cm spaced (the tracking plane); the event topology is given by S2 time-slice images taken by the SiPMs. The detector triggers on the energy information read by the PMTs and provides PMT and SiPM waveforms; the sampling time of the PMTs is $\sim$25~ns, while the SiPM charge is integrated every $\mu$s.

The largest HPXe chamber ever operated in the world before NEXT was the Gotthard TPC \cite{gotthard}, with a total mass of 5~kg of xenon at a pressure of $\sim$5~bar; the technology of the time (the mid-nineties) allowed achieving a modest energy resolution (around 7\%). Compared to other present xenon-based experiments (see Secs.~\ref{trackdetww} and \ref{scint}), with a mass in the range of hundred kilograms, NEXT features a better resolution and the extra handle of the identification of the two electrons, which could be crucial in case of discovery \cite{discoverypotential}. NEXT is ideally suited to confirm in an unambiguous way the existence of a potential signal, in particular given the discriminating power of the topological signature. The implementation of Ba-tagging capabilities in NEXT detectors is being also thoroughly analyzed.

\subsection{Detector development}
\label{nextdetdev}

The NEXT program has followed different phases for the detector development; the relevant technical features of the different detectors are described in this section.
\begin{itemize}
\item The first phase of the program included the construction, commissioning and operation of prototypes with masses around 1~kg (NEXT-MM,  NEXT-DEMO and NEXT-DBDM), which allowed to demonstrate the robustness of the selected technology, its excellent energy resolution and its unique topological signal.

\item As a second step, the NEXT-WHITE demonstrator\footnote{The name honours the memory of the late Professor James White, key scientist of the NEXT project.} (also named NEW), deploying 5~kg of xenon, is fully operational at LSC in 2019. It is the first NEXT detector with underground and radiopure operation intended to validate the HPXe-EL technology in a large-scale detector: to assess the robustness and reliability of the technological solutions; to compare in detail the background model with data, particularly the contribution to the radioactive budget of the different components; and to study the energy resolution and the background rejection power of the topological signature characteristic of a HPXe-EL TPC. It has also facilitated the commissioning of the large infrastructures (most importantly the gas system) needed for operation of the successive systems. Furthermore, NEXT-White can provide a measurement of the two-neutrino DBD mode.

\item NEXT-100 constitutes the third phase of the program. It is a radiopure detector with 100~kg of xenon at 15~bar and scaling up NEXT-White by slightly more than 2:1 in all dimensions. The needed enriched xenon is already available (in use or stored underground) at LSC. In addition to a physics potential competitive with the best current experiments in the field, NEXT-100 can be considered as a large scale demonstrator of the suitability of the HPXe-EL technology for detector masses in the tonne-scale.

\item For future NEXT phases, such a detector at the tonne scale is being considered. The so-called NEXT-HD will improve the ``high- definition'' of the technology thanks to the combination of an improved topological signature using low diffusion gas mixtures, improved energy resolution and drastically reduced radioactive budget. Furthermore, the collaboration is working on NEXT-BOLD, a detector that would implement barium tagging based in a sensor of molecular indicators capable of tracking and marking the presence of the Ba ions produced in $^{136}$Xe DBD; this could lead to an almost background-free experiment allowing a full exploration of the inverse hierarchy.
\end{itemize}

\subsubsection{First prototypes}
\label{secfp}

Small prototypes were operated at different institutions with different purposes during the first years after the formation of the NEXT collaboration, from 2008 to 2014 \cite{firstproto}.

Following the advances in TPC readouts, Micromegas have been proved to be a very competitive option for the DBD investigation using high pressure gas TPCs \cite{mmdbd}. The so-called NEXT-MM (Micromegas) prototype \cite{protommlaura,protomm,protomm2,diego} was developed at the University of Zaragoza, having an active volume with  $\sim$35~cm for drift and 28~cm in diameter. It was the largest Micromegas-read TPC operated in xenon ever constructed, made by a sectorial arrangement of the four largest single wafers manufactured with the Microbulk technique to date. It was equipped with a suitably pixelized readout and with a sufficiently large sensitive volume ($\sim$23~l) so as to contain and register long ($\sim$20~cm) electron tracks. Although the use of this technology was finally disregarded for NEXT, the detector operated continuously for 100 live days and a FWHM energy resolution inside the fiducial volume in the range from 14.6\% (at 30~keV) to 4.6\% (at 1275~keV) was obtained working at 10~atm with Xenon and trimethylamine (TMA); this extrapolates to $\sim$3\% at the peak of the neutrinoless DBD of $^{136}$Xe. As mentioned in Sec.~\ref{trackdetww}, Microbulk Micromegas are now being used at both ends of the HPXe TPC for the charge readout in the PandaX-III that will investigate also the DBD of $^{136}$Xe at the China Jin Ping Underground Laboratory \cite{pandax}; they have been selected for providing good energy and spatial resolution together with good intrinsic radiopurity and electron track reconstruction for background reduction \cite{patternrecognition}.

The NEXT prototype for Double Beta and Dark Matter (NEXT-DBDM) was a HPXe-EL TPC built and operated at the Lawrence Berkeley National Laboratory. An array of 19~VUV Hamamatsu R7378A 1 inch PMTs were placed inside a stainless steel pressure vessel capable to operate up to 20 bar pressure. The main objectives were demonstrating near-intrinsic energy resolution at energies up to 662~keV and optimizing the detector design and operating parameters. Energy resolutions of $\sim$1\% FWHM for 662~keV gamma rays from $^{137}$Cs at 10 and 15~atm and of $\sim$5\% FWHM for 30~keV fluorescence xenon X-rays were obtained, demonstrating that resolutions below 1\% FWHM for the hypothetical neutrinoless DBD peak were realizable \cite{berkeley}. A track imaging system consisting of 64~SiPMs was also installed in NEXT–DBDM and tested; first results demonstrated the key functionalities required for the NEXT-100 search.

NEXT-DEMO \cite{protodemo,valencia}, operating in Valencia at the Instituto de Fisica Corpuscular (IFIC) since 2011, was a prototype holding a mass of about 1.5~kg of natural xenon. It was conceived to fully test and demonstrate the HPXe-EL technology implementing the separated readouts for energy and tracking; the response of the detector was fully characterized \cite{protoalfa,protodemo2}. In this detector the TPC was housed in a stainless steel pressure vessel, 60~cm long and with a diameter of 30~cm, which can withstand up to 15~bar. The active volume was 30~cm long. A Polytetrafluoroethylene (PTFE) tube was inserted to improve light collection. The energy plane was equipped with 19 Hamamatsu R7378A PMTs while there were 256~Hamamatsu SiPMs at the tracking plane \cite{imaging,trackingdemo}. The TPC light tube and SiPM surfaces were coated with tetraphenyl butadiene (TPB) acting as a wavelength shifter for the VUV scintillation light produced by xenon to the visible spectrum \cite{nextcoating}. The DAQ was an implementation of the Scalable Readout System (developed by the RD51 collaboration) based on FPGA (Field-Programmable Gate Array) devices; on-line triggering based on the detection of primary or secondary scintillation light, or a combination of both, that arrives to the PMT plane was possible \cite{nextdemoreadout,nextdemotrigger}. The detector neither was radiopure nor was shielded against natural radioactivity but it was operated over long periods of time for several years showing very good stability and high gas quality, with no leaks and very few sparks. Natural xenon circulated in a closed loop through the vessel and a system of purifying filters. An excellent electron drift lifetime, of the order of 10~ms, was measured. The SiPM array was used to reconstruct event topologies and to demonstrate its power to reject background; the difference between ``electrons-like'' and ``double-electrons like'' events, which can be easily separated with a cut on the energy of the lower-energy blob, was clearly shown. The SiPM tracking plane allowed for the definition of a large fiducial region in which an excellent energy resolution of 1.82\% FWHM at 511~keV was measured, which extrapolates to 0.83\% at the transition energy of $^{136}$Xe. NEXT-DEMO has been refurbished during 2018-2019 and continues to be used by the NEXT Collaboration for R\&D related to gas mixtures and NEXT readout plane upgrades.

\subsubsection{NEXT-White}
\label{secnew}

\begin{figure}
\centerline{\includegraphics[width=10cm]{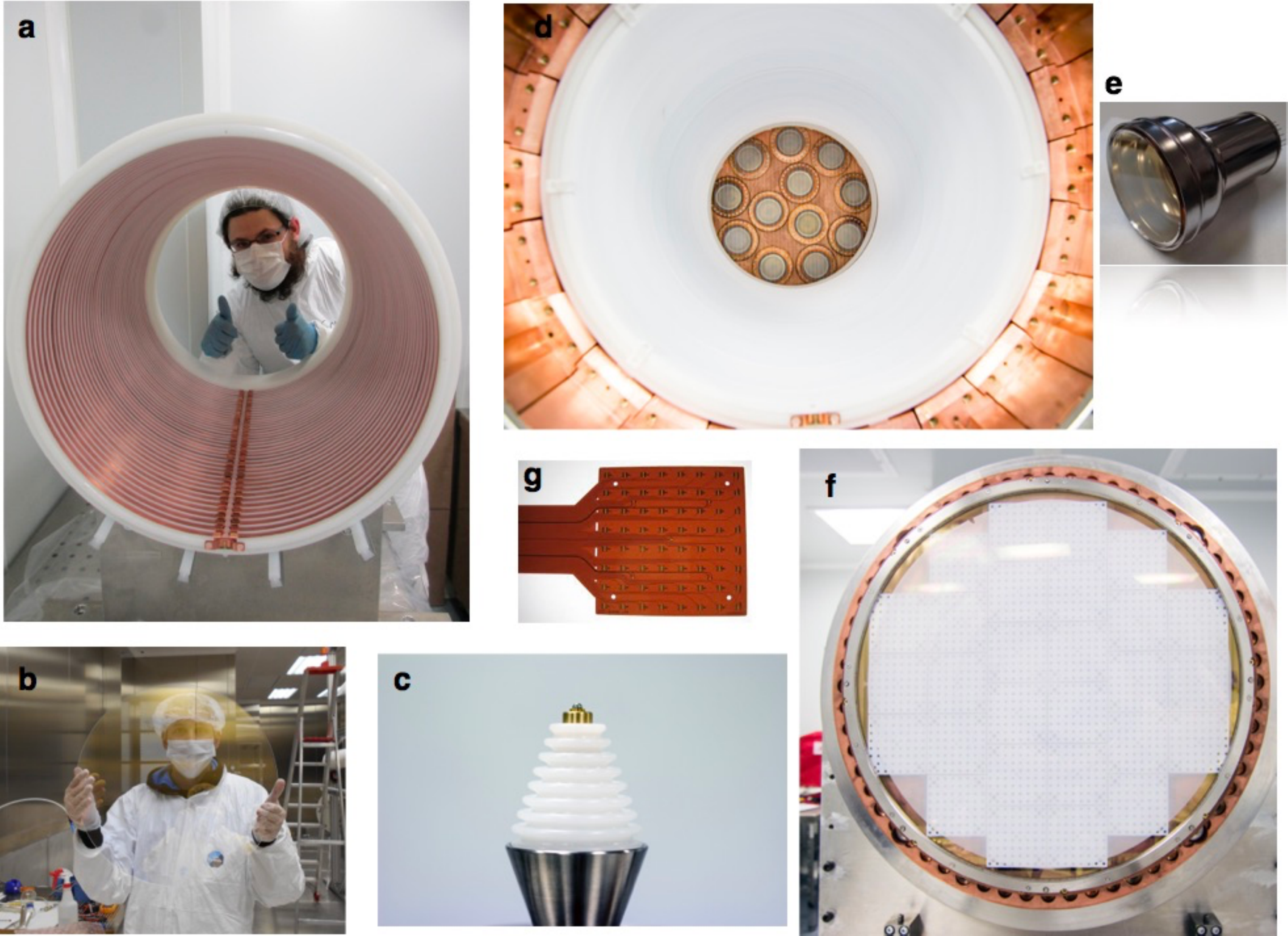}}
\begin{center}
\begin{minipage}[t]{16.5 cm}
\caption{Selection of the main subsystems of NEXT-White: a) the field cage; b) the anode plate; c) high voltage feedthrough; d) energy plane; e) PMTs used in the energy plane; f) tracking plane; g) kapton boards composing the tracking plane (courtesy of the NEXT collaboration).}
\label{newcomp}
\end{minipage}
\end{center}
\end{figure}

\begin{figure}
\centerline{\includegraphics[width=12cm]{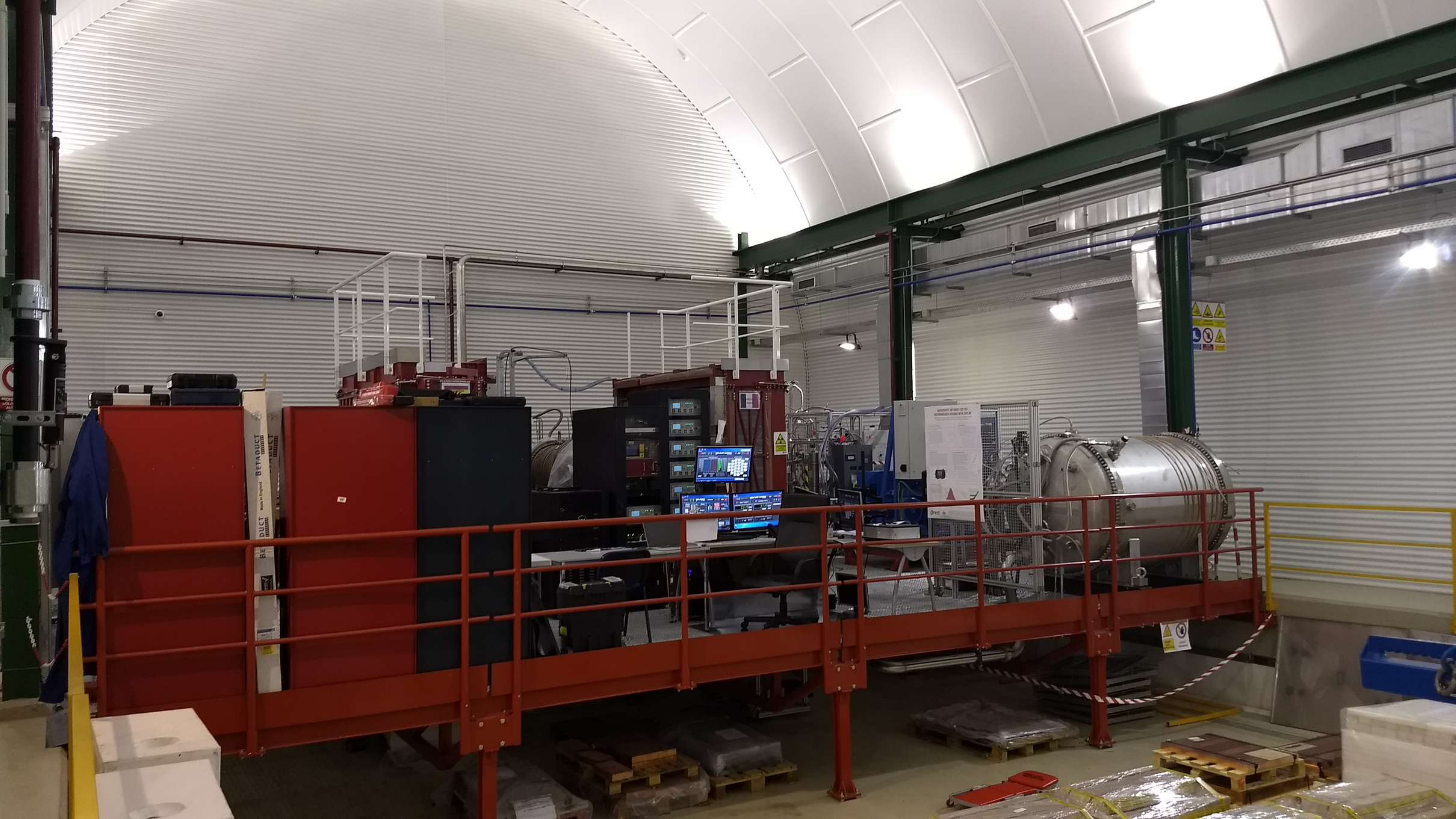}}
\centerline{\includegraphics[width=12cm]{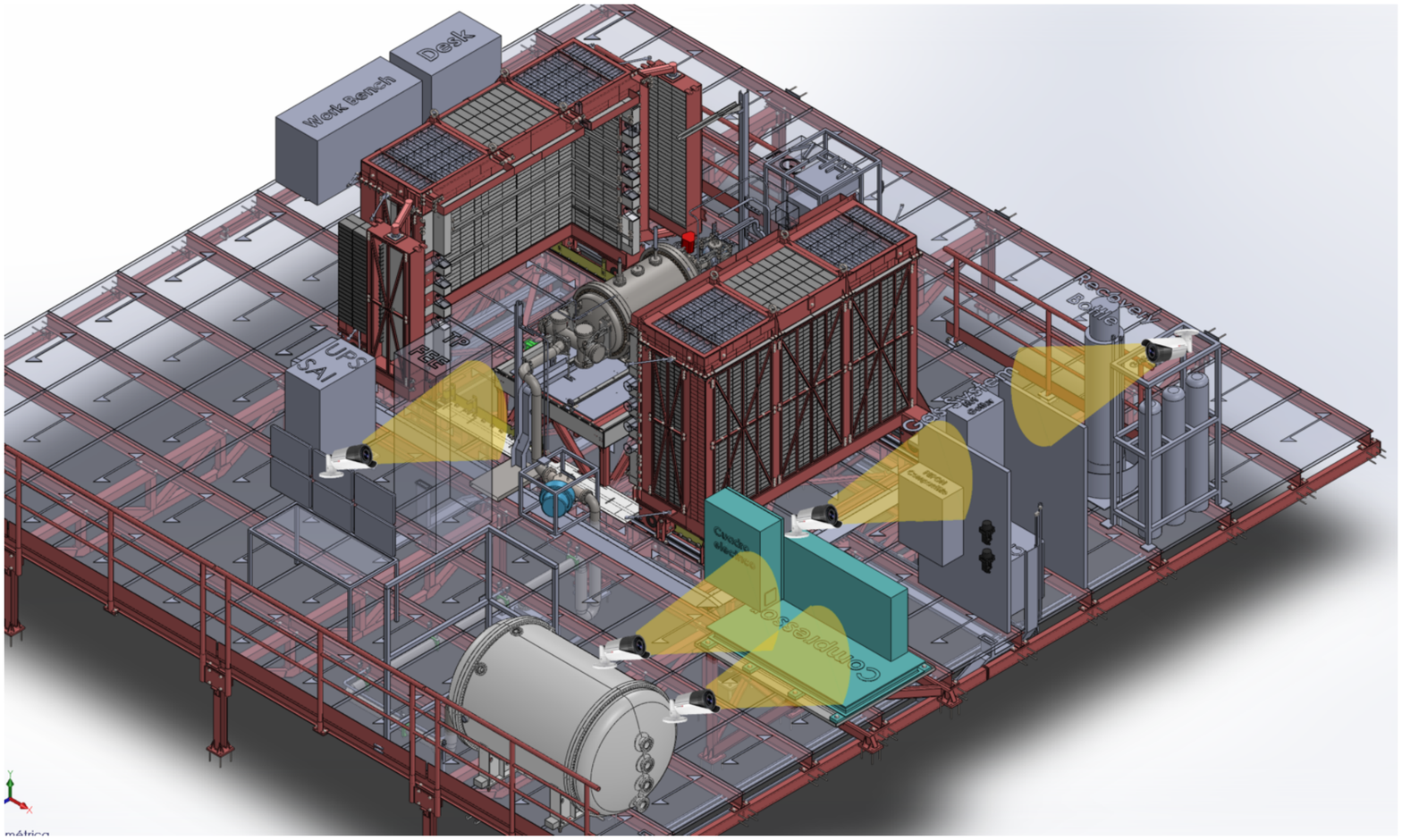}}
\begin{center}
\begin{minipage}[t]{16.5 cm}
\caption{Design of the NEXT-White detector and associated infrastructures (bottom) and a picture of the real setup at the Hall A of the LSC (top) (courtesy of LSC and NEXT collaboration).}
\label{newapp}
\end{minipage}
\end{center}
\end{figure}

At present, NEXT is running successfully at LSC since October 2016 the demonstrator called NEXT-White (or NEW) implementing the second phase of the program \cite{nextnew}. NEXT-White consists of a xenon gas TPC able to operate at 15~bar. The TPC has a length of 664.5~mm and a diameter of 454~mm, being currently the largest HPXe-EL TPC in the world. The field cage body is a High Density Polyethylene (HDPE) cylindrical shell of 21~mm thickness. The drift field, to transport ionization electrons to the anode where they are amplified, is created by applying a voltage difference between the cathode and the gate, through high voltage feedthroughs; the drift voltage is 400~Vcm$^{-1}$. Custom-made radiopure high voltage feedthroughs capable of holding high voltages in xenon have been designed and constructed. The detector active volume is 530.3~mm long along the drift direction and has a 198~mm radius. The amplification or electroluminescent region is the most delicate part of the detector, given the requirements for a high and yet very uniform electric field. The anode is defined by an Indium tin-oxyde (ITO) surface coated over a fused silica plate of 522~mm diameter and 3~mm thickness. A thin layer of TPB is vacuum-deposited on top of the ITO. The EL gap is 6~mm wide. A light tube made of teflon coated with TPB is used to maximize the light collection. The detector operates inside a pressure vessel fabricated with a radiopure stainless steel alloy (316Ti) provided by the Nironit company, placed on a seismic table and surrounded by a 6~cm-thick ultra-pure inner copper shell and a lead shield. The xenon circulates in a gas system where it is continuously purified to guarantee a long electron lifetime. The whole setup sits on top of a platform elevated over the ground in the Hall A of the LSC. The main subsystems of the NEXT-White detector can be seen in Fig.~\ref{newcomp} while the whole apparatus is shown in Fig.~\ref{newapp}.

The energy plane is instrumented with 12~Hamamatsu R11410-10 PMTs, providing a coverage of 31\%; this PMT model was chosen for their low radioactivity and good performance. Since they cannot withstand high pressure, they are protected from the main gas volume by a radiopure copper plate, 120~mm thick, which also acts as a shielding against external radiation. The PMTs are optically coupled to the xenon gas volume through 12 sapphire windows using an optical gel with a proper refractive index. The windows are coated with Poly-Ethylenedioxythiophene (PEDOT) in order to define an electric ground while at the same time avoiding sharp electric field components near the PMT windows. A thin layer of TPB is vacuum-deposited also on top of the PEDOT to shift the xenon VUV light to blue.

The tracking plane holds a sparse matrix of 1792~SiPMs from SensL (C-Series model MicroFC-10035-SMT-GP), having a size of 1~mm and being distributed on a 2-dimensional lattice at a pitch of 10~mm. There are 28 boards (DICE boards) with 8$\times$8 pixels. The material used for the DICE boards is a low-radioactivity kapton printed circuit with a flexible pigtail that passes through the copper where it is connected to another kapton cable that brings the signal up to the feedthrough. Each DICE has a temperature sensor to monitor the temperature of the gas and SiPMs and also a blue LED to allow calibration of the PMTs at the opposite end of the detector. A 60~mm-thick ultra-pure copper shell acts as a shield in the barrel region. The DICE boards, as the energy plane, are mounted on a 120~mm-thick copper plate intended to shield against external radiation. The whole tracking plane is placed 2~mm behind the end of the quartz plate that defines the anode with a total distance to the center of the EL region of 8~mm.

The electronics of NEXT-White has been designed and implemented to satisfy strict requirements \cite{newreadout,newelectronics}: an overall energy resolution below 1\% and a good radiopurity. All the components and materials were carefully screened to assure a low radioactivity level while fulfilling at the same time the required front-end electronics specifications. The final PMT channel design has been characterized with linearity better than 0.4\% and noise below 0.4~mV. The front-end electronics for the 1.8-kchannel SiPM tracking plane in the NEXT-White detector includes per-channel differential transimpedance input amplifier, gated integrator, automatic offset voltage compensation and 12-bit ADC \cite{newelectronicssipm}.

The detector was commissioned and started operation with xenon in late 2016. After the initial commissioning (Run I), extensive calibration data together with some low-background data were obtained during Run~II at a pressure of 7~bar, for 7~months in 2017. Following some detector upgrades to improve the stability and radiopurity of the apparatus (the field cage resistor chain and the PMT bases were replaced), more calibration and physics data were taken in 2018 in Runs~III and IV; the detector was operated during 9~months at a pressure of 10~bar. In particular, a low background run along several months using depleted xenon ($\sim$3\% isotopic abundance of $^{136}$Xe) was satisfactorily carried out. At Run IV the gas pressure, drift field and EL field were set to 10.1~bar, 0.6~kV$/$cm, and 1.7~kV$/$cm$/$bar, respectively; the electron drift velocity was measured to be 0.92~mm$/\mu$s. In February 2019, NEXT-White started the Run V, the first run with enriched xenon ($\sim$91\% isotopic abundance of $^{136}$Xe), having already reached the required purity and tightness conditions using the gas recovery system; this run is intended to measure the two-neutrino mode and to perform the first search for neutrinoless DBD taking data until mid 2020. The same operating conditions (gas pressure, TPC voltages) than in Run IV have been fixed. It must be noted that operation of the detector during both Run IV and V has shown excellent stability, with a very low spark rate and negligible leaks; these two operational aspects were among the critical issues to be demonstrated by NEXT-White.

\subsubsection{NEXT-100}

\begin{figure}
\centerline{\includegraphics[width=12cm]{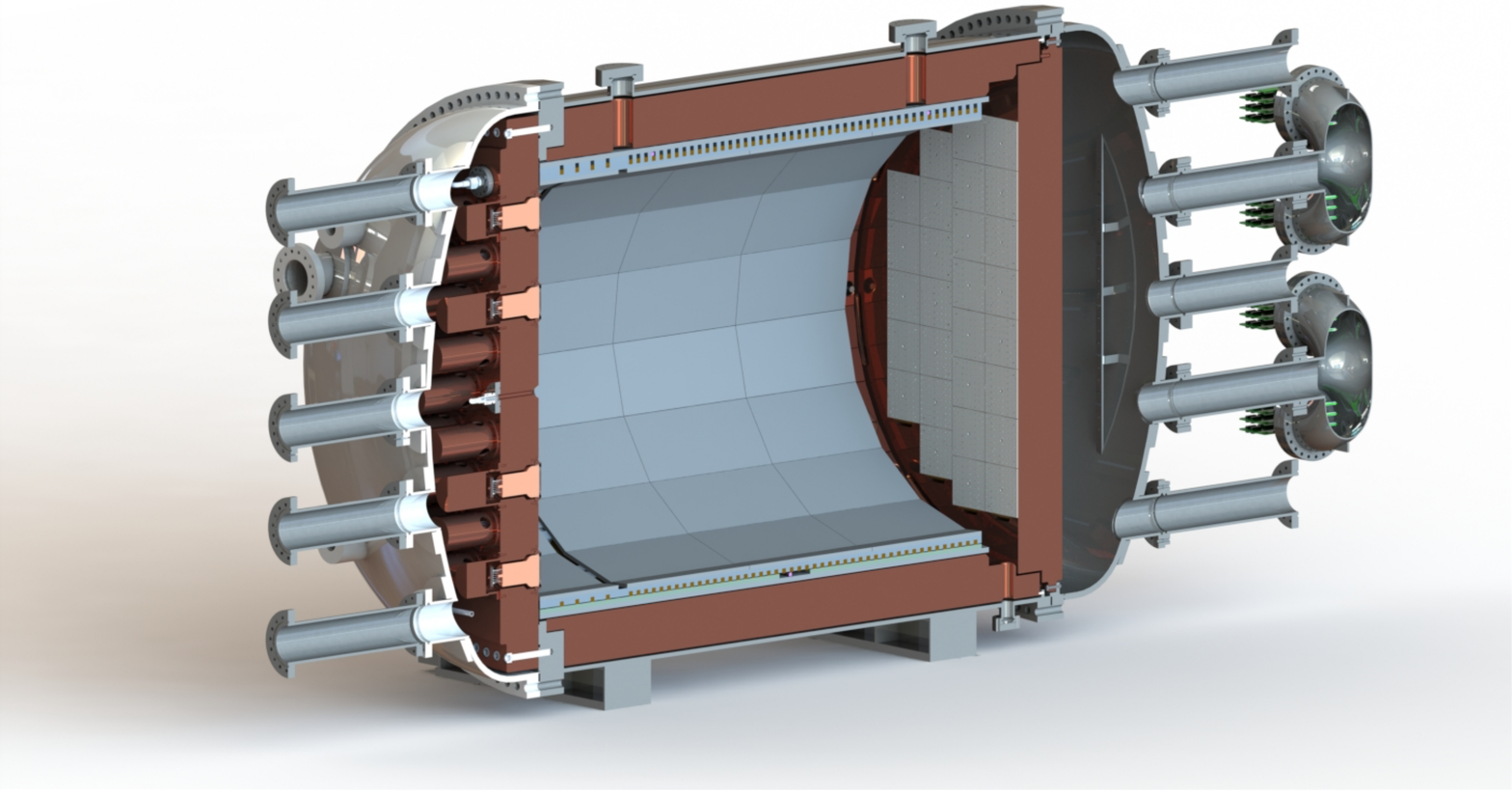}}
\begin{center}
\begin{minipage}[t]{16.5 cm}
\caption{Cross-section view of the design of the NEXT-100 detector (courtesy of the NEXT collaboration). A stainless steel pressure vessel  houses the electric-field cage and the two sensor planes (energy plane and tracking plane) located at opposite ends of the chamber. The fiducial region is a cylinder of 1050~mm diameter and 1300~mm length (defining 1.27~m$^{3}$ of fiducial volume). The active volume is shielded from external radiation by at least 12~cm of copper in all directions.}
\label{Next100}
\end{minipage}
\end{center}
\end{figure}

NEXT-100 \cite{nexttdr} constitutes the third step of the program with around 100~kg of Xe enriched at 90\% in isotope $^{136}$Xe in an asymmetric HPXe-EL TPC operating at 15~bar. Its construction is based on the successful work on prototypes developed over the past years. It is currently in phase of construction and scheduled to start commissioning in 2020. Figure \ref{Next100} shows a drawing of the NEXT-100 detector, presenting all the major subsystems, which are described in the following.

\begin{itemize}
\item The pressure vessel has been built with stainless steel (low activity 316Ti alloy) from Nironit company and designed to withstand a pressure of 15~bar. A 12~cm-thick copper layer on the inside shields the sensitive volume from the radiation which originated in the vessel and outer components. It consists of a barrel central section with two identical torispherical heads on each end, their main flanges bolted together.

\item The main body of the field cage is a 2.5~cm-thick HDPE cylindrical shell, providing electric insulation from the vessel. Three wire meshes separate the two electric field regions of the detector. The drift region, between the cathode and the first mesh of the EL region, is a cylinder of 105~cm of diameter and 130~cm of length. Copper strips are attached to the polyethylene and are connected with low background resistors to the high voltage, in order to keep the field uniform across the drift region. The EL region is 1.0~cm long.

\item The energy plane consists of a total of 60 Hamamatsu R11410-10 PMTs covering 32.5\% of the cathode area, as a compromise between the need to collect as much light as possible and the need to minimize the number of sensors to reduce cost, technical complexity and radioactivity. This 3-inch phototube model has been specially developed for radiopure, xenon-based detectors and has also good quantum efficiency and low dark count rate. The PMT modules are all mounted to a common carrier plate attached to an internal flange of the pressure vessel head.

\item The tracking function in NEXT-100 is provided by a plane of SiPMs operating as sensor pixels and located behind the transparent EL gap, inside the pressure vessel. The chosen SiPMs have an active area of 1~mm$^{2}$ each. As they have a high photon detection efficiency in the blue region, they need to be coated with the same wavelength shifter used for the windows of the copper cans. SiPMs have a low radioactivity, given their composition (mostly silicon) and very small mass. The 5600 SiPMs are mounted in an array of square boards (DICE Boards) containing each 8$\times$8 SiPM sensors with a pitch of 1.5~cm between them. The electronics for the SiPMs is also an improved version of the electronics tested in prototypes, being placed outside the chamber. The large number of channels in NEXT-100 requires the design and fabrication of large custom-made feedthroughs to extract the signals.

\item The gas system must be capable of pressurizing, circulating, purifying, and depressurizing the NEXT-100 detector with xenon, argon, and possibly other gases with negligible loss and without damage to the detector. In particular, the probability of any substantial loss of the very expensive enriched xenon must be minimized; the control system developed at IFIC guarantees gas losses lower than 10 gram per year. To keep the purity of the gas is also essential; the presence of oxygen is at the level of 0.1 parts per billion.

\item To shield NEXT-100 from the external flux of high energy gamma rays a 20~cm-thick lead castle has been built with layers of lead bricks held with a steel structure. The lead castle, as shown in Fig.~\ref{newapp}, is made of two halves mounted on a system of wheels that move on rails with the help of an electric engine. A lock system fixes the castle to the floor to avoid accidental displacements.
\end{itemize}

\subsubsection{Future NEXT phases}
\label{futnext}

The NEXT-100 detector can achieve a sensitivity comparable to the experiments of the current generation while at the same time demonstrating the potential of the technology for an almost background-free experiment. The fourth envisioned phase of the NEXT program (sometimes called NEXT-2.0), already considers a detector at the tonne scale \cite{appec}. The ``High Definition'' (NEXT-HD) option is based on the reduction of the NEXT-100 background in the region of interest by at least one order of magnitude, thus counting less than one event per tonne per year of exposure \cite{nextmoriond}. This can be obtained thanks to the combination of an optimized and improved topological signature (using low diffusion mixtures, resulting in better position resolution, and increasing the density of the tracking plane) and a reduced radioactive budget (replacing the  PMTs used by NEXT-White and NEXT-100 by large-area SiPMs). The addition of 15\% of helium is being considered to reduce diffusion. SiPMs are not only intrinsically radiopure but also resistant to pressure and able to provide better coverage. In addition, operation with cold Xenon would allow to reduce the SiPM dark current and operate a lower pressure; a vertical symmetric detector would help to simplify the design of feedthroughs and to reduce the total drift length. A first estimate of the background rate of NEXT-HD in a 29~keV region of interest is 9$\times$10$^{-4}$ counts per year and per kilogram of isotope \cite{appec}.

A fifth phase of the program is being considered, implementing Ba-tagging based on single molecule fluorescence imaging, in the approach ``Barium iOn Light Detection'' (NEXT-BOLD) \cite{nextmoriond}. No conventional radioactive process could produce a barium ion in bulk xenon; then, the detection of single barium ions emanating from the decay of $^{136}$Xe, even if complicated, is being attempted as it would help to provide almost zero background conditions for neutrinoless DBD, when combined with a good enough energy resolution at the transition energy. This ``barium tagging'' strategy has been investigated for a long time within different projects, including the NEXT collaboration, being the main difficulty the method for extraction and identification of barium ions. A new method to tag the doubly ionized barium daughter in the DBD of $^{136}$Xe in high pressure xenon gas TPCs was reported in \cite{batagging, batagging2}, based on single molecule fluorescent imaging (SMFI) techniques; individual ions were localized with superresolution ($\sim$2~nm) and detected with a statistical significance of 12.9$\sigma$ over backgrounds, confirming the very good prospects of the technique. Ba$^{++}$ ions are captured in a monolayer of fluorescent organic molecules; intense pulsed laser illuminating the molecule make the molecule give light in a different color. Fluorescent bicolor sensors are under study \cite{bicolor}. The SMFI Ba-tagging is very important since, if successful, it could permit a virtually background-free experiment at the tonne scale leading to the  exploration of the Inverse Hierarchy of neutrino masses and even beyond. The R\&D for NEXT-BOLD includes the characterization of molecules able to fluoresce in dry medium, the development of molecular monolayers, and the development of suitable laser optics \cite{nextmoriond}. The goal is to implement the SMFI method in a HPXe-EL detector and a first BOLD prototype could be installed in NEXT-100 in a few years.

\subsection{Background studies}
\label{secnextbkg}

Searching for a very rare event, NEXT requires a severe suppression of potential backgrounds; several approaches have been followed to achieve this goal. On one hand, an extensive material screening program is underway since the start of the project to quantify the radioactivity of the materials intended to be used in the experiment and select the suitable ones for all the subsystems of the set-up. On the other hand, as already discussed, the NEXT technology is able to distinguish signals from electron pairs (due to DBD) with respect to signals with single electrons (coming from background).

The radiopurity assessment program and selection process for NEXT components have been ongoing for several years. The determination of the activity levels is mainly based on gamma-ray spectroscopy using the ultra-low background germanium detectors at LSC (described in Sec.~\ref{seclscfea}, see Fig.~\ref{gelsc}) and also on other techniques like Glow Discharge Mass Spectrometry (GDMS) and ICPMS providing complementary measurements. Materials to be used in the shielding, the pressure vessel, electroluminescence and high voltage components, and the energy and tracking readout planes have been extensively analyzed; around two hundred and fifty items have been screened and all the relevant results have been made public \cite{jinstrp,aiprp,lrtrp,icheprp,trackingrp,energyrp}. These results, essential during the design of the apparatus, have been also relevant as inputs for the evaluation of the sensitivity of the NEXT-100 experiment based on Monte Carlo simulations \cite{nextsensitivity} as well as for the construction of precise background models for the NEXT-White and the NEXT-100 detectors. Adequate materials for the external and internal passive shields were identified. Stainless steel samples were analyzed for the construction of the pressure vessel; the good radiopurity found for the 316Ti stainless steel supplied by the Nironit company, of the order of tenths of mBq/kg for the isotopes at the lower part of the $^{238}$U and $^{232}$Th chains, confirmed that this material can be used for the detector vessel in combination with an inner copper shield.

The design of radiopure readout planes, in direct contact with the gas detector medium, has been especially challenging since the required components (sensors, printed circuit boards and electronic components, involving in many cases different composite materials) typically have activities too large for experiments demanding ultra-low background conditions. Exhaustive screening programs for selection of in-vessel components specifically for both the tracking and energy planes were undertaken \cite{trackingrp,energyrp}. SiPMs with low enough activity were identified. Regarding the substrate for SiPMs, printed circuit boards made of kapton and copper were chosen for their better radiopurity for $^{238}$U and $^{232}$Th in comparison with cuflon boards. Being kapton flexible, the design of all-in-one kapton boards with long flexible tails as cables allowed in addition to place connectors, having unacceptable activities behind the inner copper shielding. Solder paste and units of thermistors acting as temperature sensors and LEDs used for calibration, fulfilling NEXT requirements for the overall background level in the region of interest (see Table 4 of Ref.~\cite{trackingrp}), were also selected. Concerning the energy plane, all the available units for NEXT-100 of the selected PMT model, Hamamatsu R11410-10, were screened in 3-unit groups using the same germanium detector at LSC. Compatible activities were registered for all runs and a joint analysis of all the accumulated data allowed to quantify average activities of not only $^{60}$Co and $^{40}$K but also of the isotopes in the lower part of the $^{238}$U and $^{232}$Th chains ((0.35$\pm$0.08)~mBq/PMT of $^{226}$Ra and (0.53$\pm$0.12)~mBq/PMT of $^{228}$Th), which are similar to those measured for the new version of the PMT R11410-21 \cite{pmtxenon}. In addition, most of the components accompanying PMTs at their bases and enclosures were analyzed, including sapphire windows, optical gel, capacitors, resistors, cables, epoxy, bolts, screws and copper.

\begin{figure}
\begin{center}
\centerline{\includegraphics[width=14cm]{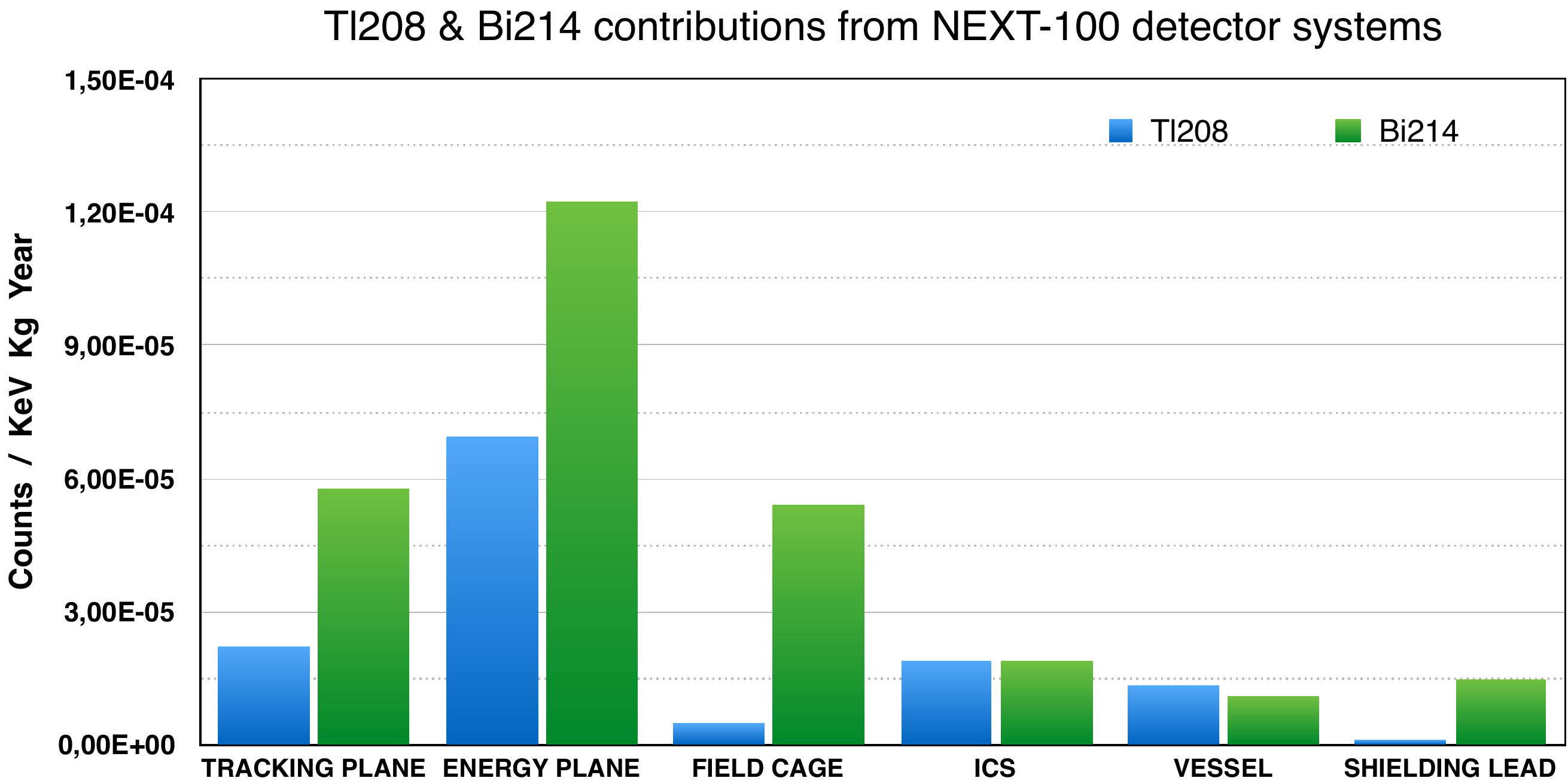}}
\begin{minipage}[t]{16.5 cm}
\caption{Background budget for NEXT-100 after event selection from $^{214}$Bi (right columns) and $^{208}$Tl (left columns) emissions coming from the different detector subsystems (as presented in \cite{thesisMunoz}).}
\label{BackgroundBudget}
\end{minipage}
\end{center}
\end{figure}

Natural radioactivity in detector materials and surroundings is the main source of background for NEXT. In particular, the peak from the expected signal of the neutrinoless DBD of $^{136}$Xe lies in between the photopeaks of the high-energy gammas emitted after the beta decays of $^{214}$Bi and $^{208}$Tl, isotopes of the progeny of  $^{238}$U and $^{232}$Th, respectively; these are relevant because of their ability to generate a signal-like track in the fiducial volume with energy around the transition energy of $^{136}$Xe. The gamma emission at 2447~keV (with 1.57\% intensity; from $^{214}$Po, daughter isotope of $^{214}$Bi) would overlap the signal even for energy resolutions as good as 0.5\% FWHM. $^{208}$Pb, decay product of $^{208}$Tl, emits a de-excitation photon of 2615~keV (with an intensity of 99.75\%); electron tracks from its photoelectric interaction can lose energy via bremsstrahlung and fall in the region of interest. Additionally, the Compton-scattered photon can generate other electron tracks close enough to the initial Compton electron to be reconstructed as a single track with energy close to the transition energy. A detailed Monte Carlo detector simulation has been carried out to assess the contribution of each detector subsystem to the overall background rate of NEXT-100 taking into account the emissions coming from the activities quantified in the material screening program; Fig.~\ref{BackgroundBudget} shows the $^{214}$Bi and $^{208}$Tl contributions. For many components the result is indeed an upper limit, as at present there are only limits to their activity. The overall background rate including the radioactive backgrounds from detector materials and components estimated for NEXT-100 is at most 4$\times$10$^{-4}$ counts keV$^{-1}$ kg$^{-1}$ y$^{-1}$ \cite{nextsensitivity}; the considered event selection requires one single track confined within the fiducial volume and a blob at both ends of the track. For a 29~keV wide energy region of interest and 91~kg $^{136}$Xe active mass, this corresponds to a background rate of 1~count per year, with the leading background sources being the PMTs and the substrates of the SiPMs. Contribution from other background sources, as the environmental gamma radiation, neutrons and muons, has been also evaluated and all other sources of background are expected to contribute each one at the level of 10$^{-5}$ counts keV$^{-1}$ kg$^{-1}$ y$^{-1}$ or below in the region of interest. In particular, considering the measured fluxes at LSC, the event rate due to either muons or neutrons is estimated to be at least two orders of magnitude smaller than that of the radioactive background from $^{214}$Bi and $^{208}$Tl \cite{nextsensitivity}.

The background activity induced by radon, mainly $^{222}$Rn, is in particular a serious concern for most rare event experiments. A thorough study has been carried out by NEXT for internal radon within the xenon recirculation loop, based on the measurements from NEXT-White using data collected between March and August 2017 in Run II \cite{nextrn}. Both radon-induced alpha particles and electrons have been studied. The specific activity of $^{222}$Rn was measured to be (38.1$\pm$2.2(stat)$\pm$5.9(syst))~mBq/m$^{3}$ through the alpha production rate induced in the fiducial volume by this isotope and its alpha-emitting progeny. Radon-induced electrons were also characterized from the decay of the $^{214}$Bi daughter ions plating out on the cathode of the TPC. The implications of these observations for two-neutrino and neutrinoless DBD searches of $^{136}$Xe, in NEXT-White and NEXT-100, have been analyzed; the $^{222}$Rn-induced expected rate in NEXT-White is about two orders of magnitude smaller than the background expected from radioactive impurities in the detector components while for NEXT-100 the projected rate should not exceed 0.1~counts/y in the neutrinoless DBD sample after cuts, at most one order of magnitude smaller than the rate estimated from radioactive impurities. Therefore, it has been concluded that radon-induced backgrounds are sufficiently low to enable a successful NEXT-100 physics program.

\begin{figure}
\centerline{\includegraphics[width=14cm]{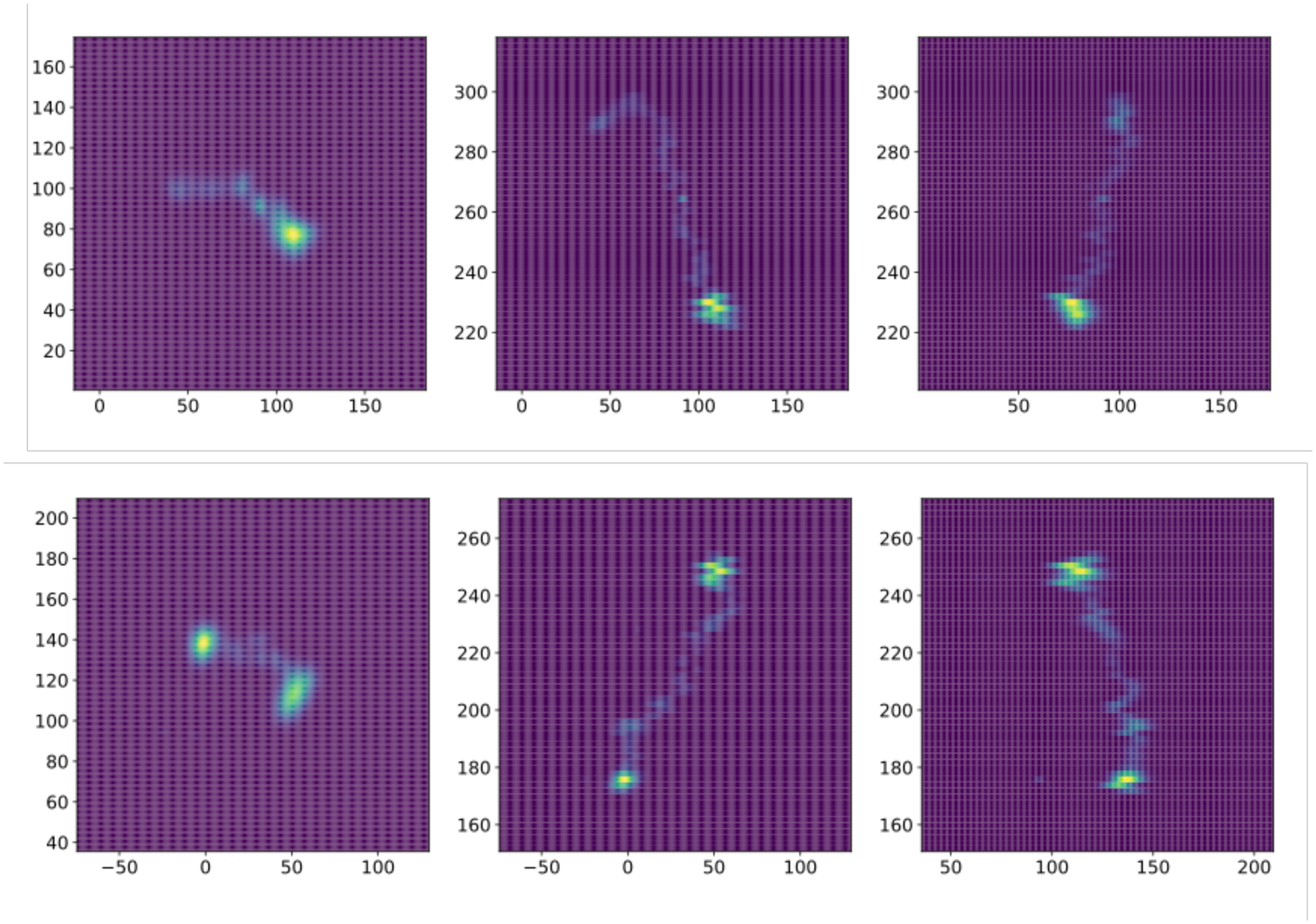}}
\begin{center}
\begin{minipage}[t]{16.5 cm}
\caption{Topological discrimination measured by NEXT-White: examples of background-like single-electron tracks (induced by Compton interactions, top) and signal-like double-electron events (induced by pair production interactions, bottom), as presented in \cite{nextmoriond}.}
\label{TopoNEXT}
\end{minipage}
\end{center}
\end{figure}

One of the main advantages of the technology chosen for NEXT is the possibility to reconstruct the topology of events with energies close to the transition energy of $^{136}$Xe, providing extra handles to reject background events. The use of this topological signature to eliminate background was pioneered by the Gotthard experiment \cite{gotthard}. For electrons moving through xenon gas the energy deposition is approximately constant along the length of the track, except for the two extremes where the electrons deposit a significant fraction of the event energy in a compact region, referred as a blob. The two electrons produced in DBD events appear as a single continuous trajectory with a blob at each end; background events from single electrons, however, typically leave a single continuous track with only one blob, as shown in Fig.~\ref{nexttrack}. As mentioned before, the main background in NEXT comes from high energy gammas emitted in $^{214}$Bi and $^{208}$Tl decays entering the active volume of the detector. In \cite{nextrecognition}, the demonstration of the use of a topological particle identification based on the expected signature of a double electron (signal) event compared to that of a single electron (background) was presented; Monte Carlo simulation and data taken with the NEXT-DEMO prototype were used. Single electron tracks originating from the photoelectric interaction of $^{22}$Na gammas and double electron tracks from the pair production of the 2.615~MeV gamma from $^{208}$Tl (with double escape peak at 1.593~MeV) were compared representing the background and the signal in a DBD. These data were used to develop algorithms for the reconstruction of tracks and the identification of the energy deposited at the end-points, providing a background acceptance of 24.3\% while maintaining an efficiency of 66.7\% for signal events. Using the data of the NEXT-White detector with a $^{228}$Th calibration source to produce signal-like and background-like events with energies near 1.6~MeV, a signal efficiency of (71.6$\pm$1.5)\% for a background acceptance of (20.6$\pm$0.4)\% has been found \cite{recognitionII}, in good agreement with Monte Carlo simulations and improving the factor of merit obtained for NEXT-DEMO; moreover, the expected performance specifically in the neutrinoless DBD region has been quantified by Monte Carlo in the same way, obtaining a signal efficiency of 71.5\% and a background acceptance of 13.6\% \cite{recognitionII}. In these analyses, the difference in the deposited energy at the beginning and at the end of an electron (or positron) track has been exploited to define a cut to separate signal from background, namely, a threshold on the lower energy extreme of a track. The same analysis will be done with the simulation of the NEXT-100 detector geometry, at its operation pressure of 15~bar. Figure~\ref{TopoNEXT} presents examples of double and single-electron tracks, corresponding to energies in the region of the double escape peak from $^{208}$Tl 2615~keV emission, as measured by NEXT-White.
The background rejection power could be even enhanced using deep learning techniques, like deep neural networks trained to classify events as signal or background thanks to the differences in the topological signatures, as shown in \cite{jinst17nn}.

All in all, it seems that the required background level in NEXT-100 is achievable thanks to passive shieldings, superb energy resolution, background discrimination techniques based on charged particle tracking and a thorough material radiopurity control.

\subsection{Results} \label{nextres}

\begin{figure}
\centerline{\includegraphics[width=10cm]{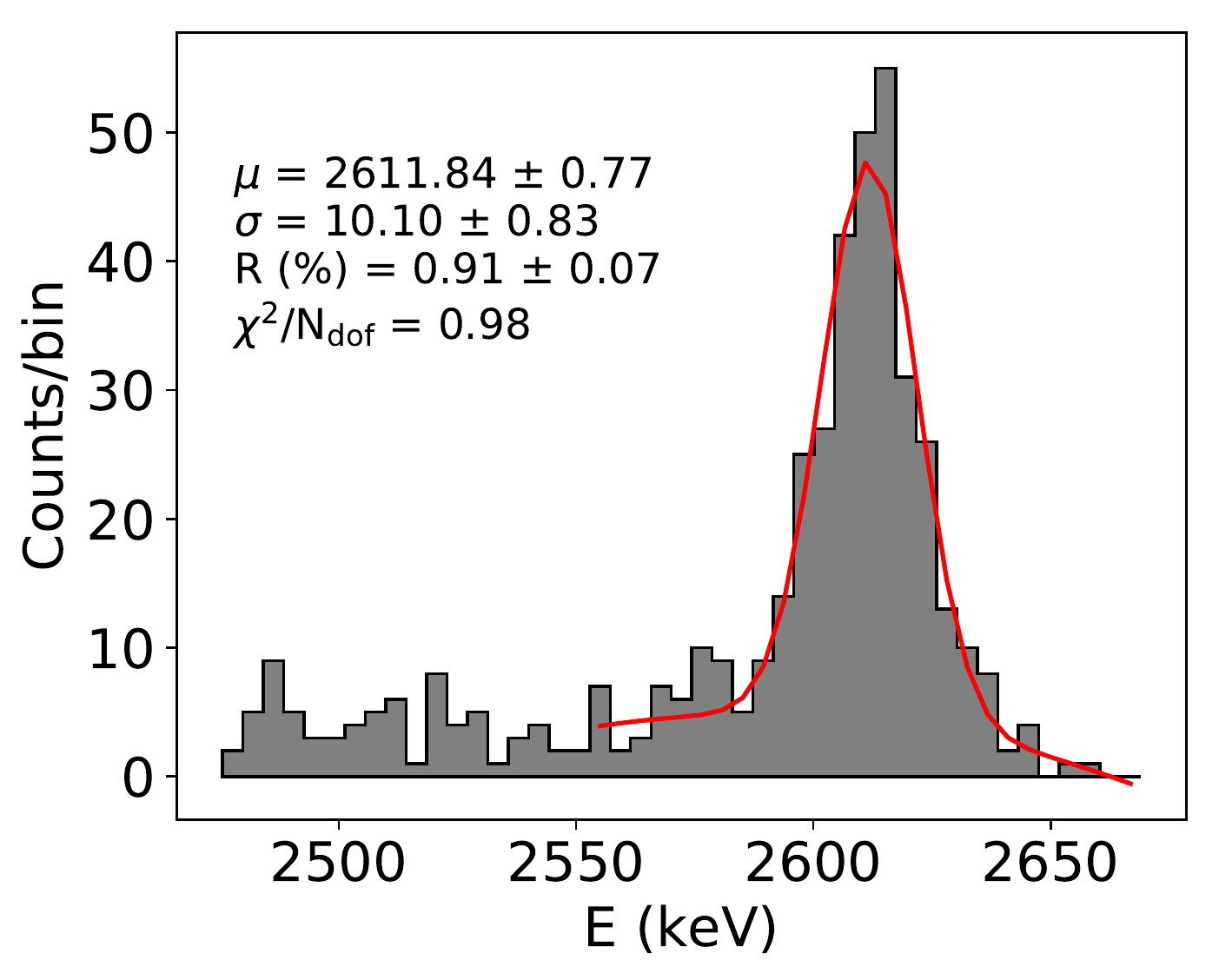}}
\begin{center}
\begin{minipage}[t]{16.5 cm}
\caption{Measurement of the energy resolution for the $^{208}$Tl emission at 2615~keV, from a calibration spectrum obtained with a $^{228}$Th source by NEXT-White, as presented in \cite{resolutionII}. Only statistical errors of the fits are reported in the plot.}
\label{ResNEXT}
\end{minipage}
\end{center}
\end{figure}

As presented in Sec.~\ref{secnew}, the NEXT-White detector is fully operative in Canfranc. The analysis of the available data has already released very relevant results \cite{nextnew}:
\begin{itemize}
\item Excellent stability, with data being taken without interruption over many months with very few sparks.
\item A robust procedure to calibrate the detector, correcting for finite (and variable) drift electron lifetime as well as for geometrical effects. This has been developed from krypton calibrations (using $^{83m}$Kr decays) performed during Run II \cite{jinst18calibration}.
\item A measured energy resolution of (0.91$\pm$0.12)\% FWHM at 2615~keV, following the analysis of calibration data with a $^{228}$Th source taken at 10.3~bar \cite{resolutionII} (see Fig.~\ref{ResNEXT}). Indeed, this result establishes NEXT-White as the xenon-based detector with the best energy resolution. Previously, an excellent linearity and an energy resolution that extrapolated to values well within range of the target value of 1\% FWHM at the transition energy, had been obtained from the measurements with krypton and also $^{137}$Cs and $^{232}$Th sources at lower pressure \cite{jinst18resolution}.
\item A characterization of the drift velocity and the longitudinal and transverse diffusion (parameters of great importance for the tracking capabilities) from the krypton data, finding a good agreement (at the 5\% level or better) with Magboltz simulations \cite{jinst18drift}.
\item Demonstration of the event identification capabilities by discriminating single and double electron tracks, as discussed in Sec.~\ref{secnextbkg}.
\end{itemize}

\begin{figure}
\centerline{\includegraphics[width=16cm]{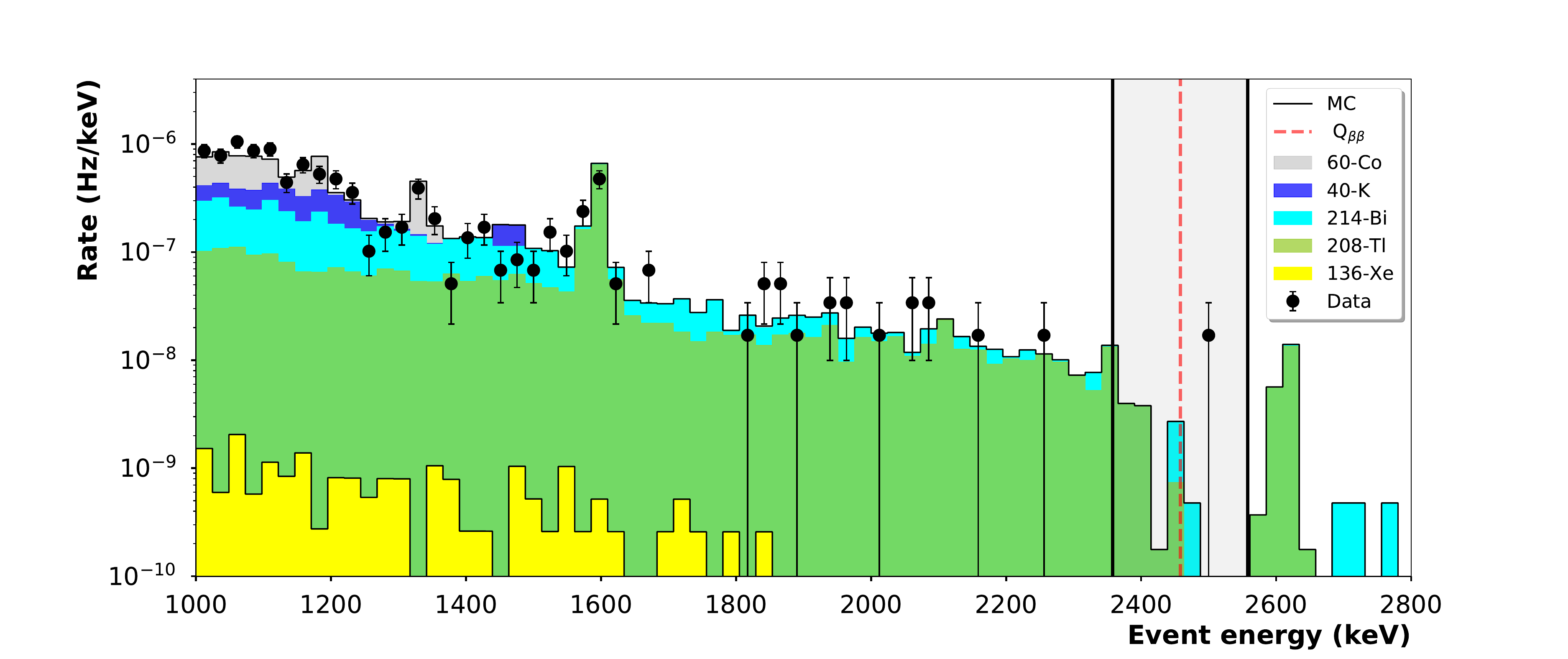}}
\begin{center}
\begin{minipage}[t]{16.5 cm}
\caption{Energy spectrum measured in NEXT-White Run IVc after topological cuts compared with simulated background including different contributions (as presented in \cite{nextradiogenicbkg}).}
\label{NextSpectrum}
\end{minipage}
\end{center}
\end{figure}

Another important result is a precise study of the radiogenic backgrounds in NEXT, from the data of the Run IV of NEXT-White and a detailed Monte Carlo simulation based on radiopurity measurements of all the relevant components \cite{nextradiogenicbkg}. During that run, data were taken at three different conditions: before the radon abatement system of LSC started flushing air inside the outer lead castle (Run IVa, 41.5~days), after that system started operation (Run IVb, 27.2~days) and after including additionally an inner lead castle surrounding the pressure vessel (Run IVc, 37.9~days). From the comparison of the data in the three different conditions, it can be concluded that the radon abatement system allows for operation of the NEXT detector in an environment virtually free of airborne Rn. The measured background rate in Run IVc for energies above 1000~keV has been (0.84$\pm$0.02)~mHz, which is a factor 1.71 higher that the predicted from simulation. After applying a set of topological cuts, the measured rate is reduced to (0.25$\pm$0.01)~mHz. Figure \ref{NextSpectrum} presents the energy spectrum measured in NEXT-White Run IVc after applying the topological cuts compared with the simulated background, including different contributions considering the normalization factors deduced from a fit to data in the 1-3~MeV range; it illustrates a good agreement between measured data and MC despite the limited statistics. In addition, in \cite{nextradiogenicbkg}, using the background model the sensitivity of NEXT-White to the DBD with neutrino emission half-life has been estimated to be (3.5$\pm$0.6)$\sigma$ after one year of data taking. For the neutrinoless channel, the expected background in a 200~keV window centered at the transition energy is (0.75$\pm$0.12$_{stat}$ $\pm$0.02$_{syst}$)~counts in 37.9~days, while 1~event was observed in the Run IVc data. Overall, these results derived from NEXT-White Run IV data and the satisfactory comparison with the model allow to validate the background assumptions used to estimate the physics case of the NEXT-100 experiment. Concerning NEXT-100 design, the anode (tracking plane) region has been identified as the detector area where improvements could be particularly beneficial to further improve the background levels.

One of the final goals of the NEXT-White detector, once operating with enriched xenon, is to provide a measurement of the two-neutrino DBD mode of $^{136}$Xe. Although work is in progress, from the first data taken at Runs IV-V, the very preliminary estimate of the half-life for this decay is compatible with published data within 1$\sigma$ \cite{paueps2019}. The full operation of NEXT-White during 2019 and 2020 will allow for the measurement of the two-neutrino DBD half-life implementing for the first time the analysis with the NEXT technology.

\begin{figure}
\centerline{\includegraphics[width=10cm]{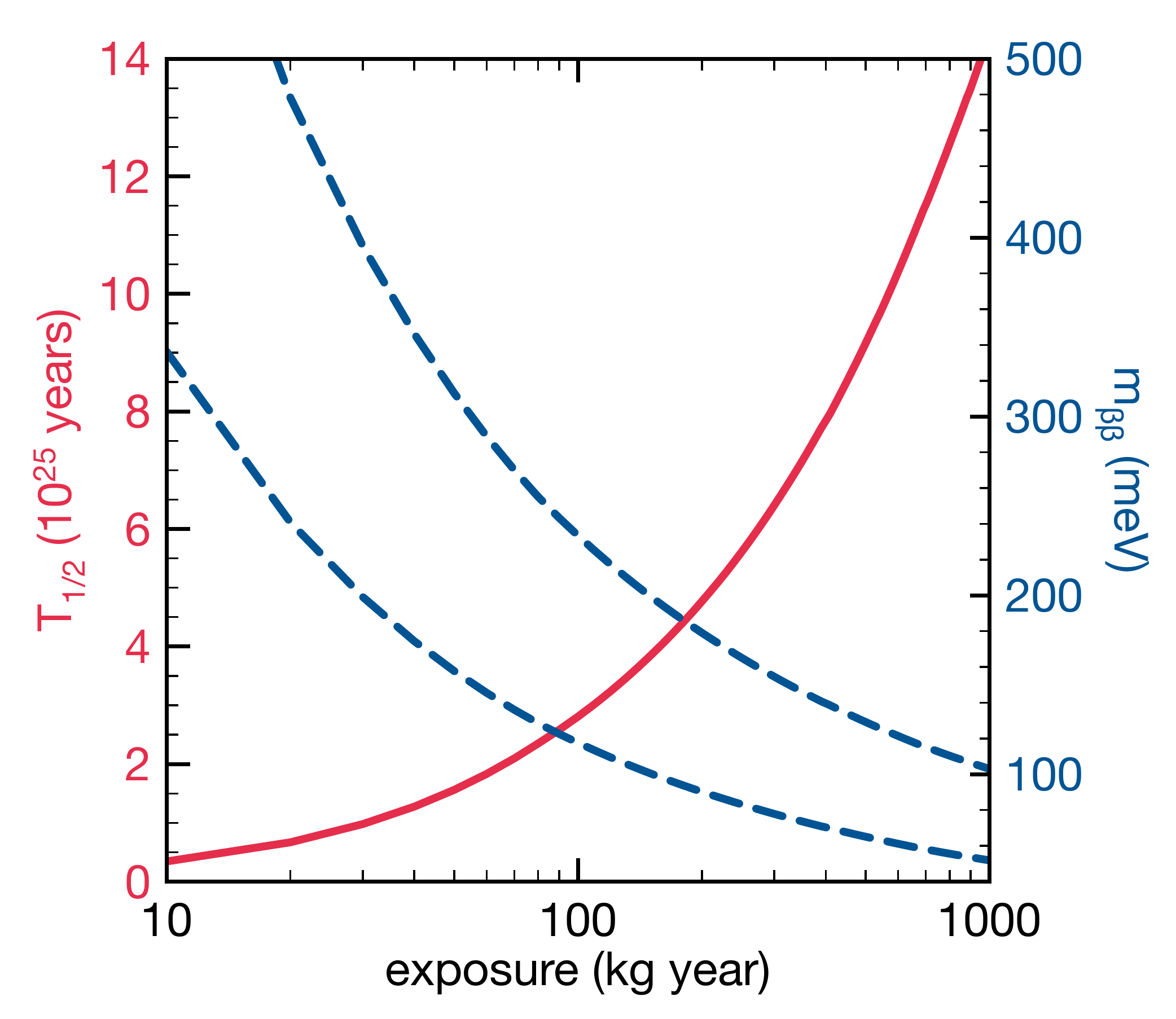}}
\begin{center}
\begin{minipage}[t]{16.5 cm}
\caption{Sensitivity of NEXT-100 in terms of the accumulated exposure for an estimated background rate of 4$\times$10$^{-4}$ counts keV$^{-1}$ kg$^{-1}$ y$^{-1}$ (as presented in \cite{nextsensitivity}). The solid curve represents the half-life sensitivity, while the dashed curves correspond to the neutrino effective mass sensitivity for the largest and smallest nuclear matrix elements considered.}
\label{Next100sensi}
\end{minipage}
\end{center}
\end{figure}

While the design and construction of the NEXT-100 are being completed, its sensitivity (see Sec.~\ref{det}) has been evaluated based on the results derived from the prototypes and the background level expectation \cite{nextsensitivity}. As discussed in section \ref{secnextbkg}, material screening measurements and a detailed Monte Carlo detector simulation predicted a background rate for NEXT-100 of at most 4$\times$10$^{-4}$~counts keV$^{-1}$ kg$^{-1}$ y$^{-1}$. With a signal detection efficiency of 28\% and the estimated background rate, the sensitivity of NEXT-100 is plotted in Fig.~\ref{Next100sensi} in terms of the accumulated exposure. The NEXT-100 detector will reach a sensitivity to the neutrinoless DBD half-life of 1$\times$10$^{26}$~years for an exposure of 500~kg$\cdot$year \cite{revHPXe}. Although this is the same sensitivity already achieved by KamLAND-Zen (described in Sec.~\ref{scint}), the capability of NEXT-100 to explore a nearly background-free experiment at the 100~kg scale, with the potential to improve its radioactive budget, resolution and topological signature at the tonne scale, is the strongest asset of the experiment.
In the longer term, for the expected background in the HD option, the experiment could reach for the neutrinoless DBD a discovery sensitivity of 1.7$\times$10$^{27}$~y after 10~years, allowing to explore neutrino effective masses in the range 13-57~meV \cite{appec}. For the BOLD option, the half-life sensitivity would be 8$\times$10$^{27}$~y for an exposure of 10~t$\cdot$y \cite{nextmoriond}; the goal is to achieve a full exploration of the Inverse Hierarchy.

Other interesting results have been derived beyond the horizon of the NEXT-100 detector, aimed in many cases to get almost background-free conditions for future detectors at the tonne-scale. Important progress has been made, for instance, for the Ba-tagging implementation, as already discussed in Sec.~\ref{futnext}.

The operation of the xenon detector in the presence of different additives has been thoroughly studied within the NEXT collaboration. As already commented in Sec.~\ref{secfp}, the response of Xe/TMA mixtures was characterized using the NEXT-MM prototype \cite{protomm,protomm2,diego} and the capability of TMA ((CH$_{3}$)$_{3}$N) molecules to shift the wavelength of Xe VUV emission (160-188~nm) to a longer, more manageable, wavelength (260-350~nm) has been also investigated in \cite{nima18tma}. The discrimination efficiency through pattern recognition of the topology of primary ionization trails can be improved reducing the electron diffusion by the addition of a small fraction of a molecular gas to pure xenon; however, this makes the EL yield drop, compromising the achievable energy resolution. In \cite{plb17co}, the electroluminescence yield of xenon with sub-percent levels of CO$_{2}$ was measured, demonstrating that the EL production is still high in these mixtures. Moreover, the effect of adding also other molecular gases to xenon (CO$_{2}$, CH$_{4}$ and CF$_{4}$ have been considered) on the EL yield and energy resolution has been studied with data using a gas proportional scintillation counter and microscopic simulation \cite{jhep19el}; the main conclusion is that CH$_{4}$ shows the best performance and stability as molecular additive to be used in the NEXT-100 TPC, with an extrapolated energy resolution which is only slightly worse than the one obtained for pure xenon. The possibility of reducing the diffusion of the drifting electrons while keeping nearly intact the energy resolution of a pure xenon EL TPC by a substantial addition of helium, around 10-15\%, has been also studied, firstly based on state-of-the-art microscopic simulations \cite{nima18xe} and then from measurements using a small gas proportional scintillation chamber; in particular, the EL yield of Xe–He mixtures is reduced by $\sim$ 2\%, 3\%, 6\% and 10\% for 10\%, 15\%, 20\% and 30\% of He concentration (volume), respectively (a reduction which is even lower than the expectation from simulation results), without degrading the energy resolution \cite{xeheII}. In addition, a simulation framework for microscopic simulation of the electron transport in xenon-based optical TPCs in the presence of molecular additives has been developed and cross-checked with data using mixtures with CO$_{2}$ and CF$_{4}$ \cite{nima18aditives}.

Many other results, dealing with different areas from technical issues to analysis methods and particular Physics aspects, have been obtained in the framework of the NEXT collaboration. For example, an improved measurement of electron-ion recombination in high-pressure xenon gas was derived \cite{serra}. Ionization and scintillation produced by nuclear recoils in gaseous xenon were studied in a HPXe-EL TPC, demonstrating the ability to discriminate between electronic and nuclear recoils using the ratio of ionization to primary scintillation, which is relevant for dark matter direct detection \cite{renner}. The mechanical and electrical effects of high pressure gas environments on insulating polymers (like PTFE, HDPE, PEEK, \dots) in argon and xenon have been analyzed in \cite{jinst18insulation}. Different reconstruction algorithms have been explored, like the Maximum Likelihood Expectation Maximization method, a generic iterative algorithm to find maximum-likelihood estimates of parameters that has been applied to reconstruct a transverse projection of the event using the photosensor signals integrated over time \cite{jinst17algo}; this algorithm could help to obtain nearly optimal energy resolution for events distributed over the full active volume of the TPC.

Summarizing, the results of the NEXT program using HPXe-EL TPCs have already shown that the approach of having separated energy and tracking readout planes using different sensors allows to combine the measurement of the topological signature of the event for background discrimination with the energy resolution optimization, two features of great value for neutrinoless DBD searches. Very good energy resolution, $\sim$1\% FWHM, has been measured at the transition energy of $^{136}$Xe and the topological signature of electrons has been clearly established confirming that a background level at the order of 10$^{-4}$ counts keV$^{-1}$ kg$^{-1}$ y$^{-1}$ can be at reach. Therefore, NEXT has a large discovery potential and the capability to offer a technique that can be scaled, at a very competitive cost, to the tonne scale.

\subsection{TPCs and Tracking Double Beta Decay experiments worldwide}
\label{trackdetww}

The ``Enriched Xenon Observatory'' (EXO), located at the Waste Isolation Pilot Plant (WIPP), New Mexico, US, has investigated the DBD of $^{136}$Xe using also a TPC but with liquid xenon. In EXO-200, with 200 kg of xenon enriched to 80.6\% in $^{136}$Xe, the ionization signal and the 178~nm scintillation light are read by crossed wire planes and avalanche photodiodes, respectively, placed at the two sides of the symmetric cylindrical chamber. The sides of the chamber are covered with Teflon sheets that act as VUV radiation reflectors, improving the light collection. The z-coordinate is obtained by measuring the difference in the arrival time between the ionization and scintillation signals. The combined charge and light information allows also to improve the energy resolution for the expected signal. This detector was built after an extremely careful selection of low radioactive materials \cite{exoscreening} and has derived very relevant results \cite{exonature}. The measured half-life for the DBD with two neutrinos emission confirmed a value shorter than expected \cite{exo2nu} and for the neutrinoless channel the upper limit set is $T_{1/2}^{0\nu}>3.5 \times 10^{25}$~y (90\% C.L.) \cite{exoprl}. Results for Majoron-emitting modes \cite{exomaj}, decays to excited states of the daughter nuclei \cite{exoexc}, for the DBD of $^{134}$Xe \cite{exo134} and for nucleon decays \cite{exonuc} have also been obtained. The next-generation detector in this project is the nEXO observatory, using 5~tonnes of isotopically enriched liquid-xenon in a TPC; an improvement of two orders of magnitude in sensitivity over current limits is foreseen \cite{nexo}, following a predicted median background rate of 3.6$\times$10$^{-4}$~counts kg$^{-1}$ y$^{-1}$ per FWHM in the inner 2000~kg of LXe and a FWHM energy resolution of 2.35\%. Works are also underway for imaging individual Ba atoms resulting from the DBD in xenon for barium tagging \cite{exotag}.

Following the approach of NEXT, two other projects are proposing the use of HPXe TPCs for the neutrinoless DBD investigation of $^{136}$Xe. In the first phase of the PandaX-III (``Particle And Astrophysical Xenon Experiment III'') experiment, the plan is to operate a TPC with 200~kg of xenon gas enriched to 90\% in $^{136}$Xe at 10~bar \cite{pandax}. The use of TMA as quencher is foreseen, although  then, the starting time of events (t$_{0}$), related to the z-axis position, cannot be directly obtained. Microbulk Micromegas, a fine pitch micro-pattern gas detector, will be used for charge readout (rather than wires and pads) in order to register tracks of DBD events with good energy and spatial resolution; an energy resolution of 3\% at the transition energy of the DBD of $^{136}$Xe is expected. The Micromegas readout can allow a reduction of background associated with the PMTs used in other projects (which could also be obtained by using only SiPMs). The additional background suppression based on the tracking capabilities of the detector is under thorough study \cite{pandaxtopo} and work is also underway for prototypes. The second phase, a ton-scale experiment, is expected to comprise five of such TPCs. The AXEL (A Xenon ElectroLuminescence) project in Japan is based on a high pressure xenon gas TPC for DBD searches too, exploring a new electroluminescence light collection system to achieve high energy resolution in a large detector \cite{axel}; a 10~l size prototype has been developed so far and a larger one, with 180~l, is in preparation.

As mentioned in Sec.~\ref{trackdet}, other approaches have been followed to get the event tracks in DBD detectors, like the one used by the NEMO experiment. The third phase of NEMO finished the data taking at the Modane Underground Laboratory in 2011. NEMO3 took data using a cylindrical detector containing about 10 kg of emitters in different sectors: $\sim$6.91 kg of $^{100}$Mo, $\sim$0.93 kg of $^{82}$Se and smaller amounts of $^{150}$Nd (36.6~g), $^{130}$Te, $^{116}$Cd, $^{96}$Zr and $^{48}$Ca (6.99~g). The DBD process with emission of neutrinos has been studied with unprecedented statistics for several isotopes ($^{100}$Mo, $^{116}$Cd, $^{82}$Se and $^{96}$Zr). For the neutrinoless channel, no positive signal has been found, and the following limits have been obtained using more than 5~y of data for the half-lives and the effective Majorana neutrino mass (assuming the light neutrino exchange mechanism) at 90\% C. L.: $T_{1/2}^{0\nu}>2.0 \times 10^{22}$~y and $m_{\beta\beta}<6.0-26$~eV for $^{48}$Ca \cite{nemoca}, $T_{1/2}^{0\nu}>2.5 \times 10^{23}$~y and $m_{\beta\beta}<1.2-3.0$~eV for $^{82}$Se \cite{nemose}, $T_{1/2}^{0\nu}>1.1 \times 10^{24}$~y and $m_{\beta\beta}<0.33-0.62$~eV for $^{100}$Mo \cite{nemomo} and $T_{1/2}^{0\nu}>2.0 \times 10^{22}$~y and $m_{\beta\beta}<1.6-5.3$~eV for $^{150}$Nd \cite{nemond}. The mass ranges reflect nuclear matrix element values from different calculations. Half-life limits for decays to excited states of the daughter nuclei as well as constraints on lepton number violating parameters for other neutrinoless DBD mechanisms, such as right-handed currents, Majoron emission and R-parity violating supersymmetry modes, have also been set.

The project SuperNEMO keeps the same approach of the NEMO experiment, increasing the mass of the sources ($^{82}$Se, $^{150}$Nd) to the scale of hundreds of kg, which requires a change in the detector geometry; about 20 modules are envisaged, having each one a planar design with the thin foil of enriched material containing the DBD emitters in the middle of a gas tracking chamber covered by plastic scintillator blocks acting as calorimeters. The first module, the SuperNEMO Demonstrator, has been installed in the Modane laboratory. Details of both the tracking and calorimeter systems can be found at \cite{nemotrack,nemocal}. Improving also some experimental parameters, the sensitivity to the effective neutrino mass could reach $\sim$50 meV and new Physics models of neutrinoless DBD could be probed \cite{supernemoepjc}.

\section{CROSS} \label{crosssec}

The ``Cryogenic Rare-event Observatory with Surface Sensitivity'' (CROSS) is a new project devoted to DBD searches which is underway intended to achieve unprecedented sensitivity thanks to novel background rejection techniques. It will be based on arrays of tellurium oxide (TeO$_{2}$) and lithium molybdate (Li$_{2}$MoO$_{4}$) bolometers enriched in the DBD isotopes $^{130}$Te and $^{100}$Mo, respectively. The CROSS collaboration is composed by researchers from CNRS/IN2P3 and IRFU/CEA (France), INFN (Italy), INR (Ukraine), ITEP (Russia) and LSC (Spain). The concept, the planned set-up in Canfranc and the expected sensitivity of this experiment will be  summarized in this section.

\subsection{Concept}
The chosen approach in CROSS offers important advantages: the high double beta transition energy of the DBD emitters, the easy crystallization processes of TeO$_{2}$ and Li$_{2}$MoO$_{4}$, and the superior bolometric performance of these compounds in terms of energy resolution (they have demonstrated to provide $\sim$1.5 per thousand FWHM energy resolution in the region of interest) and intrinsic purity. The enrichment, purification and crystal growth are mostly based on already established protocols. As mentioned in Sec.~\ref{bolo}, different strategies have been explored to reject backgrounds in bolometers \cite{giulianiijmpa}; the simultaneous detection of light and heat and the comparison of the respective signal amplitudes, represent a powerful tool for $\alpha/\beta$ discrimination. The key idea of this project is to provide the bolometric detection technique with an effective PSD capability, obtained by exploiting Solid-State-Physics phenomena in superconductors, to reject events from surface radioactive impurities (a limiting background in large-scale bolometric searches) by discrimination of pulse shapes. The surfaces of the crystals will be coated by an ultrapure superconductive aluminium film, which will act as a pulse-shape modifier by changing the pulse development in case of close-to-the-surface energy depositions, exploiting phonon-physics effects in aluminium. The use of two types of phonon sensors is foreseen: a neutron transmutation doped (NTD) Ge thermistor (following the scheme adopted in Cuoricino and CUORE), which is mainly sensitive to thermal phonons, and a NbSi thin film, which is faster and exhibits a significant sensitivity to athermal phonons.

The two technological advancements introduced by CROSS (the sensitivity to surface events and the additional phonon readout based on superconductive NbSi films) have been thoroughly studied at CSNSM (Orsay, France) \cite{cross,cross2}.
It has been shown that ultra-pure superconductive Al films deposited on the crystal surfaces act successfully as pulse-shape modifiers, both with fast and slow phonon sensors. Rejection factors higher than 99.9\% of $\alpha$ surface radioactivity have been demonstrated in a series of prototypes based on crystals of Li$_{2}$MoO$_{4}$ and TeO$_{2}$. In particular, four crystals of these materials with natural isotopic composition (20$\times$20$\times$10~mm$^{3}$ or 20$\times$20$\times$5~mm$^{3}$ in size, with masses from 12 to 25~g), equipped with NTD Ge thermistors, were used in the tests. The thickness of the Aluminium layers evaporated on a crystal side was 1 or 10~$\mu$m. Several PSD parameters, like the rise time, have been tested. An excellent discrimination factor for surface events with practically full acceptance of bulk events has been obtained for the 1~$\mu$m thick Al film; as the thin layer option has important advantages (shorter evaporation time, better film adhesion and much less total mass of material), it has been chosen as the baseline solution for CROSS detectors. It has been also shown in \cite{cross} that point-like energy depositions can be identified up to a distance of $\sim$1~mm from the coated surface; the depth-extended PSD capability of the technique is promising for $\beta$-surface-event discrimination, given the range of relevant $\beta$ electrons but additional R\&D work on this is still necessary. Some tests with NbSi sensors have been made, showing a rise time behaviour opposite to that of NTD Ge thermistors, as surface events are slower when detected by the NbSi sensor; the NbSi option could provide a further mean to enhance the surface sensitivity in the CROSS technology. Following the promising results achieved for the PSD in these first tests, the near future goals include to fully coat the crystals with Al film and also to test other coating materials like Palladium.

\subsection{Set-up at Canfranc}

\begin{figure}
\begin{center}
\includegraphics[width=7cm]{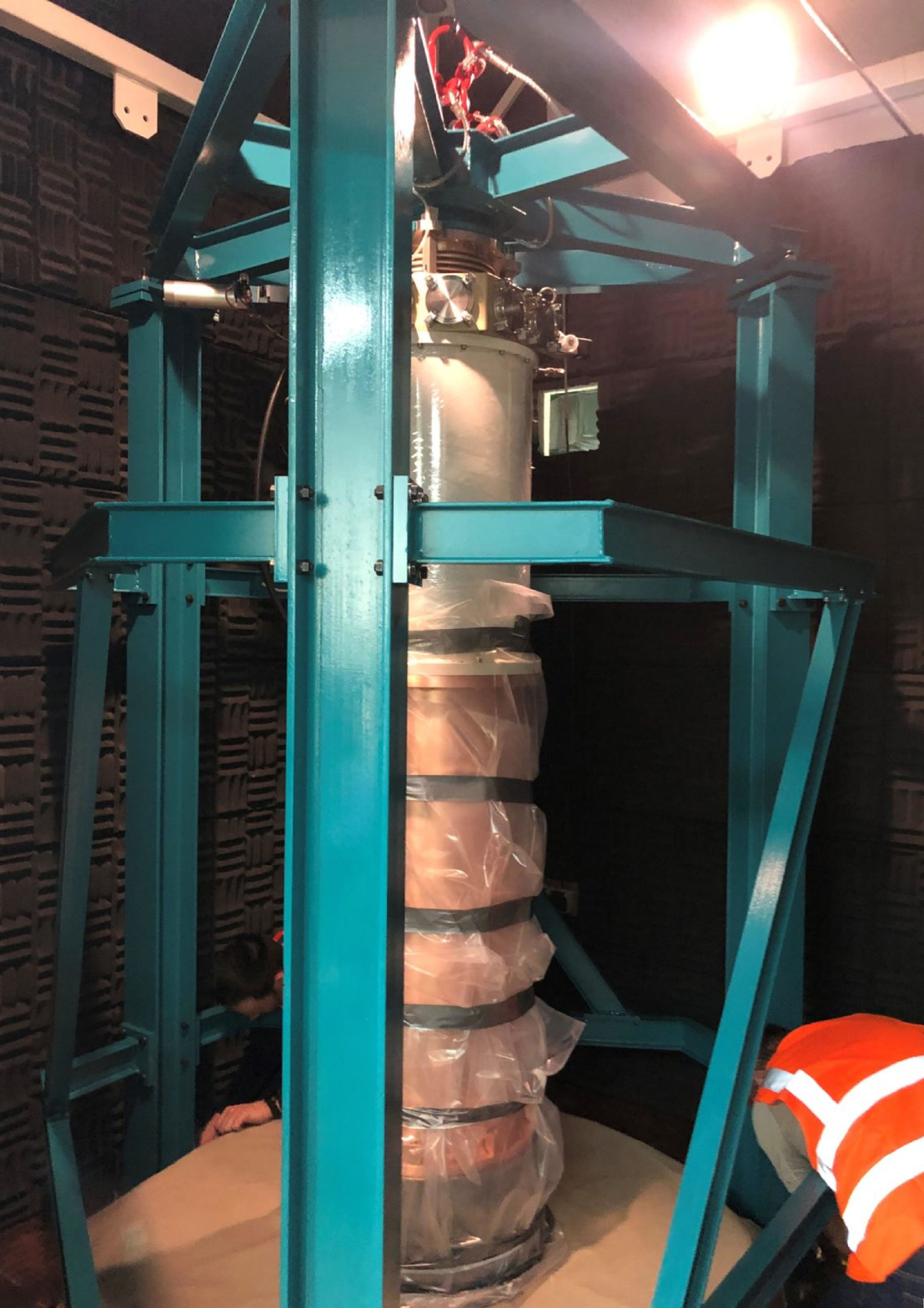}
\hskip 0.1 cm
\includegraphics[width=6cm]{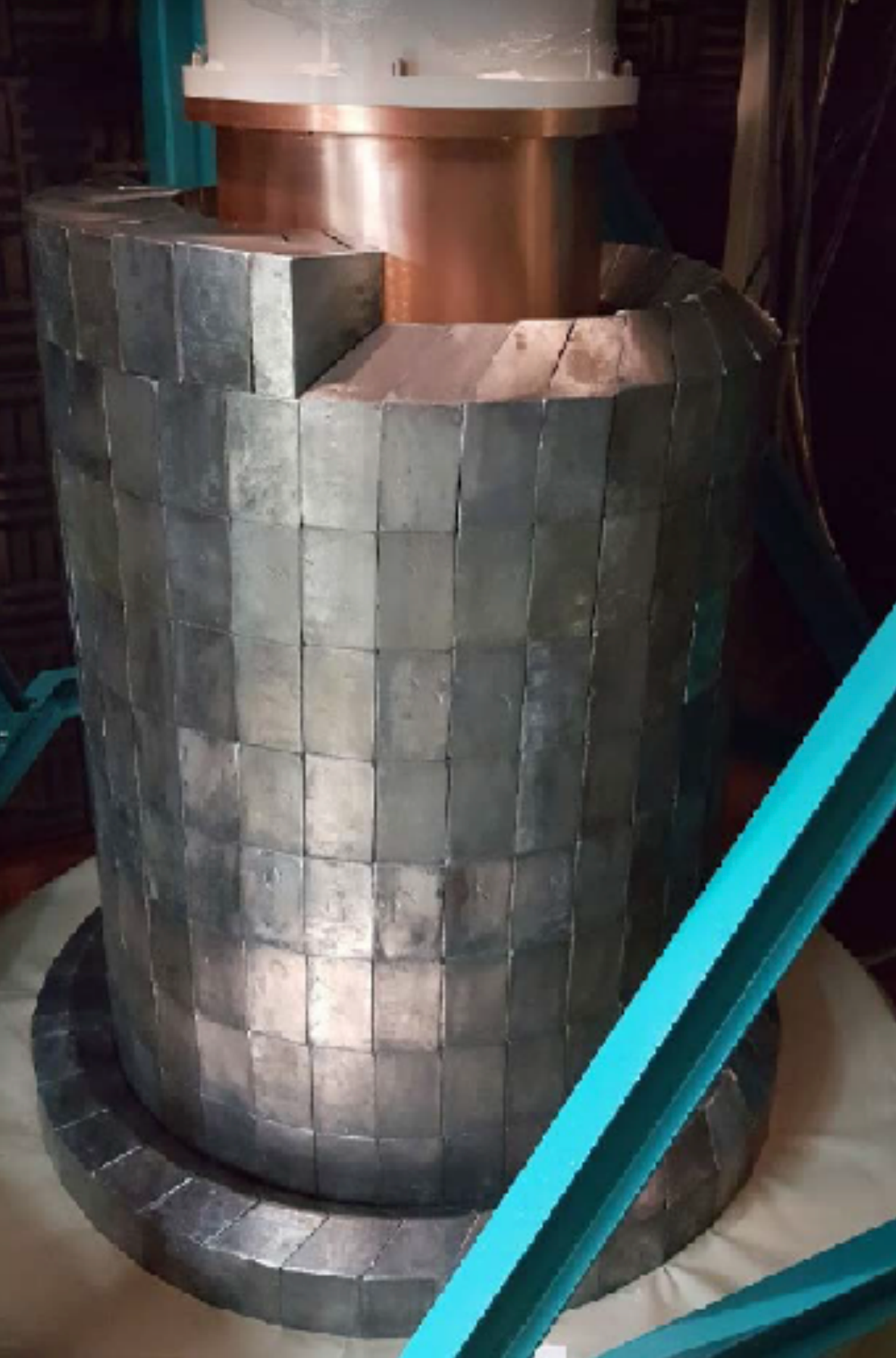}
\begin{minipage}[t]{16.5 cm}
\caption{Images of the dilution refrigerator (left) and the partial lead shielding (right) installed at LSC for the CROSS demonstrator (courtesy of LSC and CROSS collaboration).}
\label{crossrefri}
\end{minipage}
\end{center}
\end{figure}

An array of detectors provided with these advanced features will be installed at LSC as a medium scale demonstrator. Its purpose is to test the technique with high statistics, to prove the stability and the reproducibility of the CROSS methods and to demonstrate the potential of CROSS in terms of background. But this pilot experiment will be a very sensitive neutrinoless DBD search too, competitive in the international context. A dedicated dilution refrigerator with an experimental volume of $\sim$150~l was installed in Canfranc in April, 2019 inside the hut previously used by the ROSEBUD experiment \cite{rosebud} (see Fig.~\ref{crossrefri}). This machine provides a base temperature of $\sim$10~mK, being optimized for the operation of macro-bolometers. Tests have been satisfactorily made with several crystals, made of CdWO$_{4}$, Li$_{2}$MoO$_{4}$ and TeO$_{2}$, and new ones are foreseen in 2020.

The first demonstrator of the CROSS technology will be installed in this cryogenic set-up, with 32 Li$_{2}$MoO$_{4}$ crystals grown with molybdenum enriched in $^{100}$Mo at $>$95\% level. In a longer term, the operation of 90 enriched Li$_{2}$MoO$_{4}$ and TeO$_{2}$ bolometers is foreseen. Each crystal is a cube with 45~mm side and a mass of 0.28~kg. The crystals have been produced according to a protocol developed in the LUMINEU experiment, which ensures excellent bolometric performance and high radio-purity \cite{lumineu,cupidmo}. The crystals will be arranged in a tower of eight floors with four crystals each, with almost full visibility between adjacent elements, in order to improve the background rejection through anti-coincidences. The total $^{100}$Mo mass will be of 4.7~kg, corresponding to 2.9$\times$ 10$^{25}$ $^{100}$Mo nuclei. A $^{130}$Te section is foreseen in the CROSS demonstrator too, but the design of this part is less advanced; cubic crystals with 60~mm side will be tested.

The detector configuration and the assembly procedures will be as close as possible to those extensively tested in previous bolometric experiments. Crystal holders will be made by ultra-low-radioactivity copper and small PTFE elements will be used to hold the crystals. The cryostat will be shielded with lead, copper and polyethylene against environmental gamma and neutron backgrounds. A muon veto will be set up by covering with scintillating elements the walls of the hut housing the cryostat. An anti-vibration suspension for the detectors is foreseen too. The LSC offers an anechoic room, placed in a Faraday cage, with a separate room for the pumping system. The underground clean room for detector assembly and the radon abatement system of the laboratory are also useful for CROSS preparation and operation. The adoption of PSD techniques has important implications in the design of both the front-end and back-end electronics; a high resolution digitization system is being designed for the CROSS experiment based on a custom solution comprised of an analog-to-digital board interfaced to a FPGA module \cite{crossdig}. The DAQ system is particularly adapted to the requirements of fast scintillating bolometers (faster signals, higher pile-up, higher resolution, continuous acquisition or spread in detector characteristics).

\subsection{Sensitivity}
The preliminary background studies for the CROSS demonstrator to be operated in LSC point to levels in the range of 10$^{-2}$-10$^{-3}$~counts keV$^{-1}$ kg$^{-1}$ y$^{-1}$, taking into account that the cryostat has been fabricated with a careful selection of the materials and that the shielding was designed to achieve very low background in the experimental volume.
In these conditions, the 2-year sensitivity to the effective Majorana neutrino mass, considering computations of nuclear matrix elements taken from different models, is in the range $\sim$100–200~meV \cite{cross}. The 5-year sensitivity goes down to 68-122~meV in the most favorable conditions of background (10$^{-3}$~counts keV$^{-1}$ kg$^{-1}$ y$^{-1}$), being the corresponding expected half-life limit at 90\% C.L. of 2.8$\times$10$^{25}$~years for $^{100}$Mo \cite{cross}.
This shows the potential of the CROSS technology, considering the low size and the demonstration character of the proposed set-up.

This project will set the grounds for large-scale experiments searching for neutrinoless DBD with almost zero background (less than 0.5~counts/y in 1~tonne of isotope in the region of interest).
When applied to large arrays containing neutrinoless DBD candidate masses of the order of hundreds of kilograms, the CROSS
technology has the potential to explore fully the inverted-ordering region and to detect a neutrinoless DBD even in case of direct ordering, provided that the lightest neutrino mass is higher than $\sim$10–20~meV \cite{cross}.

\subsection{Bolometric Double Beta Decay experiments worldwide}
\label{boloww}

From the first MIBETA results in the nineties \cite{ALE00c} to the recent ones of the ``Cryogenic Underground Observatory for Rare Events'' (CUORE) \cite{cuore18}, passing through CUORICINO \cite{prccuoricino} and CUORE-0 \cite{prlcuore0}, the mass of TeO$_{2}$ used at the LNGS has been increasing steadily improving the experiment sensitivity. Cubic crystals with a side of 5~cm are operated at a temperature of $\sim$10~mK. The CUORICINO and CUORE-0 experiments accumulated 40.7~kg and 39~kg of natural TeO$_{2}$, respectively. CUORE is the culmination of a long-standing effort, using 741~kg of tellurite (206~kg of $^{130}$Te) in 988 crystals grouped in 19 towers. The challenge of cooling such a big mass was enormous; it has been accomplished using the technology of cryogen-free dilution refrigerators. Following their first published results \cite{cuore18}, an effective energy resolution of (7.7$\pm$0.5)~keV FWHM and a background in the region of interest of (0.014$\pm$ 0.002) counts keV$^{-1}$ kg$^{-1}$ y$^{-1}$ were achieved. This background level confirms a reduction of more than one order of magnitude with respect to CUORICINO. From a total exposure of 372.5 kg$\cdot$y, the latest limit on the neutrinoless DBD of $^{130}$Te is $T_{1/2}^{0\nu}\geq 3.2 \times 10^{25}$~y (90\% C.L.); this results in a limit on the effective Majorana neutrino of $m_{\beta\beta} \leq 75-350$~meV, depending on the nuclear matrix elements considered \cite{cuore19}. From CUORE-0 data, limits for the DBD of $^{130}$Te to the first 0$^{+}$ excited state of $^{130}$Xe \cite{cuore0exc} and for the neutrinoless EC$\beta^{+}$ decay of $^{120}$Te \cite{cuore0EC} have been also derived.

Following the work of LUCIFER \cite{lucifer}, CUPID-0 is the first large array of scintillating ZnSe cryogenic calorimeters implementing particle identification. The heat-light readout providing a powerful rejection tool has allowed them to reduce the background level in the region of interest down to 3.5 10$^{-3}$~counts keV$^{-1}$ kg$^{-1}$ y$^{-1}$. As no signal has been observed, they have constrained the $^{82}$Se half-life as $T_{1/2}^{0\nu}\geq 3.5 \times 10^{24}$~y (90\% C.L.), which corresponds to an effective Majorana neutrino mass $m_{\beta\beta} \leq 311-638$~meV at 90\% C.L. depending on the nuclear matrix element calculations \cite{cupid0}. For the investigation of $^{100}$Mo, the compounds ZnMoO$_{4}$, Li$_{2}$MoO$_{4}$ and CaMoO$_{4}$ are being used by LUMINEU and CUPID-Mo \cite{lumineu,cupidmo}, CLYMENE \cite{clymene} and AMoRE \cite{amore} experiments. The CUPID-Mo detection technology has given the most accurate determination of the half-life for the two-neutrino channel to date, confirming, with a statistical significance larger than 3$\sigma$, that the single-state dominance model is favored over the high-state dominance model \cite{cupidmo2nu}.

CROSS is deeply connected with CUPID \cite{cupid0,lumineu,cupidmo,cupid1,cupid2,cupidpreCDR}, which plans to use the current CUORE infrastructure at LNGS in preparation for a tonne-scale experiment and has been selected as one of the most competitive projects, together with LEGEND and NEXT, by the Double Beta Decay APPEC Committee \cite{appec}. CUPID, irrespectively of the chosen isotope, foresees in principle the use of light detectors. But the CROSS method will allow getting rid of the light detectors used up to now to discriminate surface alpha particles, simplifying a lot the bolometric structure being devoid of an additional device and achieving the additional advantage to reject also beta surface events, which become an ultimate background source if only alpha particles are tagged. The intrinsic modularity and the simplicity of the read-out will make CROSS easily expandable. A mixed configuration with Al films replacing the reflector used to increase the light collection is a realistic option for the first phase of CUPID, in order to have PSD of surface events and to reduce background (as radioactivity of reflecting foils is an issue). The full CROSS technology could be considered for a second phase of CUPID or for another similar-scale bolometric experiment.

\section{The BiPo-3 detector for radiopurity measurements}
\label{biposec}

BiPo-3 is a unique set-up installed at LSC to measure $^{208}$Tl and $^{214}$Bi contamination at $\mu$Bq/kg level on planar geometries. It is the final development of the BiPo project, promoted by the SuperNEMO collaboration devoted to the investigation of DBD (see Sec.~\ref{trackdetww}) in order to screen mainly the source foils containing the DBD emitters. Due to its high sensitivity, BiPo-3, an IN2P3/SuperNEMO facility, has become a general low background detector of interest for the measurement of the radiopurity of large surfaces used in underground experiments. It has been already used to analyze planar samples for experiments other than SuperNEMO, from the University of Zaragoza and presently CIEMAT (Centro de Investigaciones Energ\'{e}ticas, Medioambientales y Tecnol\'{o}gicas). Here, the motivation and principle of such a facility will be  presented, the detector set-up of BiPo-3 described and the main results obtained summarized.

\subsection{Motivation and principle}

The ultra-low background conditions required in the search for DBD, as for other rare event processes, demand levels of radiopurity in the components which in some cases cannot be achieved even by very sensitive techniques typically used in this context like gamma spectrometry performed using HPGe detectors operated deep underground; then, specific, non-conventional measurements have to be developed to quantify extremely low activity values. This was the case of the DBD source foils to be used in the SuperNEMO demonstrator. The best detection limit that could be reached with germanium detectors for $^{208}$Tl is around 50~$\mu$Bq/kg, which is about one order of magnitude less sensitive than the required value. A dedicated detector was conceived to screen these foils, and also other components, based on non-destructive techniques. The design goal of this detector was to reach for bulk activity the sensitivity levels required by SuperNEMO (2~$\mu$Bq/kg for $^{208}$Tl and 10~$\mu$Bq/kg for $^{214}$Bi) in a measurement of a few months for a whole source foil being 40~mg/cm$^{2}$ thick (corresponding to 80~$\mu$m for a selenium foil).

\begin{figure}
\centerline{\includegraphics[width=10cm]{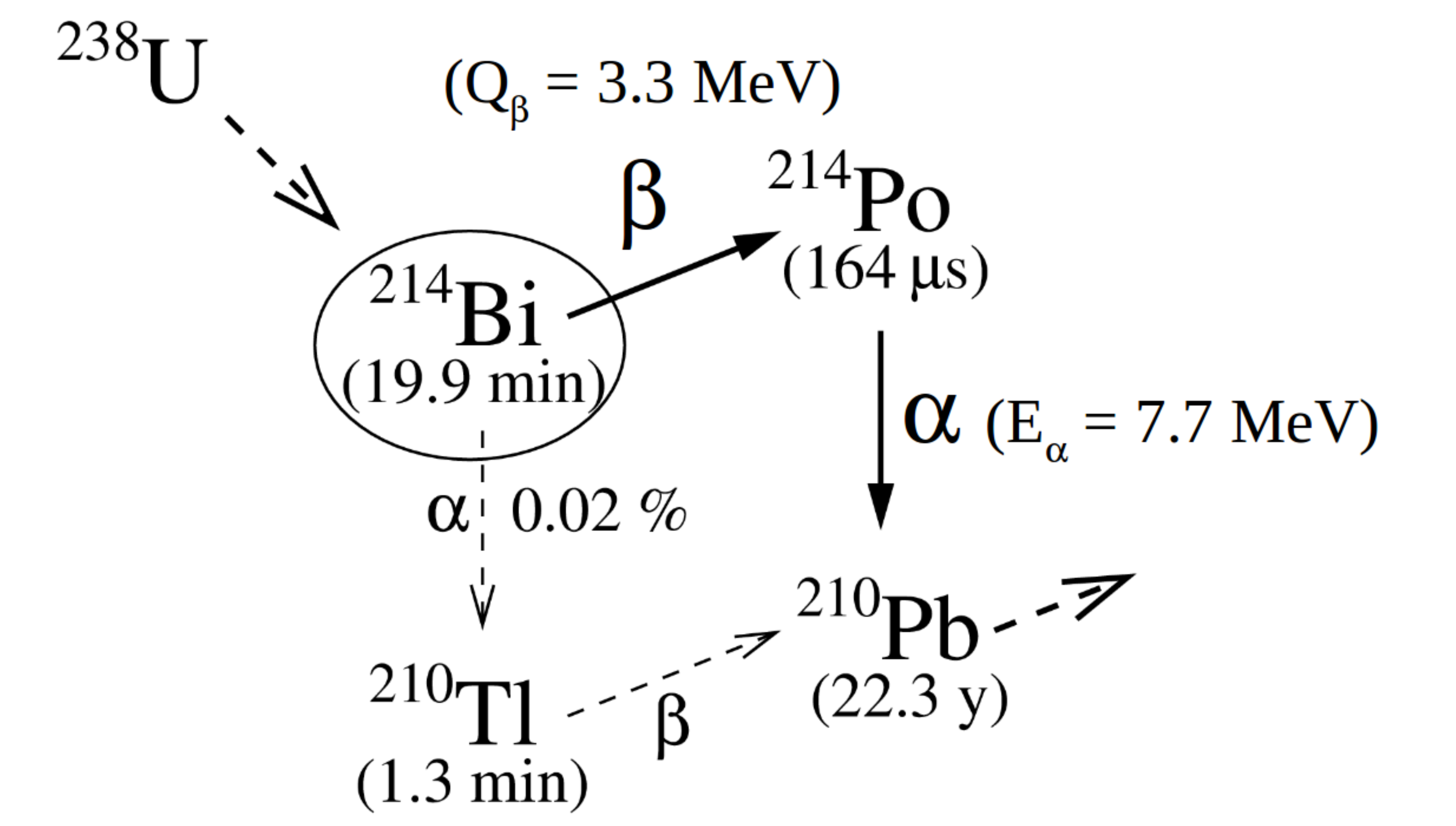}}
\vskip 0.5 cm
\centerline{\includegraphics[width=10cm]{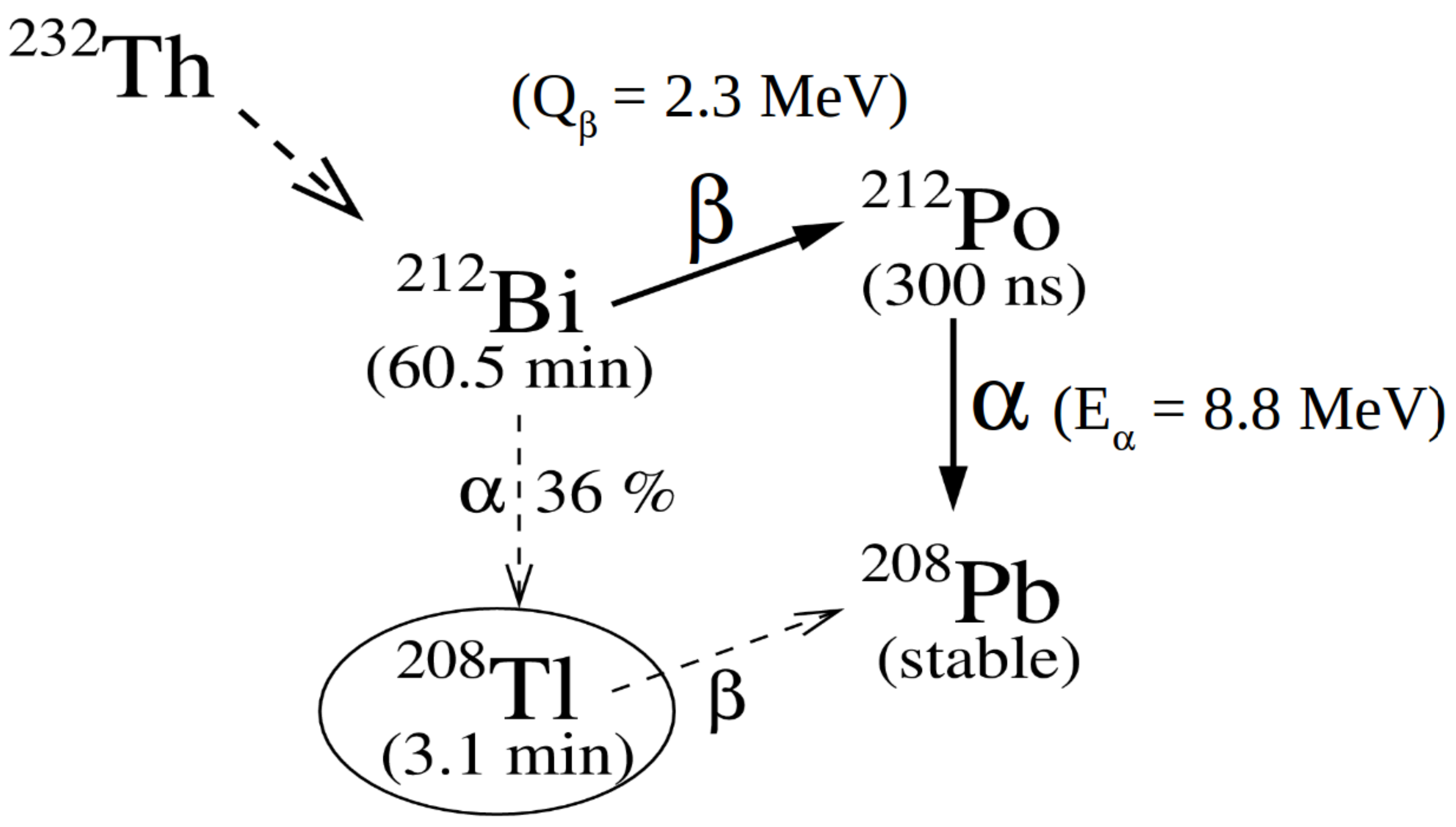}}
\begin{center}
\begin{minipage}[t]{16.5 cm}
\caption{The $^{214}$Bi$\rightarrow^{214}$Po and $^{212}$Bi$\rightarrow^{212}$Po decay sequences considered for the determination of the $^{214}$Bi and $^{208}$Tl activity in the BiPo detectors.}
\label{BiPoSchema}
\end{minipage}
\end{center}
\end{figure}

\begin{figure}
\centerline{\includegraphics[width=14cm]{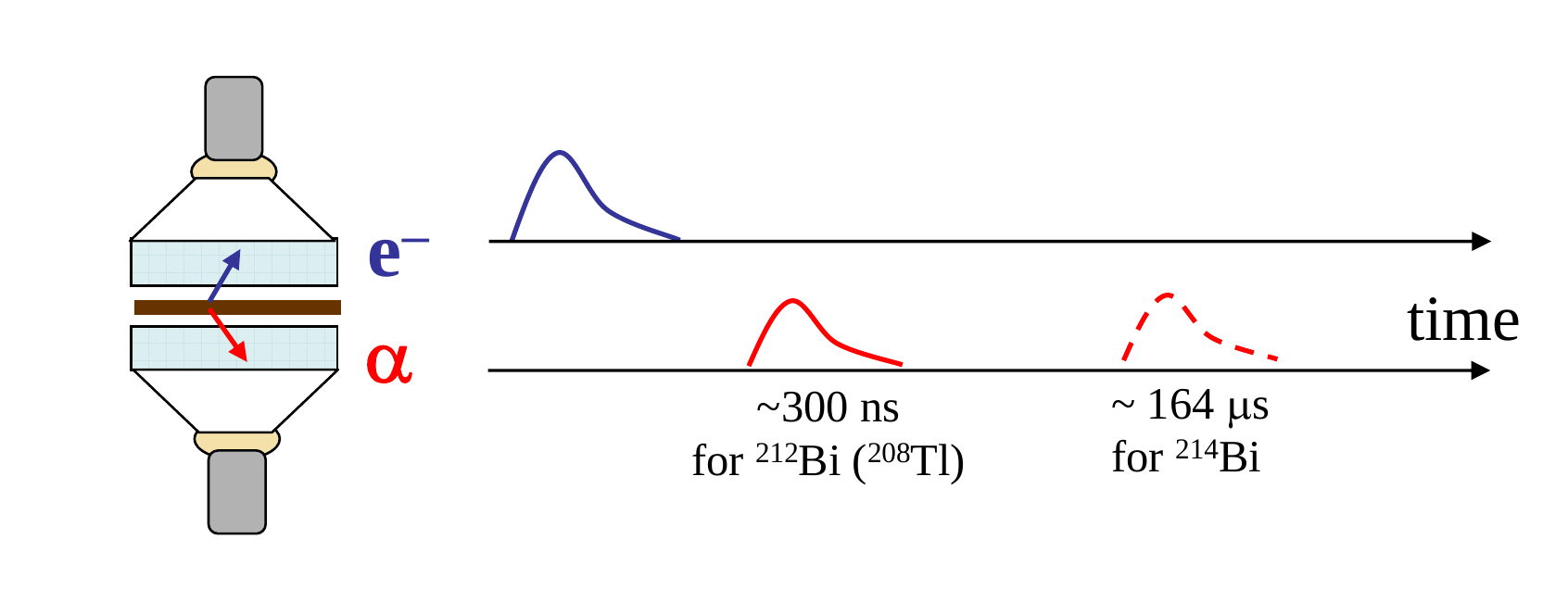}}
\begin{center}
\begin{minipage}[t]{16.5 cm}
\caption{Principle of the BiPo detectors to identify BiPo events: electron and alpha emissions from the sample are detected in opposite scintillators following a particular time sequence (courtesy of the SuperNEMO collaboration).}
\label{BiPoPrinciple}
\end{minipage}
\end{center}
\end{figure}

To determine the contamination levels of $^{208}$Tl and $^{214}$Bi, the concept of the BiPo detectors is to detect the so-called BiPo process appearing in the $^{232}$Th and $^{238}$U chains \cite{bipojinst2008,biponima2013}. This process gives a delayed coincidence between an electron and an alpha particle. In the case of $^{214}$Bi, it decays $\beta^{-}$ to $^{214}$Po, which is an $\alpha$ emitter with a half-life of 164~$\mu$s. The $^{208}$Tl contamination level is determined by the detection of the BiPo process of its parent, $^{212}$Bi, which decays in 64\% of all cases via $\beta^{-}$ emission to $^{212}$Po, an $\alpha$ emitter with a half-life of 300~ns. Figure \ref{BiPoSchema} shows these cascade processes including the time information of the decays and the energy of the emissions. Placing the sample foil between two thin layers of ultra-radiopure plastic scintillators, it is possible to register the BiPo events by detecting the electron energy deposition in one of the scintillators and the delayed alpha signal in the opposite one. Searching for this ``back-to-back'' topology allows to avoid some backgrounds. An array of scintillators coupled to PMTs on both sides (or face-to-face) allows to cover a large surface. The identification of these events, by checking the time of the delayed alpha particle allows to quantify the isotope contamination. This principle of detection is illustrated in Fig.~\ref{BiPoPrinciple}. An important advantage of the BiPo technique is that it allows a direct determination of the activity of the involved isotopes, which are typically a dangerous source of background; other assessment techniques, like those based on mass spectroscopy, quantify the chain progenitors, having different activity if secular equilibrium is broken.

Due to the characteristic signature of the BiPo events, the background sources capable to mimic the expected signal are not abundant, but they have been controlled and reduced as much as possible. Random coincidences due to gamma background could simulate a signal event if they fall within the BiPo time window. Scintillator surface contamination in the sensitive volume could produce BiPo events close to the source foil, being impossible to distinguish them from the signal ones. Scintillator bulk contamination could also be dangerous, but it can be discarded by selecting back-to-back events. The installation of the detector inside a passive shielding, in a radioclean environment and controlling the radiopurity of the detector components, especially of the scintillators, helps to minimize the effect of these background sources.

\subsection{Detector set-up}

Before the installation of the BiPo-3 detector at LSC, other two BiPo detectors based on the same approach but with different designs were constructed and operated at the ``Laboratoire Souterrain de Modane'', in France. Before describing the set-up of BiPo-3, those of the two previous prototypes are briefly presented.

A modular prototype, the so-called BiPo-1 detector \cite{biponima2010}, was running in Modane since 2008 for a few years. It consisted of 20~capsules (with total surface of 0.8~m$^{2}$). Each submodule contained two face-to-face 20$\times$20$\times$0.3~cm$^{3}$ polystyrene-based scintillator plates, coupled by UV Polymethyl Methacrylate (PMMA) light guides to low radioactivity 5-inch PMTs and enclosed inside a gas and light tight box. A deposit of 200~nm of ultra-pure aluminium was evaporated on the surface of the scintillators facing the source foil in order to optically isolate one scintillator from the other. PMT signals were sampled by MATACQ VME digitizing boards (12~bit resolution) at 1~GS/s in 2.5~$\mu$s. The acquisition was triggered each time a PMT pulse reached $\sim$100~keV; the delayed hit search was performed off-line by the analysis of the signals.
A more compact design than that of BiPo-1, using less PMTs and offering spatial resolution was studied, implemented and run also in Modane in the BiPo-2 prototype, consisting of two plastic scintillators BC408 (75$\times$75$\times$1~cm$^{3}$ with total surface of 0.56~m$^{2}$), coupled each to 10 low radioactive 3-inch PMTs.
The main goals of these prototypes were to validate the detection principle and the face-to-face geometry, as well as to measure the surface contamination of the scintillators confirming that the radiopurity levels to achieve the design goals were reached.

\begin{figure}
\begin{center}
\includegraphics[width=14cm]{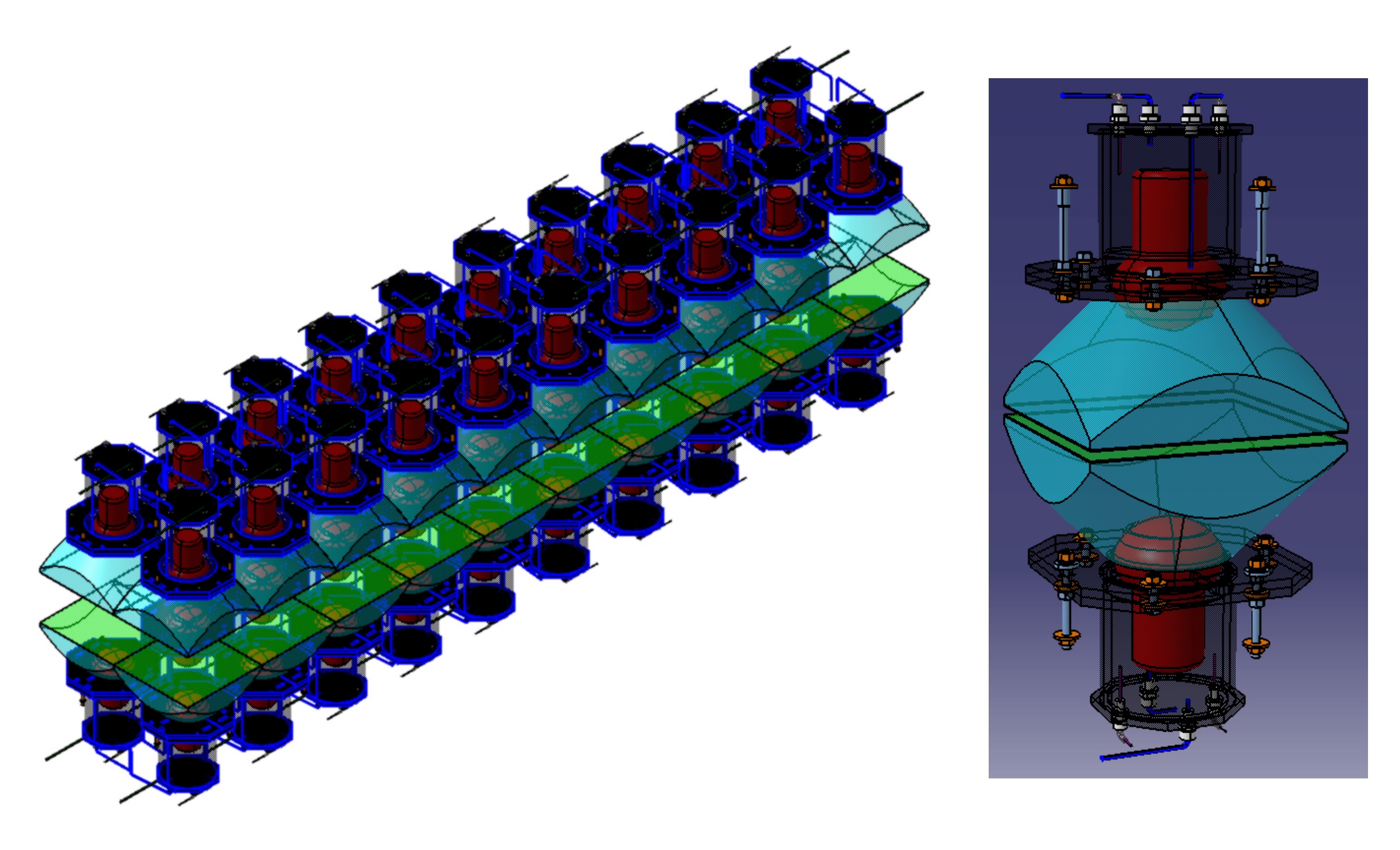}
\begin{minipage}[t]{16.5 cm}
\caption{Design of the assembly of the 40 optical sub-modules inside a BiPo-3 module (left) and of an optical sub-module (right) with the two thin scintillators face-to-face, coupled with PMMA optical guides to low-radioactivity 5-inch PMTs (courtesy of the SuperNEMO collaboration).}
\label{bipo3design}
\end{minipage}
\end{center}
\end{figure}

\begin{figure}
\begin{center}
\includegraphics[width=12cm]{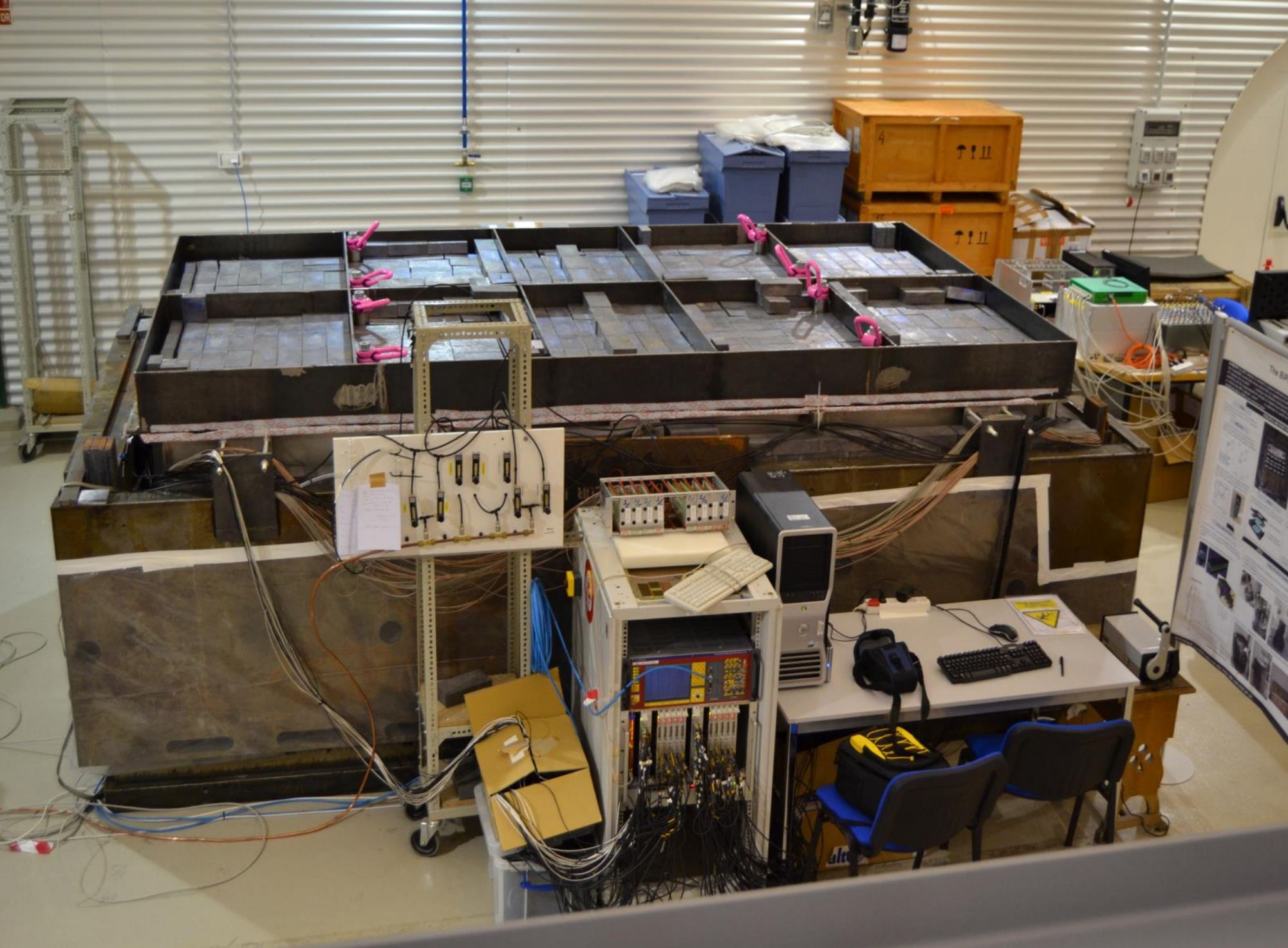}
\begin{minipage}[t]{16.5 cm}
\caption{Picture of the BiPo-3 detector installed in the Hall A of LSC (courtesy of LSC and SuperNEMO collaboration).}
\label{fotobipo}
\end{minipage}
\end{center}
\end{figure}

Following the satisfactory development and operation of the first BiPo detectors, the design and construction of the BiPo-3 \cite{bipojinst2017}  final set-up started. This detector consists of two modules containing 40 optical submodules (positioned in two rows), having a sensitive surface of 3.6~m$^{2}$. The geometry of the detector was optimized to reach the sensitivity levels required for the SuperNEMO source foils in the shortest time possible. Each capsule has two organic plastic polystyrene-based scintillators (30$\times$30$\times$0.3~cm$^{3}$ each) coupled again to low radioactive 5-inch PMTs (model Hamamatsu R6594-MOD) through PMMA optical guides to register the energy and the time of the energy depositions (see Fig.~\ref{bipo3design}). As in BiPo-1, the surface of the scintillators facing the source foil is covered with a 200~nm-thick layer of evaporated ultra radiopure aluminium. The light guide is first wrapped in a Tyvek layer in order to diffuse and collect the light into the PMT and then covered with a black polyethylene film to avoid any optical crosstalk between sub-modules. The modules are held in a pure iron structure and closed by 2~cm-thick iron plates. The materials for the optical submodules and the structure were carefully chosen in terms of radiopurity. The total mass of one BiPo-3 module is approximately 700~kg. The two BiPo-3 modules are placed inside a radon-tight stainless steel tank with dimensions 3.9$\times$2.1$\times$1.4~m$^{3}$ containing also the lead shielding made of low activity lead bricks for a total thickness of 10~cm. Pure iron plates (18~cm-thick) are added under the tank and on its lateral sides. Figure \ref{fotobipo} shows a picture of the BiPo-3 set-up in the Hall A of LSC. Different volumes of the modules are flushed using nitrogen to remove the radon contamination; the total volume of the BiPo-3 detector to be flushed is about 2~m$^{3}$. Twenty MATACQ boards are used to sample the PMT signals and the acquisition system is analogue to that employed for BiPo-1 \cite{bipojinst2017}.

The BiPo-3 modules were installed in Canfranc along 2012 and routine calibrations and background measurements were carried out during the first months. Energy calibrations are performed using different radioactive sources. The insertion of the samples inside the BiPo-3 detector is performed in the LSC clean room; the samples are placed directly on the surface of the lower scintillators and the upper scintillators are moved down until they are in contact with the samples.

\subsection{Results}

\begin{figure}
\begin{center}
\includegraphics[width=14cm]{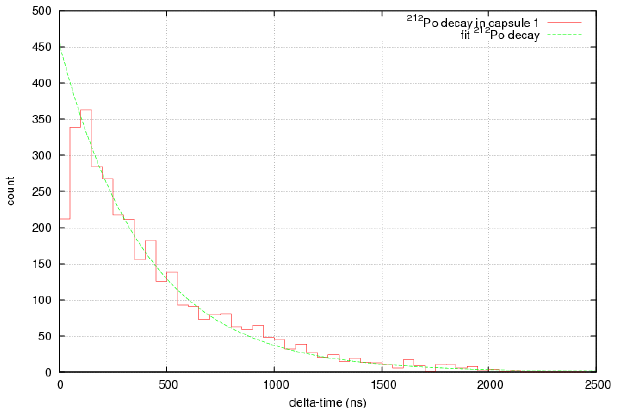}
\begin{minipage}[t]{16.5 cm}
\caption{Example of the distribution of time intervals between $\beta$-$\alpha$ signals in BiPo-1 background measurements and the corresponding exponential fit (courtesy of the SuperNEMO collaboration). $^{212}$Bi half-life (299~ns) is well reproduced.}
\label{timedistribution}
\end{minipage}
\end{center}
\end{figure}

A number of selection cuts are applied to identify BiPo events, as a time window; an example of a measured time interval distribution between $\beta$-$\alpha$ signals for $^{212}$BiPo events is shown in Fig.~\ref{timedistribution} from BiPo-1 data.
Detection efficiencies were calculated by Geant4 Monte Carlo simulations for each sample, and typically are of a few per cent (27\% for scintillators surface when there is no foil). The detection technique and its efficiency were firstly validated in the BiPo-1 detector by measuring a 150~$\mu$m-thick aluminium foil of known activity; for the $^{212}$Bi$\rightarrow^{212}$Po sequence, the activity previously measured by low background HPGe detectors was (0.19$\pm$0.04)~Bq/kg while the result derived from BiPo-1 detector after 160 days of data collection was (0.16$\pm$0.005(stat)$\pm$0.03(syst))~Bq/kg, in good agreement \cite{biponima2010}.
The first phase of the data collection in BiPo-1 was dedicated to the measurement of the $^{208}$Tl activity; from the results corresponding to more than one year of background data, a surface activity of the scintillators of (1.5$\pm$0.3(stat)$\pm$0.3(syst))~$\mu$Bq/m$^{2}$ of this isotope was determined, confirming the capability to screen the selenium DBD foils of SuperNEMO \cite{biponima2010}.

Before starting the analysis of the selenium foils, a characterization program of the BiPo-3 detector, similar to the one developed in BiPo-1, was carried out \cite{bipojinst2017}.
Event identification protocols and detection efficiency were studied, finding a surface background detection efficiency (without any
sample in-between) of (30$\pm$2.7)\% for $^{212}$BiPo events and (26$\pm$2.3)\% for the $^{214}$BiPo ones. Background measurements of several months were performed for each one of the two modules with the final shielding configuration during 2012 and 2013, deriving a $^{208}$Tl activity of (0.9$\pm$0.2)~$\mu$Bq/m$^{2}$ and a $^{214}$Bi activity of (1.0$\pm$0.4)~$\mu$Bq/m$^{2}$ on the surface of the scintillators. Random coincidences between opposite scintillators are other source of background; although negligible for the $^{208}$Tl measurement, they are relevant for $^{214}$Bi. The validation of BiPo-3 operation was carried out also with a calibrated 88~$\mu$m-thick aluminium foil, measured successively in the two BiPo-3 modules. For $^{208}$Tl, the final activity derived from BiPo-3 was (70.5$\pm$0.7(stat)$\pm$8.5(syst))~mBq/kg, in agreement, within uncertainties, with the activity reported by a HPGe measurement, (83$\pm$4(stat)$\pm$15(syst))~mBq/kg; for $^{214}$Bi, the activities determined for each module were (11.9$\pm$2.8(stat)$\pm$1.4(syst))~mBq/kg and (8.5$\pm$1.8(stat)$\pm$1.0(syst))~mBq/kg, again in agreement with the activity reported by the HPGe measurement, (9.2$\pm$3.3(stat)$\pm$0.9(syst))~mBq/kg.

The analysis method used for the background and the aluminium foils measurements was also applied for the $^{212}$Bi and $^{214}$Bi contamination measurements in the first samples of SuperNEMO selenium DBD source foils, studying the spectra of the prompt and delayed energy and the distribution of the time delays. The SuperNEMO foils were in the form of strips, 270~cm long and 13.5~cm wide. To produce enriched $^{82}$Se foils, thin and chemically purified $^{82}$Se powder is mixed with polyvinyl alcohol (PVA) glue and then deposited between 12~$\mu$m-thick Mylar foils. These raw materials (PVA and Mylar) were first measured separately with the BiPo-3 detector to consider their possible contribution; their corresponding $^{208}$Tl and $^{214}$Bi activities were constrained or quantified precisely. The first SuperNEMO enriched selenium foils were measured during 2014 and 2015 and others were analyzed afterwards. For the first four foils, with a total $^{82}$Se mass of 0.52~kg and a total surface area of 1.4~m$^{2}$, the obtained results are the following: (24$^{+20}_{-15}$)~$\mu$Bq/kg of $^{208}$Tl and $<$140~$\mu$Bq/kg (90\% C.L.) of $^{214}$Bi.
Other $^{82}$Se foils were developed  with different purification methods and production techniques in order to reach the required radiopurity. From the obtained results, the sensitivity at 90\% C.L. of the BiPo-3 detector for the measurement of the SuperNEMO $^{82}$Se foils is $<$2~$\mu$Bq/kg for $^{208}$Tl and $<$140~$\mu$Bq/kg for $^{214}$Bi after 6 months of measurement. This confirms that the design goals required for SuperNEMO have been perfectly fulfilled for $^{208}$Tl.

In addition, other samples used by fundamental Physics experiments, like readouts used in Micromegas TPCs, have been already measured with the BiPo-3 detector. Several sheets of different kapton-based materials, related to the construction of Micromegas readout planes in different stages were assayed, setting upper limits to the activities of both $^{208}$Tl and $^{214}$Bi \cite{trexdbd}; the results for Microbulk Micromegas developed at CERN, with limits at the order or below 0.1~$\mu$Bq/cm$^{2}$, have been particularly relevant in the context of the TREX (``TPCs for Rare Event eXperiments") project. Presently, the BiPo-3 detector is being used also for the analysis of samples of the DarkSide experiment, devoted to the direct detection of WIMPs operating a two-phase argon TPC.

In summary, profiting from previous developments, the BiPo-3 detector was installed at LSC to reach sufficient sensitivity to measure the radiopurity of the selenium DBD source foils planned for the SuperNEMO demonstrator. Now, it has become a generic low-radioactivity planar detector, which can measure the natural radioactivity in  $^{208}$Tl and $^{214}$Bi of thin materials (with surface density less than about 50~mg/cm$^{2}$) with an unprecedented sensitivity. It is available to measure samples for various purposes since 2017.

\section{Summary} \label{sumsec}

The DBD is a rare nuclear process observed for several nuclei in the two-neutrino channel. The neutrinoless mode, implying the violation of leptonic number conservation, has not been evidenced, but it could be a unique probe of the nature of neutrinos. Its identification would be outstanding in neutrino Physics, since this would confirm that neutrinos are Majorana particles, inform on the absolute neutrino mass scale and shed light on CP violation. Neutrinoless DBD is presently the only practical way to discover if the neutrino is its own antiparticle. Its observation would have important consequences not only for Nuclear and Particle Physics but also for Astrophysics and Cosmology. DBD experiments with various emitters use semiconductors, scintillators, gas and liquid chambers and bolometric detectors, based on different techniques. Each approach has pros and cons, which makes different experiments necessary (also due to the nuclear uncertainties). The extremely low expected rate of the signal imposes underground operation and the development of specific background rejection techniques.

The Canfranc Underground Laboratory, located under the Spanish Pyrenees, has taken part in the DBD searches during the last three decades. Some small experiments were carried out very successfully at their modest first facilities, while now, being an international, renowned underground laboratory, LSC hosts important projects which can give very significant results in the field. The DBD research performed or underway at LSC by international collaborations comprises a wide range of experimental techniques, from the well-established germanium detectors, to xenon TPCs and low temperature calorimeters. In addition, the BiPo-3 detector built to reach sufficient sensitivity to measure the radiopurity of the DBD source foils for the SuperNEMO demonstrator and operated at LSC, has now become a generic low-radioactivity planar detector offering an unprecedented sensitivity.

IGEX was one of the two world-leading experiments studying the DBD of $^{76}$Ge in the nineties. IGEX operated several natural and enriched HPGe detectors, including three germanium crystals of $\sim$2~kg each, enriched to 86\% in $^{76}$Ge, mounted in ultralow radioactive background cryostats made of electroformed copper. After operation in Homestake and Baksan, the final location of the large IGEX detectors was Canfranc. Following the successful implementation of PSD techniques, the background level achieved was 0.1~counts keV$^{-1}$ kg$^{-1}$ y$^{-1}$ in the region of the transition energy. The final results derived for the half-life of the two modes of DBD are reported in Table~\ref{sumtab}. The corresponding upper limits to the effective neutrino mass ranged from 0.33 to 1.35~eV, depending on the nuclear matrix element considered. Some of the IGEX detectors were used in Canfranc to perform also direct detection of dark matter and to study the neutron background in the laboratory. After decommissioning the experiment, the three large germanium crystals have been used in the GERDA experiment.

The NEXT program is developing the technology of high-pressure xenon gas TPCs with electroluminescent amplification for neutrinoless DBD  searches of $^{136}$Xe. The first phase of the program included the construction, commissioning and operation of various prototypes with masses of around 1~kg, which demonstrated the robustness of the technology, its excellent energy resolution and its unique topological signal. The NEXT-White demonstrator, deploying 5~kg of xenon, implements the second stage of the program and, in 2020, is taking data smoothly in Canfranc. Using NEXT-White, a record energy resolution of 1\% FWHM has been demonstrated at the transition energy of $^{136}$Xe and the topological signature of electrons has been clearly established for background discrimination. NEXT-100 constitutes the next step, being a radiopure detector deploying 100~kg of xenon at 15~bar and scaling up NEXT-White by slightly more than 2:1 in linear dimensions; it could start operation at LSC at the end of 2020. For entering the new phases of the program with tonne-scale, ultra-low background detectors, different approaches are being considered, exploiting to the limit the distinctive capabilities of the NEXT technology (NEXT-HD) and also exploring the daughter Barium ion detection (NEXT-BOLD); the goal is to explore half-lives for the neutrinoless DBD larger than 10$^{27}$~y, as presented in Table~\ref{sumtab}.

CROSS plans to install at LSC arrays of TeO$_{2}$ and Li$_{2}$MoO$_{4}$ bolometers enriched in $^{130}$Te and $^{100}$Mo, respectively. The key idea of the CROSS project is to reject surface events, being a dominant background source in bolometric experiments, by PSD obtained by exploiting Solid-State-Physics phenomena in superconductors. It has been shown using small prototypes in Paris that thin ultra-pure superconductive Al films deposited on the bolometer surfaces act successfully as pulse-shape modifiers. In this way, light detectors used up to now to discriminate surface $\alpha$ particles can be avoided, simplifying a lot the bolometric structure. The simplicity of the read-out makes the CROSS technology easily expandable. The current program is focused on an intermediate experiment to be installed in Canfranc in an existing dedicated facility. The goal is to test the technique with high statistics and to prove the stability and the reproducibility of the CROSS methods; but at the same time, as shown in  Table~\ref{sumtab}, this experiment will be competitive in the investigation of the neutrinoless DBD of the used isotopes.

It can be seen that LSC strongly endorses the recommendations made in 2019 by the Double Beta Decay APPEC Committee \cite{appec}, like ``The search for neutrinoless double beta decay is a top priority in Particle and Astroparticle Physics'', ``A multi-isotope program at the highest level of sensitivity should be supported in Europe in order to mitigate the risks and to extend the physics reach of a possible discovery'' and ``The European underground laboratories should provide the required space and infrastructures for next generation double beta decay experiments and coordinate efforts in screening and prototyping''.

The experiments operating all over the world, with increasing size and sophistication, have not found up to now any evidence of neutrinoless DBD. The challenge for the future is the construction of detectors characterized by a tonne-scale size and an incredibly low background, to fully probe the inverted-hierarchy region of the neutrino masses and unveil some of the unknown neutrino properties. Experiments at LSC will keep on playing an important role in this effort.

\begin{table}
\begin{center}
\begin{minipage}[t]{16.5 cm}
\caption{Summary of the DBD processes investigated at LSC and the obtained results or expected sensitivity in short-term projects. The sensitivity numbers correspond to the conditions described in Secs. \ref{nextres} and \ref{crosssec}, for NEXT-100 (500~kg$\cdot$y) and CROSS (5~y and 4.7~kg of isotope), respectively. }
\label{sumtab}
\end{minipage}
\vskip 0.5 cm
\begin{tabular}{lcccc}
\hline
Isotope & Process & Collaboration & Result  & Reference \\ \hline

$^{76}$Ge &  $(2\beta^{-})_{0\nu}$ &   IGEX & $T_{1/2}^{0\nu}>1.57 \times 10^{25}$~y (90\% C.L.) & \cite{igexfinal} \\
$^{76}$Ge &  $(2\beta^{-})_{2\nu}$ &   IGEX & $T^{2\nu}_{1/2}=(1.45 \pm 0.20) \times 10^{21}$ y & \cite{dbdmorales,amnu98} \\
$^{78}$Kr  & $(EC \beta^{+})_{0\nu}$ &  Zaragoza/INR Moscow & $T_{1/2}\geq5.1 \times 10^{21}$~y (68\% C.L.)& \cite{krypton} \\
&   $(EC \beta^{+})_{2\nu}$  & Zaragoza/INR Moscow & $T_{1/2}>1.1 \times 10^{20}$~y (68\% C.L.)   & \cite{krypton} \\
& $(2\beta^{+})_{0\nu+2\nu}$ & Zaragoza/INR Moscow & $T_{1/2}\geq2.0 \times 10^{21}$~y  (68\% C.L.)  & \cite{krypton} \\
\hline
Isotope & Process & Collaboration & Sensitivity & Reference \\ \hline
$^{100}$Mo &  $(2\beta^{-})_{0\nu}$ & CROSS & $T_{1/2}^{0\nu}>2.8 \times 10^{25}$~y (90\% C.L.) & \cite{cross} \\
$^{136}$Xe &  $(2\beta^{-})_{0\nu}$ & NEXT-100 & $T_{1/2}^{0\nu}>1.0\times 10^{26}$~y (90\% C.L.) & \cite{revHPXe} \\
\hline
\end{tabular}
\end{center}
\end{table}

\section*{Acknowledgements}
I would like to dedicate this article to the memory of three promoters of the Canfranc Underground Laboratory and pioneers of the searches for Double Beta Decay carried out there, who passed away too early but left us an invaluable legacy: Prof. Angel Morales ($\dag$ 2003), Prof. Julio Morales ($\dag$ 2009) and Prof. Jos\'{e} Angel Villar ($\dag$ 2017). I deeply acknowledge Frank Avignone, Eduardo Garc\'{i}a, Andrea Giuliani, Juan Jos\'{e} G\'{o}mez-Cadenas, Pia Loaiza, Christine Marquet, Carlos Peña-Garay and Michel Sorel for their the help and advice and for allowing the use of materials from the Canfranc Underground Laboratory and from the CROSS, IGEX, NEXT and SuperNEMO collaborations. I very much thank the anonymous referees for their helpful comments.

\end{document}